\documentclass[12pt]{article}
\usepackage{setspace}
\usepackage{placeins}
\usepackage{amsmath}
\usepackage{amssymb}
\usepackage{amsthm}
\usepackage{ascmac}
\usepackage{physics}
\usepackage{graphicx}
\usepackage{amsfonts}
\usepackage{makecell}
\usepackage{float}
\usepackage{cancel}
\usepackage{cases}
\usepackage[english]{babel}
\usepackage[hang,flushmargin]{footmisc}
\usepackage{stmaryrd}
\usepackage[top=25truemm,bottom=25truemm,left=20truemm,right=20truemm]{geometry}
\usepackage{bbm}
\usepackage[utf8]{inputenc}
\usepackage{newtxtext}
\usepackage{newtxmath}
\usepackage{threeparttable}
\usepackage{placeins}
\usepackage{makecell}
\usepackage{ulem}
\usepackage{mathtools}
\usepackage{booktabs}
\usepackage{caption}  
\usepackage{longtable}

\usepackage{hyperref}
\hypersetup{
    colorlinks=true,
    linkcolor=black,
    citecolor=black,
    filecolor=black,
    urlcolor=black,
    pdfborder={0 0 0},
    linktoc=all
}

\usepackage{natbib}
\setcitestyle{authoryear,open={[},close={]}}  
\begin{document}

\begin{titlepage}
    \title{Part-Time Penalties and Heterogeneous Retirement Decisions}
    \author{Kanta Ogawa\footnote{A Ph.D. student at Graduate School of Economics, The University of Tokyo. \\
    Contact: ogawa-kanta319@g.ecc.u-tokyo.ac.jp, kantaogawa319@gmail.com \\
    I would like to express my sincere gratitude to Professor Kosuke Aoki at the University of Tokyo for his invaluable guidance as my master's thesis advisor, and to Professor Sagiri Kitao at the National Graduate Institute for Policy Studies for her insightful advice. I also extend my thanks to the Recruit Works Institute, the Social Science Japan Data Archive, the Center for Social Research and Data Archives at the Institute of Social Science, The University of Tokyo, and the Panel Data Research Center at the Institute for Economic Studies, Keio University, for providing the data used in this research.}}
    \date{\today}
    \maketitle
    \begin{abstract}
        \noindent
           Older male workers exhibit diverse retirement behaviors across occupations and respond differently to policy changes, influenced significantly by the part-time penalty—wage reduction faced by part-time workers compared to their full-time counterparts. Many older individuals reduce their working hours, and in occupations with high part-time penalties, they tend to retire earlier, as observed in data from Japan and the United States. This study develops a general equilibrium model that incorporates occupational choices, endogenous labor supply, highlighting that the impact on the retirement decision is amplified by the presence of assets and pensions. Using the Japanese Panel Study of Employment Dynamics, I find that cutting employees' pension benefits reduce aggregate labor supply in occupations with high part-time penalties, reducing overall welfare across the economy.  Furthermore, a commonly used policy measure—extending the pension eligibility age—is also found to decrease both output and welfare.  In contrast, this paper suggests that increasing income tax credits and exempting pension benefits from income taxation can boost labor supply across all occupations. These policies enhance welfare by raising disposable income relative to the reservation wage.
            \vspace{0in}
           \vspace{0in}

        \bigskip
        \end{abstract}
        \setcounter{page}{0}
        \thispagestyle{empty}
        \end{titlepage}
        \pagebreak \newpage

\section{Introduction}
\begin{figure}[H]
        \centering
        \includegraphics[width=0.70\linewidth]{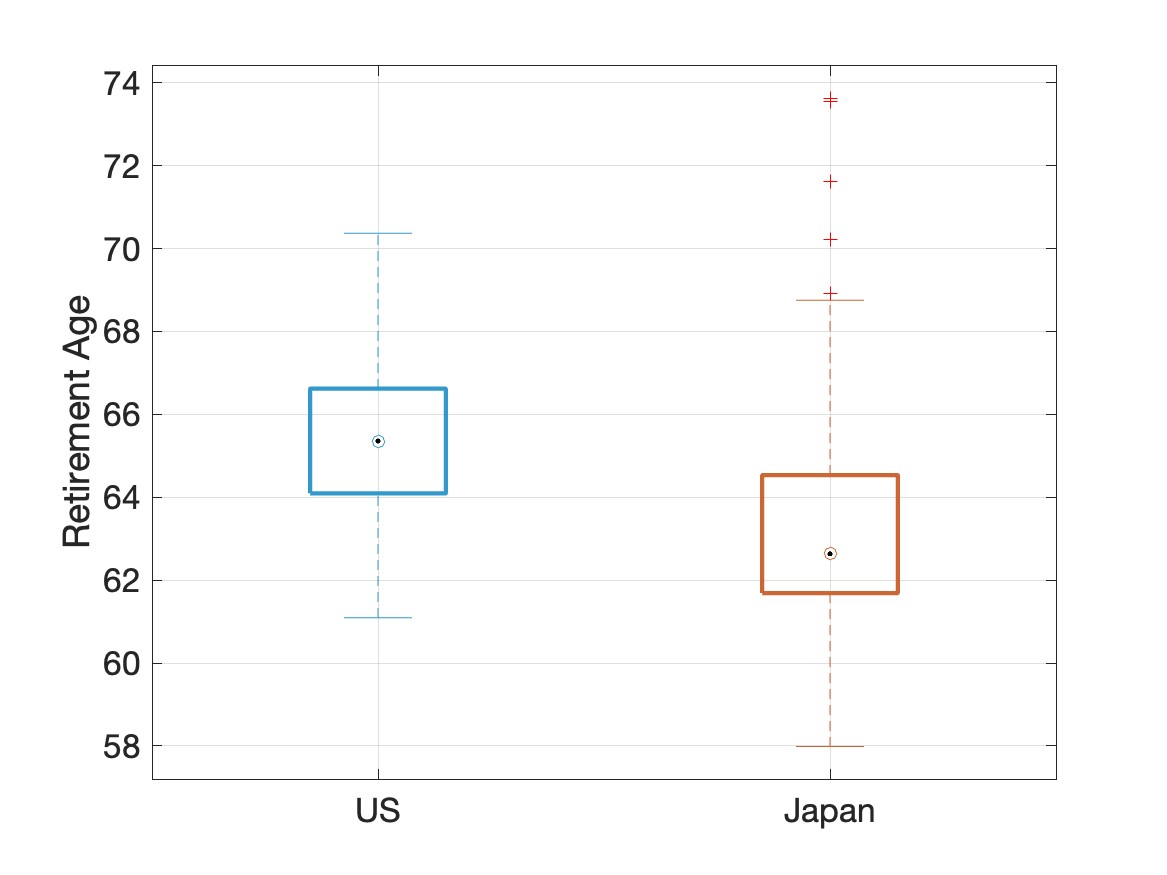}
        \caption{Mean Retirement Timings of Each Occupation: Males, 2015-2019}
        \label{fig:boxplot_ret}
    \end{figure}

Retirement timings vary significantly across occupations. While some professions tend to see workers retiring at sixty, others have mean retirement ages extending beyond seventy, as shown in \autoref{fig:boxplot_ret}. Despite this striking heterogeneity, the factors driving such diverse retirement decisions remain unclear. Understanding these underlying factors is a requisite step to examine the impact of policy measures aimed at increasing the labor supply of older workers across the occupational breakdown.

\begin{figure}[H]
        \centering
        \includegraphics[width=0.70\linewidth]{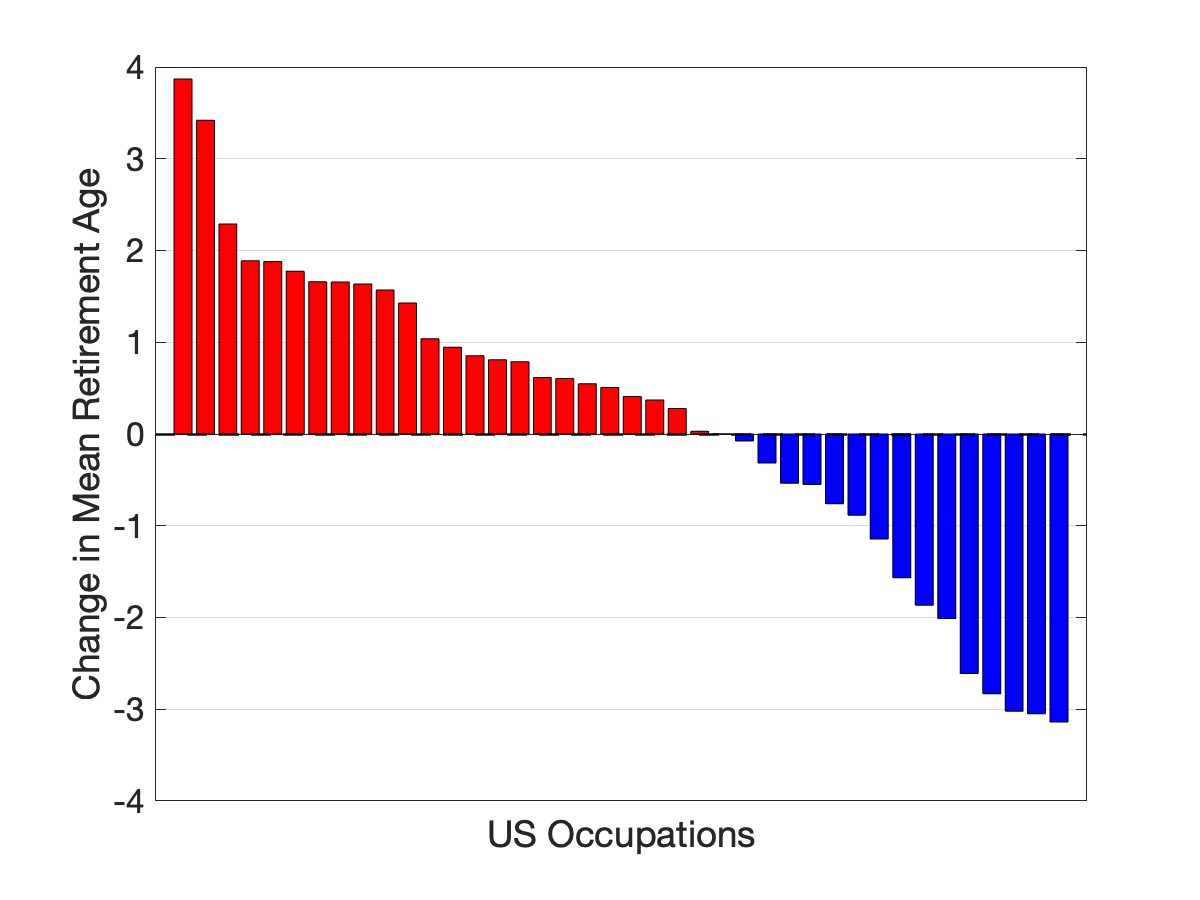}
        \caption{Change in Mean Retirement Age by Occupation: Males Born 1918–1924 vs. 1938–1944 in US}
        \label{fig:Change_ret_age}
    \end{figure}

In aging economies, policymakers attempt to maintain the labor force size amidst demographic shifts. Policy reforms, such as raising the pension eligibility age and reducing pension benefits, have been introduced to encourage older workers to remain in the labor force. However, these reforms can yield strikingly heterogeneous effects across occupations. While some professions can respond positively by extending work participation, others may exhibit negligible or even negative responses, complicating the efficacy of such measures. \autoref{fig:Change_ret_age} illustrates the change in mean retirement age across occupations, comparing male cohorts born between 1918–1924 and those born between 1938–1944, using IPUMS-CPS data [\cite{ipums_cps_2024}]. The sample includes individuals who retired between the ages of 55 and 79, with data spanning from 1970 to 2024. While the earlier cohorts faced a normal retirement age of 65 under the pension system, the later cohorts experienced a gradual increase in the retirement age from 65 to 66 following the pension reforms enacted in the U.S. in 1983. Occupations are sorted by the magnitude of retirement age change, revealing substantial variation. While some occupations exhibit an increase in retirement age, a smaller subset of professions shows a decline. Although this trend may partially reflect broader time trends, it underscores the differential occupational responses to an extension of pension eligibility age.

This issue is particularly pressing as labor force is projected to decline further: working-age population in OECD countries is expected to shrink by 11\% by 2062 compared to 2022 \cite{OECD2023pension}. Given the widespread nature of population aging and its uneven impact on labor supply across occupations, it is plausible that policymakers will need to design retirement policies tailored to specific occupational groups. In this context, I investigate how each occupation responds to changes in retirement policy, including an increase in the pension eligibility age and a reduction in pension benefits.

\begin{figure}[H]
        \centering
        \includegraphics[width=0.70\linewidth]{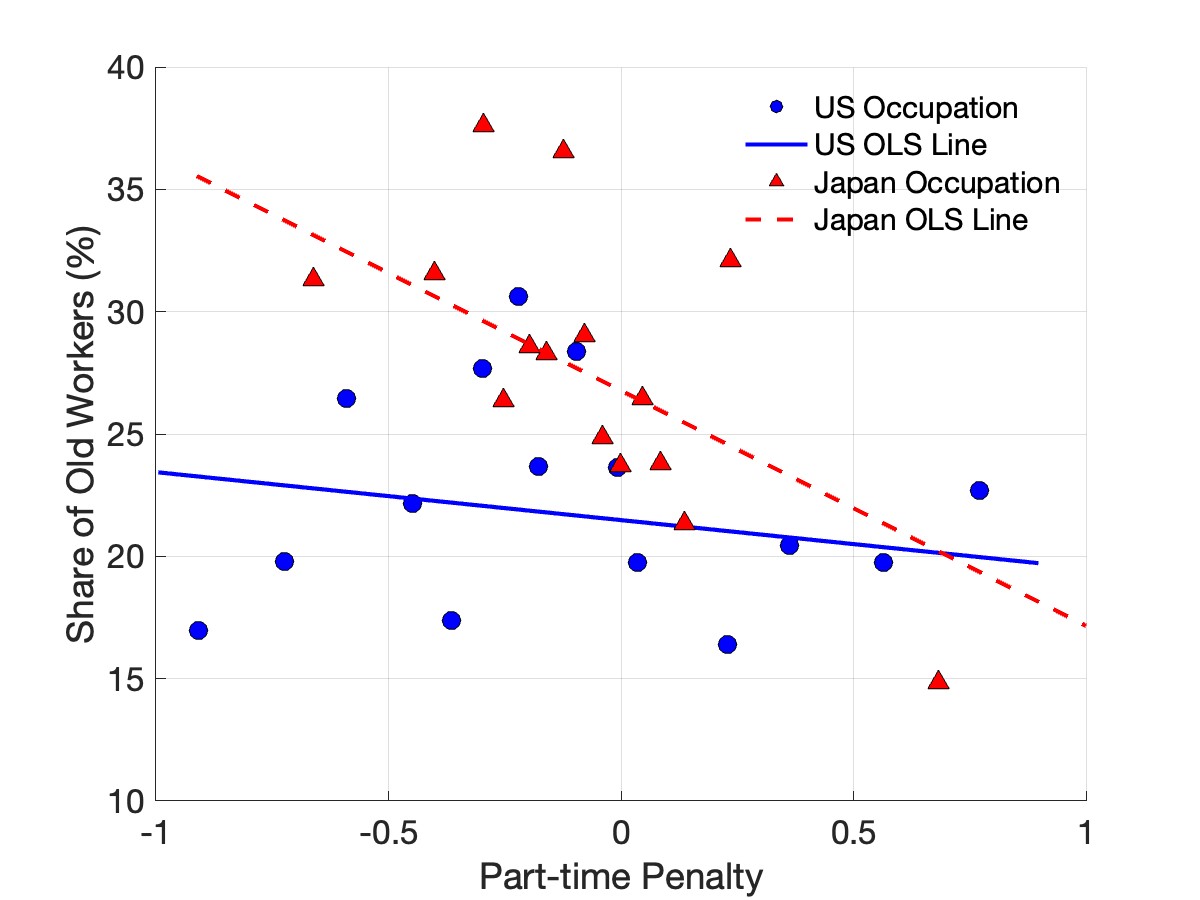}
        \caption{Binscatter of Part-time Penalty and Share of Old Workers: Males, 2015-2019}
        \label{fig:PT_OLD_Share}
    \end{figure}

 This paper is the first to propose a framework to analyze how different occupations respond to policy changes, highlighting part-time penalties—wage reductions experienced by part-time workers compared to their full-time counterparts—as a key source of this heterogeneity. As illustrated in \autoref{fig:PT_OLD_Share}, there is a negative correlation between part-time penalties and share of old workers: a proportion of male workers aged 60 and over among those aged 40 and over. The figure presents a binscatter summarizing occupational trends. The denominator includes both middle-aged and older workers to mitigate the influence of trends in occupational choice. Occupations with smaller part-time penalties tend to have higher rates of old workers as workers face less significant wage reductions when reducing working hours to spend more leisure time as they get older. In other words, in such occupations, workers are likely to retire earlier than in other occupations. This mechanism plays a crucial role in explaining the divergence in retirement timings across occupations.

  As \cite{Blundell_French_Telow_2016} provides a cursory overview of the general factors driving retirement discussed in the literature, much of the existing research has extensively examined retirement decisions in terms of health and social security systems. However, there is a notable oversight regarding the significant wage decreases faced by part-time older workers. While \cite{rogerson2013nonconvexities} argue that part-time penalties discourage older individuals from working part-time and often lead to permanent retirement without experiencing part-time roles, their analysis focuses on the general phenomenon. In contrast, my paper examines the varying degrees of part-time penalties across occupations and their role in shaping differences in retirement ages, incorporating features that enhance the understanding of this heterogeneity.

 To begin with, this study accounts for the occupational heterogeneity of retirement decisions, building on the framework of \cite{goldin2014grand}, which is developed to explain the narrowing gender wage gap. Following  \cite{jang2022nonlinear} and \cite{erosa2022hours}, which formalizes her concept within an equilibrium, the key differences between nonlinear and linear occupations are defined as part-time penalties, experience premiums, occupation-specific productivity, and age penalties. These characteristics characterizes the occupations in the model. While the classification primarily hinges on part-time penalties, the other three factors also play significant roles in explaining economic outcomes and worker behavior.

Furthermore, this paper first uncovers that part-time penalties play a more significant role than they may initially seem, as they interact with assets and pension benefits. The gist of the mechanism is as follows; as \cite{goldin2014grand} notes, occupations with high part-time penalties are typically high-skilled, offering greater compensation. Workers in these occupations tend to accumulate larger assets and expect more generous pension benefits, raising their reservation wage. Older individuals often experience increasing labor disutility due to declining health, the desire to spend more leisure time with their spouses, or the pursuit of hobbies, making them more inclined to reduce their working hours\footnote{Another important consideration is highlighted by \cite{French_and_Jones_2012}, which demonstrates that older individuals have higher labor elasticities compared to middle-aged workers.}. In these circumstances, workers in high part-time penalty occupations face significant wage reductions\footnote{\cite{Daniel_French_2004} further demonstrates that transitioning to part-time jobs results in wage reductions for individuals in their early sixties.
}, making their potential earnings more likely to fall below their elevated reservation wage. Without switching to occupations with smaller part-time penalties, they are likely to exit the labor market permanently, as shown in \autoref{fig:job_switch} in Appendix. Permanent exits are most frequent among those aged 55 to 79, followed by job switches within the same occupation category. Faced with significant part-time penalties\footnote{
 \cite{Ameriks_et_al_2020} examines a similar issue from a different angle, noting that the scarcity of jobs with flexible working conditions discourages older individuals from continuing to work.}, these workers are highly likely to choose permanent retirement. In contrast, workers in occupations with smaller part-time penalties typically continue working, as the wage reduction upon transitioning to part-time work is smaller.

Building on this concept, this research constructs a general equilibrium model of overlapping generations with endogenous labor supply, capturing both extensive and intensive margins with regard to labor, as well as occupational choices. Agents make decisions regarding consumption and savings, balancing the desire to leave a bequest or prepare for longevity while facing a survival shock each period. People unexpectedly become eligible to receive pension benefits at either age 60 or 65. This quantitative framework evaluates the impacts of policy reforms on different generations. In contrast to the literature\footnote{
While \cite{French_2005} and \cite{Fan_Seshadri_Taber_2022} estimate life-cycle models to analyze retirement behavior, their approaches focus on a partial equilibrium. Similarly, \cite{imrohorouglu2012social} demonstrates that social security reforms significantly affect the extensive and intensive margins of older individuals but do not incorporate occupational choices or part-time penalties.
}, my model highlights how changes in retirement behavior can significantly affect the welfare of other generations through shifts in labor supply, saving behavior, and prices, considering a general equilibrium effect.

To classify occupations and compute moments for quantitative analysis, this study utilizes the Japanese Panel Study of Employment Dynamics (JPSED). The JPSED covers more than 200 occupations and provides detailed personal information on each worker, including birth year, sex, education, work history, family status, and more. These rich variables enable precise regressions for classifying occupations. Additionally, the Japanese Household Panel Survey (JHPS/KHPS) supplements the analysis with asset data, which is not available in the JPSED.These datasets allow the study to focus on one of the most rapidly aging populations and to derive policy prescriptions that may be informative for other countries facing similar demographic shifts in the near future. To ensure that similar retirement behaviors are observed in other countries, the IPUMS-CPS (\cite{ipums_cps_2024}), which provides data for the United States, is also employed.

 Nonlinear and linear occupations are classified by regressing hourly wages on a quartic polynomial of working hours, controlling for factors such as age, birth year, family status, and others. The analysis reveals that nonlinear occupations tend to have a lower proportion of older workers, whereas linear occupations exhibit higher rates of older individuals.

The calibration analysis identifies part-time penalties as the primary source of nonlinearity in the model, followed by varying experience premiums across occupations, which reflect the increase in compensation from working additional periods. Moreover, counterfactual experiments are conducted to assess the impacts of policies aimed at increasing the labor supply of older individuals, and I find some interesting outcomes. The results indicate that eliminating the earnings test in pension rules\footnote{
Eliminating the earnings test has been shown to effectively increase the labor supply of older workers, particularly older males, in some countries (U.S.\cite{blinder1980reconsidering}; \cite{friedberg2000labor}; \cite{song2007new}; \cite{haider2010elderly}; \cite{gelber2013earnings}, Canada: \cite{baker1999retirement}, U.K.: \cite{disney2002labour}).} raises the intensive margin of older males by only $2.539\%$, while having minimal impact on welfare. This is consistent with empirical analysis in Japan[\cite{shimizutani2008labor}], although research in other countries also demonstrate this policy change increases extensive margin of older males. Extending pension eligibility\footnote{Numerous analyses also examine the extension of retirement ages across different countries and verify the resulting increase in the extensive margin of older workers (U.S.: \cite{pingle2006social}; \cite{mastrobuoni2009labor}, U.K.: \cite{blundell2003fiscal}; \cite{cribb2013incentives}, Austria: \cite{staubli2013does}; \cite{atalay2015impact}, Switzerland: \cite{hanel2012timing}; \cite{lalive2015does}). } and cutting pension benefits\footnote{Several studies indicate that past pension reforms, including benefit reductions, have increased the labor force participation rate of older individuals (\cite{Anderson1999pension}; \cite{gustman2009changes}; \cite{blau2010can}; \cite{brown2013link}). } increases labor supply but slightly reduces output, leading to a welfare loss for all generations. When the pension eligibility age is extended, the capital supply decreases by $2.366\%$ in contrast to \cite{imrohorouglu2012social} as workers adjust their retirement timing and experience flatter income profiles over time. In contrast, cutting pension benefits has varying effects across occupations: it reduces the labor supply in nonlinear occupations, as the working-age population becomes less motivated to increase working hours to boost future pension benefits. At the same time, it stimulates older workers in nonlinear occupations to remain in the workforce, highlighting the heterogeneous occupational responses to such policy changes. In this case, the former effect outweighs the latter.

I propose several unconventional policies—such as increasing tax credits and exempting pension benefits from income taxation—which are effective in boosting labor supply across both types of occupations, thereby raising output and improving welfare. Notably, these policies reduce tax revenue by less than 3\% in general equilibrium.

  Section 2 discusses empirical facts, and section 3 elaborates on the model. Section 4 presents the calibration results. Section 5 details the counterfactual experiments, and section 6 concludes this paper.

    \vspace{-1em}
     \FloatBarrier
    \vspace{-1em}
    \FloatBarrier
   \footnotetext{The part-time penalty is rescaled such that the average hourly wage is standardized to one.  \\
   "Share of Old Workers" represents the proportion of male workers aged 60 and older among those aged 40 and older in each occupation.}

\section{Empirical Analysis}

     \subsection{Data}
     The Japanese Panel Study of Employment Dynamics (JPSED), compiled by the Recruit Works Institute and released by the University of Tokyo, provides the primary data for occupational classification and model calibration. Spanning the years 2015 to 2022, the dataset includes over 100 individual attributes and approximately 50,000 observations per year, allowing for granular analysis across a range of personal and occupational characteristics.

     For occupational classification, the analysis includes all valid observations of individuals aged 25 to 79 from 2015 to 2022, regardless of sex, resulting in a total sample size of 237,897. Older individuals and females are included in this regression to ensure a sufficient number of part-time workers. Among males under age 60, the vast majority are full-time workers, making it difficult to estimate the part-time penalty. To address heterogeneity, the regression controls for sex, age, survey year, marital status, presence of children, and other personal characteristics.

     For model calibration, the sample is restricted to males aged 25 to 104 from 2015 to 2019 to avoid the heterogeneous impact of the COVID-19 pandemic on retirement behavior, resulting in a final sample of 93,297 observations.

    Household asset data is supplemented with the Japanese Household Panel Survey (JHPS/KHPS), with KHPS starting in 2004 and JHPS in 2009. Asset moments are calculated using data from 2012 to 2019, including financial and housing assets, while pre-2015 data increases observations for older individuals. KHPS and JHPS provide approximately 3,000 and 2,500 annual observations, respectively, ensuring alignment with model calibration.

     For robustness checks, IPUMS-CPS data from 2009 to 2024 is used to examine whether a similar pattern is observed in the U.S. The dataset, which includes approximately 130,000 to 220,000 individuals, provides detailed information on personal attributes and work-related characteristics. Occupations are classified based on this data, allowing for a comprehensive analysis of labor market trends and retirement behavior. The data is used as cross-country, and the sample size is 170,071, and the number of occupations is 173.

     \subsection{Occupational Classification Strategy}

     I classify occupations using regression analysis, addressing limitations in methods used by \cite{erosa2022hours} and \cite{jang2022nonlinear}, which categorize occupations based on male working hours. This approach is unsuitable for Japan, where most males aged 25–59 work full-time, leading to unintended results. For example, truck drivers and barbers, with longer working hours, often exhibit linear wage-hour relationships, while researchers and IT engineers, with shorter hours, show nonlinear patterns. To resolve this, I adopt a regression-based method, controlling for factors like age, sex, and family status, using data from both sexes aged 25–79 to capture more part-time workers.

    I conduct the classification in the following procedure:
     \begin{enumerate}
         \item For each of the more than 200 occupations, I estimate the relationship between hourly wages and weekly working hours using the following regression model. Specifically, I regress hourly wages on a quartic polynomial of weekly working hours, controlling for individual characteristics and fixed effects. Occupations with fewer than 200 observations are excluded from the analysis. After limiting the sample, the number of occupations total to 135.

         \begin{align*}
         y_{i,j} = \beta_{0,j} + \beta_{1,j} h_{i,j} + \beta_{2,j} h_{i,j}^2 + \beta_{3,j} h_{i,j}^3  + \gamma_j X_{i,j} + \epsilon_{i,j}
         \end{align*}

         Here, $y_{i,j}$ denotes the hourly wage of individual $i$ in occupation $j$, and $h_{i,j}$ represents weekly working hours. The vector $X_{i,j}$ includes control variables such as age (as a polynomial), sex, education, marital status, child status, residential area, and time-fixed effects. The coefficients $\beta_{j} = (\beta_{0,j},\beta_{1,j},\beta_{2,j},\beta_{3,j})$ are estimated separately for each occupation $j$, allowing for occupation-specific wage-hour relationships.

         \item  I calculate the residualized hourly wage difference between individuals working 10 hours per week and those working 40 hours per week. I define this wage difference standarized by the average hourly wage of males aged 25-79 as a part-time penalty for occupational classification.
         Part-time penalty in occupation j is calculated as follows:
            \begin{align*}
            \text{Part-time Penalty}_{j} = \Big(\beta_{1,j} (40-10) + \beta_{2,j} (40-10)^2 + \beta_{3,j} (40-10)^3  \Big)/\text{Average Hourly Wage}
            \end{align*}

         \item I classify the top 50 \% of occupations with the largest part-time penalties as nonlinear, and the remainder as linear. Notably, nearly all occupations classified as nonlinear exhibit positive part-time penalties, whereas linear occupations typically show negligible or negative penalties.
     \end{enumerate}

     For example, IT engineers, researchers, pharmaceutical sales representatives, and banking sales representatives are classified as nonlinear occupations. In contrast, construction workers, cooks, and character and CG designers are classified as linear occupations. A detailed table showing the mapping between each occupation and its nonlinear/linear classification is provided in Appendix \autoref{tab:Occ_list_JP}.

     A finer occupational classification is used here, as broader categories encompass a wide variety of jobs with differing part-time penalties. For example, the nature of sales work varies considerably across industries: pharmaceutical sales is classified as nonlinear, whereas insurance sales is linear.

\begin{figure}[H]
       \centering
       \phantomsection  
       \includegraphics[width=0.70\linewidth]{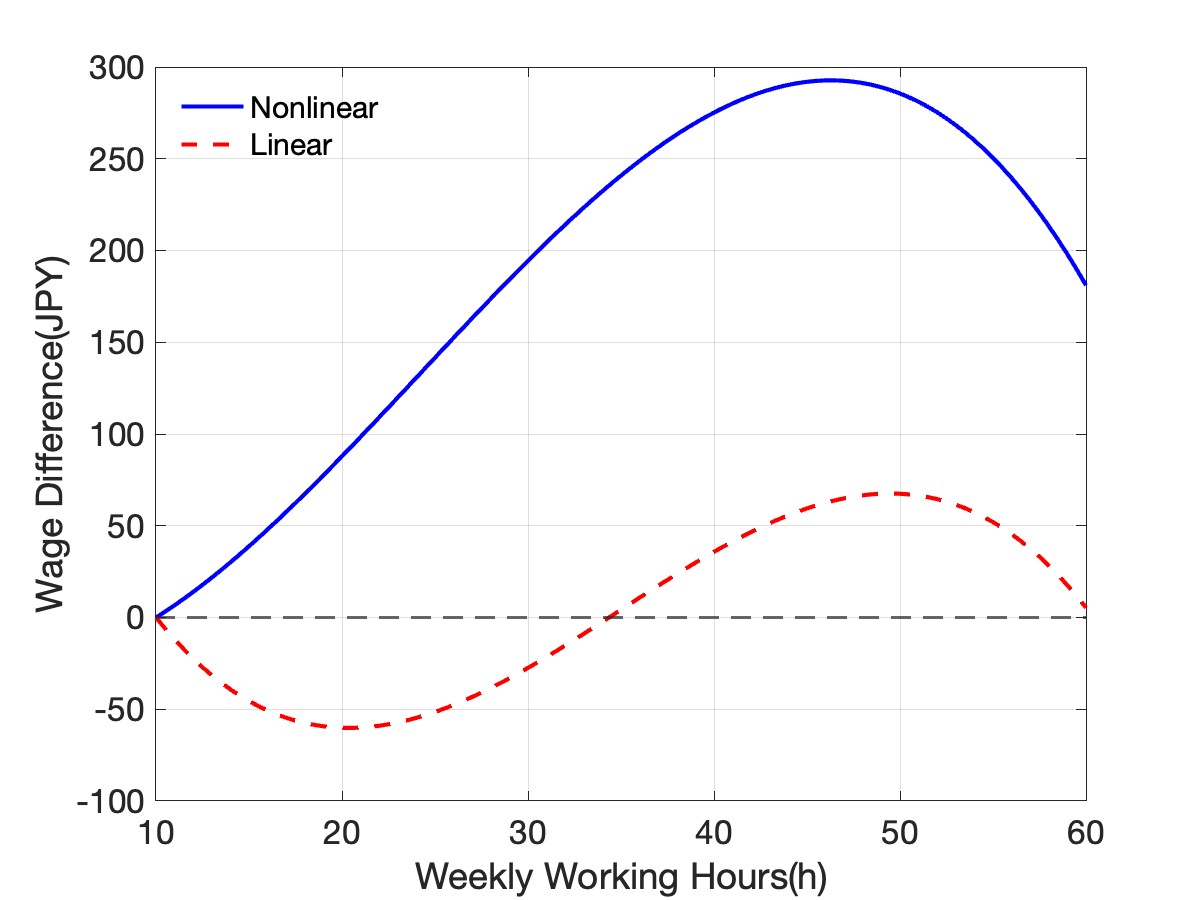}
       \caption{Hourly Wage Change over Working Hours in Japan(nonlinear vs. linear), 2015-2019}
       \label{fig:nonlinear_linear_wage_diff_male}
   \end{figure}

    \autoref{fig:nonlinear_linear_wage_diff_male} illustrates the change in hourly wage, normalized to the hourly wage at 10 hours per week, between nonlinear and linear occupations. The figure shows that hourly wages increase more rapidly with working hours rise in nonlinear occupations, which aligns with the original concept of these occupational categories. Unlike \cite{goldin2014grand}, who assumes no wage changes in linear occupations, I allow for minor wage increases.

\begin{figure}[H]
       \centering
        \phantomsection  
       \includegraphics[width=0.70\linewidth]{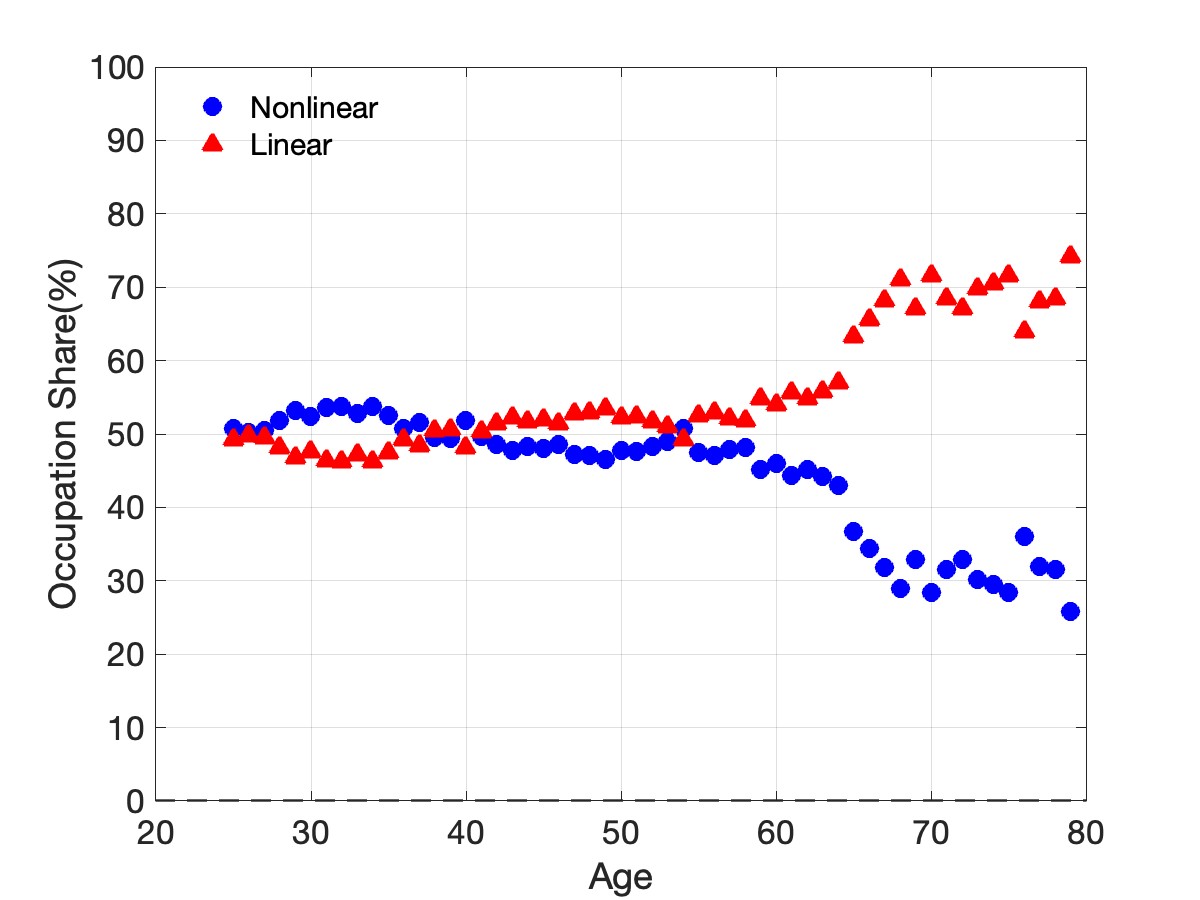}
       \caption{Change of Conditional Occupation Share over Age in Japan (nonlinear vs. linear), 2015-2019}
       \label{fig:nonlinear_linear_occ_share}
   \end{figure}

   The data supports the hypothesis that workers in nonlinear occupations face high part-time penalties, leading to earlier retirement compared to those in linear occupations. As shown in \autoref{fig:nonlinear_linear_occ_share}, the share of nonlinear occupations among working males steadily declines after age 60, while the share of linear occupations rises. \autoref{tab:wwh_dist} shows that non-working rates increase sharply from 4.74\% (ages 25–59) to 77.10\% (ages 70–79). Before age 60, most workers in both occupation types work full-time. After age 60, the share of workers decreases more sharply in nonlinear occupations, while linear occupations see a smaller decline as workers continue with reduced hours.

   The same phenomenon can be observed in the United States, as shown in Appendix 2. Thanks to greater data availability, the U.S. sample begins in 2009, allowing for a larger number of observations than in Japan, where the data starts in 2015. However, the mechanism is less apparent in the U.S. than in Japan. This is likely because workers in linear occupations in the U.S. may exit the labor force for reasons other than part-time penalties, such as health problems, which appear to play a more significant role than in Japan.

\begin{figure}[H]
    \centering
    \includegraphics[width=0.70\linewidth]{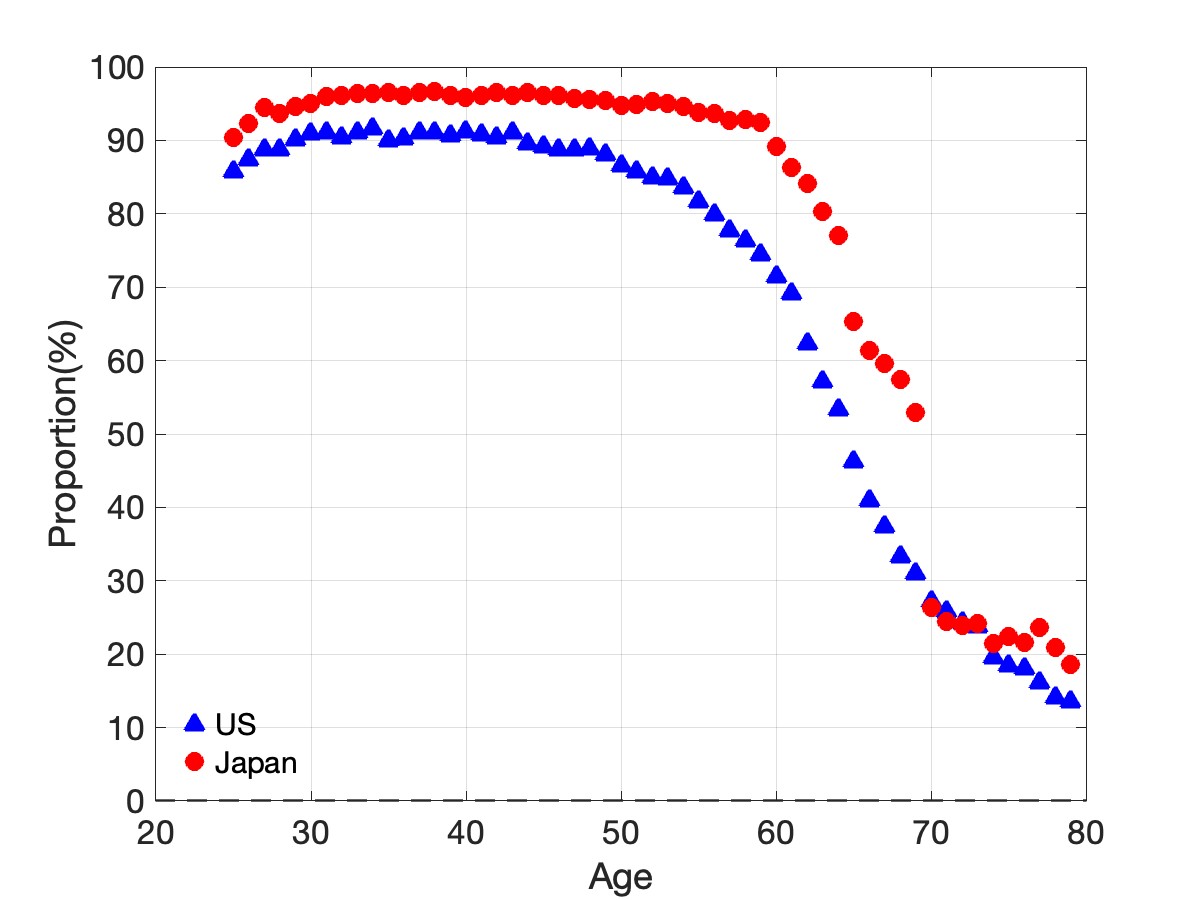}
    \caption{Labor force participation rate, cross-country comparison: Male, 2015-2019}
    \label{fig:LFP_both}
\end{figure}

   As shown in \autoref{fig:LFP_both}, the labor force participation rate among working-age males in the U.S. is around 90\% but begins to decline after age 50. In contrast, the rate in Japan is approximately 95\% and remains high until around age 60. At age 60, the labor force participation rate is 71.42\% in the U.S., compared to 89.16\% in Japan. This discrepancy may reflect differences in health conditions and access to healthcare. These factors are likely to be more prevalent among workers in linear occupations, which primarily consist of low-skilled jobs. In the U.S., such workers may be less able to afford medical expenses and are more likely to experience adverse health conditions, potentially leading to earlier labor force exit in these roles.

   \FloatBarrier

   \vspace{-1em}
   \FloatBarrier
   \vspace{-1em}
   \FloatBarrier
\begin{table}[H]
   \centering
   \caption{Working Hours Distribution by Age in Japan, 2015-2019: Proportion($\%$)}
   \label{tab:wwh_dist}
   \scalebox{0.7}{
       \begin{threeparttable}
           \def\sym#1{\ifmmode^{#1}\else\(^{#1}\)\fi}
\begin{tabular}{lccccccccc}
    \hline\hline
    & \multicolumn{9}{c}{\textbf{Age}} \\  
     & \multicolumn{3}{c}{25-59} & \multicolumn{3}{c}{60-69}& \multicolumn{3}{c}{70-79}\\ 
    \hline 
    & \multicolumn{9}{c}{\textbf{Occupation}} \\
    \textbf{Annual Working Hours(h)} & Nonlinear & Linear & All & Nonlinear & Linear & All & Nonlinear & Linear & All\\ 
    \hline
     $0$ & &  & 4.743 & & & 30.56 & & & 77.10\\
      $(0,1000)$ & 0.2373 & 0.5963 & 0.8336 & 1.499 & 3.278 & 4.777 & 1.411 & 3.632 & 5.043 \\
      $[1000,1500)$ & 0.5223 & 1.325 & 1.847 &2.662 & 5.871 & 8.533 & 1.439 & 3.699 & 5.138 \\
      $[1500,2000)$ & 5.194 & 4.753 & 9.767 & 7.268 & 8.567 & 15.84 &1.653 & 3.198 & 4.851\\
    $[2000,2500)$ & 28.61 & 25.58 & 54.19 &13.42 & 18.82 & 32.24 & 2.071 & 4.140 & 6.211\\
    $2500 \leq$ & 12.71 & 15.74 & 28.45 & 2.505 & 5.549 & 8.054 &0.4214 & 1.236 & 1.657\\ \hline
    $0 <$ & 47.27 & 47.99 & 95.26 & 27.35 & 42.09 & 69.44 &6.574 & 15.91 & 22.48 \\
    \hline\hline
\end{tabular}

       \end{threeparttable}
   }
\end{table}
\footnotetext{ The table provides an unconditional proportion of workers in each category. For example, between the ages of 25 and 59, 4.743\% of individuals do not work, while 95.26\% are employed. 47.27\% of agents work in nonlinear occupations, and 0.2373\% of individuals work less than 1,000 hours per year in nonlinear occupations . }

\section{Model}

This section presents the details of the model. The occupational choice model developed by \cite{jang2022nonlinear} is integrated into a retirement decision framework to account for the heterogeneity in the proportion of older workers across occupations. A distinctive feature of this model is its ability to capture both the extensive and intensive margins of labor supply, while also modeling intergenerational competition for occupational positions— aspects often overlooked in the existing literature, which tends to focus either solely on labor force participation or on individuals around age sixty. These components are important when we consider labor supply of old individuals.  The latter part of this section formally defines the concept of a stationary equilibrium.

\subsection{Demographics}
t denotes age. A continuum of males is born each year at age 25 (t=1) and lives until age 104 (t=80). A fraction of these individuals begins receiving pension benefits at age 60 (t=36), while the remainder start at age 65 (t=41). Pension eligibility is determined at age 60 as a random shock, remaining unknown until that point. All agents retire from the labor force by age 80 (t=56). Each agent faces a survival shock in every period. Upon death, their bequests are evenly distributed among the remaining survivors.

\subsection{Preferences}

Each agent has preferences over consumption and labor supply, which are denoted by $c$ and $h$ respectively. The utility functin is conditional on age and an idiosynacratic intercept of labor disutility function, $\phi$. This paper adopts a separable utility function, following \cite{Fan_Seshadri_Taber_2022}, in contrast to the nonseparable utility specification used in \cite{French_2005} and \cite{french2011effects}. While those studies focus primarily on individuals near retirement age or within the working-age population, this paper extends the analysis to a broader age range—individuals aged 25 to 104. Even after retirement, individuals continue to spend on healthcare and long-term care, despite a potential sharp decline in overall consumption. A separable utility function is therefore well-suited to capture these life-cycle patterns and better reflect the agents' choices of consumtption and hours worked.
 They decide whether to work and, if working, select occupations. Utility is derived from consumption, while disutility arises from labor supply, which consists of two components: a fixed cost of working, $\xi$, and labor disutility, $\Phi$, which starts to increase at $R+1$. All agents have this utility function:
\begin{equation*}
   u (c, h;\phi,t) = \frac{c^{1-\sigma}}{1-\sigma} - \Phi(t) \frac{h^{ 1+ \frac{1}{\gamma}}}{1 + \frac{1}{\gamma}} - \xi \mathbbm{1}\{h > 0\}
\end{equation*}
where the coefficient of labor disutility is given by $\Phi_{i}(t) =\phi_{i} +\kappa (t - R) \mathbbm{1}\{t > R\}$\footnote{I set R=35, which denotes 59 years old in real terms, which means that the labor disutility starts to increase when the agent turns 60 years old.}, and the fixed cost of working is expressed by $\xi$.

\par
Agents also have a bequest motive, and all bequests are equally distributed among surviving agents. This is a key driver of saving behavior among older individuals. This model focuses on accidental bequests\footnote{I refer to \cite{denardi2004wealth} to formulate this bequest motive.} and excludes the inheritance of earnings ability and inter-vivo transfers. The utility derived from leaving a bequest is modeled as:
\begin{equation*}
     \mu(a^{\prime}) = \mu_{1}\bigg(1 + \frac{b(a^{\prime})}{\mu_{2}}\bigg)^{1 - \sigma}
\end{equation*}

where $b(a^{\prime})$ denotes the after-tax bequest. $\mu_1$ represents the agent's concern for leaving bequests, and $\mu_2$ indicates the extent to which bequests are considered luxury goods.

 \subsection{Pension}
 Social security also has a significant impact on retirement decisions. Pension benefits are composed of two terms: the national pension, $\underbar{b}$, which is distributed equally to all agents, and employees' pension insurance, which is based on the agent's past earnings. The mean of the agent’s past labor earnings, $e$, is updated each period using the following equation,subject to an upper bound on labor earnings, $\bar{e}$, when calculating pension benefits. Until age 70, $e$ is updated according to the rules of employees’ pension insurance.
\begin{align*}
    ssb(e) &= \underbar{b} + \rho  e \\
    e &= \frac{e_{-1} \times (t-1) + \min \{\lambda, \bar{e}\}}{t}
\end{align*}
, where $\lambda$ denotes today's post-tax labor income.

\subsection{Efficiency Labor}
    Each agent is compensated by firms based on their efficiency labor, which depends on several factors. A working agent provides an effective labor supply to the representative firm, and wages are paid per unit of effective labor. If an agent of age $t$ and experience $x$ works in occupation $j$ for hours $h$ per week, their income is given by:
    \begin{align*}
        w_{j} \underbrace{m_j(\eta_1) g_{j}(h) z_{j}(x,j_{-1},t)f_{j}(t)}_{\text{Effective Labor Supply}}
    \end{align*}

   \par
   First, the worker selects an occupation based on the occupation-specific productivity, $m(\eta_1)$. A worker draws $\eta_1$, an idiosyncratic value, at birth, which remains constant throughout their lifetime. If a worker is well-suited for occupation $j$, they typically remain in the same job until retirement. The occupation-specific productivity is defined as:
    \begin{equation*}
        m(\eta_j) = \begin{cases}
            e^{\eta_1} & (j = \text{NL}) \\
            1 & (j = \text{L})
        \end{cases}
    \end{equation*}

    NL and L refer to nonlinear and linear occupations, respectively.

    \par
    Second, the part-time penalty is governed by the function $g_j(h)$, which depends on labor supply. As an agent works more hours, their productivity increases, with the degree of this increase varying across occupations. In general, productivity rises more significantly in nonlinear occupations compared to linear ones. The following functional form is assumed when calibrating the parameters:
    \begin{align*}
         g_j(h) = h^{1 + \theta_j}
    \end{align*} where $\theta_j > -1$.

    \par Moreover, productivity also depends partly on experience within the occupation. A worker accumulates one unit of experience for each period of work\footnote{The process of accumulating experience does not require full-time work, as only 8.9\% of males between 25 and 59 work less than 35 hours per week in the data. This full-time work constraint would be necessary if the focus were on analyzing the gender wage gap.}. The experience premium depends on the worker's previous occupation, denoted by $j_{-1}$.

       \begin{equation*}
         z_{j}(x,j_{-1},t) =  \begin{cases}
             1 + \underbrace{\Omega_{j}\min\{x,\bar{x}_j\}}_{\substack{\text{Experience} \\ \text{Premium}}}  & (j = j_{-1})\\
             1  & (j \neq j_{-1})
         \end{cases}
    \end{equation*}

   Experience is updated according to the rule: $x = x_{-1} + 1$ if $j=j_{-1}$ and $x = 0$ if $j\neq j_{-1}$, where $x_{-1}$ represents prior experience. In other words, if the worker remains in the same occupation, they accumulate one additional unit of experience; if they switch to a different occupation, their occupational experience resets to zero. Each unit of an experience adds $\Omega_j$ units of an experience premium until reaching the upper bound, $\bar{x}_j$\footnote{Based on the data, I set $x_{NL} = 35$ and $x_{L} = 29$, which correspond to ages 59 and 53, respectively. These values represent the experience levels at which hourly wages peak in nonlinear and linear occupations. The peak wage ages are interpreted as the maximum effective experience in each occupations.}, which varies across occupations.
    As long as the worker remains in the same occupation, they continue to accumulate experience. However, if the worker switches to a different occupation, their experience resets, starting from $x=0$.

    \par
    Lastly, once an agent reaches age $R+1$\footnote{This is equal to the age assigned for labor disutility function, $\Phi$.}, they incur an age-penalty, representing wage reduction typically associated with the demotion after retirement age. Workers are often reassigned to lower positions, leading to a significant reduction in their wages. This penalty is independent of the worker’s experience and continues to increase until they reach age  $\bar{R}_j(> R)$\footnote{I set $\bar{R}_j = 37$ (61 years old) to match the actual wage decline for old workers. In the data, the sharp wage decline occurs between 60 and 61 years old.}. It should be noted that this component is not a central focus of the paper; rather, it is introduced to match the lifecycle profile of hourly wages by age in the data.

    \begin{align*}
        f_{j}(t) = \exp \bigg({-\pi_j \Big(\min\{t,\bar{R}_j\} - R \Big)\mathcal{I}_{t > R} }\bigg)
    \end{align*}

    \subsection{Household Problem}

    Using these features, I construct a household problem that accounts for both the extensive and intensive margins of labor, allowing workers to choose their occupations. Productivity increases with age until experience reaches $\bar{x}_j$ after which it begins to decline at age $R+1$. Pension eligibility begins unexpectedly at age 60 for some agents, while others start receiving benefits at age 65, with benefit levels determined by their historical earnings. This stochastic timing of pension eligibility serves to better align the model with observed data, capturing the gradual exit of older workers from the labor force. All agents are assumed to retire by age 80 and survive up to age 104, facing age-dependent survival risks.

   \par First, I describe the problem faced by agents between the ages of 25-79. Each period, agents decide on their consumption, next-period assets, labor supply, and, if working, their occupations in the current period. In this problem, a worker decides whether to work in period $t$, given the state $(a,x_{-1}, j_{-1},\phi,\eta_1,t,e_{-1},p)$. Here, $a$ denotes the current assets, and $x_{-1}$ represents the years of experience in the current occupation. The variable $j_{-1}$ determines the experience premium in combination with $x_{-1}$, because if a worker switches occupations, their experience is reset to zero, and they must start from scratch. The variable $\phi$ represents an idiosyncratic coefficient for labor disutility function, $\Phi$, and $\eta_1$ is a parameter in the nonlinear occupation-specific productivity, which determines the worker’s suitability for each occupation. The variable $t$ represents the worker’s age, and $e_{-1}$ is the mean of the worker’s past earnings, which determines the amount of pension benefits. The variable $p$ represents pension eligibility: if $p = 1$, the agent is eligible for pension benefits. All younger agents are ineligible, meaning $p = 0$. A fraction of the population starts receiving a pension at age sixty, while others become eligible at age sixty-five. Agents do not know their exact pension eligibility age until they turn sixty and begin receiving pension benefits if eligible.

   \par
   The post-income tax function, $\mathcal{Y}(\cdot)$, takes three inputs: financial before-tax income, labor income, and pension benefits. I replicate Japan's 2019 tax system, as there were no significant tax reforms during the period used for calibration.
   \par
   People make these decisions simultaneously every period, solving the following maximization problem:

     \begin{equation*}
        V^{Y}( a,x_{-1}, j_{-1},\phi,\eta_1,t,e_{-1},p) \\
        = max \Big\{N( a, \phi,\eta_1, t,e_{-1},p), W( a, x_{-1},j_{-1}, \phi,\eta_1, t,e_{-1},p) \Big\}
    \end{equation*}

    Here, $N(\cdot)$ and $W(\cdot)$ correspond to the value functions of not working and the value of working, respectively. The decision to work or not is represented by $n\in \{NW,W\}$, where NW indicates not working and W indicates working.

    \par
    Next, if working, an agent selects an occupation. $J_j$ is the value of working in occupation $j$, where $j = NL$ and $j = L$ represent nonlinear and linear occupations, respectively. For convenience, I also denote $j=NW$ to represent a non-worker.

    \begin{equation*}
      W( a,x_{-1}, j_{-1}, \phi,\eta_1, t,e_{-1},p) = max \Big\{J_{NL}( a, x_{-1},j_{-1}, \phi,\eta_1, t,e_{-1}, p) ,
       J_{L}( a,  x_{-1}, j_{-1}, \phi,\eta_1, t,e_{-1},p)\Big\}
    \end{equation*}

    The value function of occupation $j$ is clearly defined by:

     \begin{align*}
         &J_{j}( a, x_{-1},j_{-1}, \phi,\eta_1, t,e_{-1}, p) \\
         &= max_{\substack{c, a^{\prime} \geq 0,
        h \in [0, 1] }} \bigg\{
        u(c, h) + \Big(1-S(t)\Big)\mu(a^{\prime})
        + \beta S(t) \mathbb{E}\Big[V^{Y}( a^{\prime},x, j,\phi,\eta_1, t+1, e, p^{\prime} )\bigg]
        \Big\}
    \end{align*}

   \par
   subject to:
   \begin{equation*}
       c + a^{\prime} = a + Tr + B + \mathcal{Y} \Big(ra,w_{j}m_j(\eta_1) g_{j}(h) z_{j}(x,j_{-1},t)f_{j}(t), \mathcal{I}_{p = 1} ssb(e)\Big)
    \end{equation*}
    , where $\beta$ denotes a discounted factor.

    In the budget constraint, $Tr$ and $B$ denote the public lump-sum transfer and bequest from the deceased, respectively. They survive to the next period with a probability $S(t)$ and die with a probability $1-S(t)$, leaving a bequest. The agent earns income from assets, labor, and, if eligible, a pension. They allocate disposable income—after paying social security contributions and taxes on labor and financial income—towards consumption and asset accumulation.

    Alternatively, if not working, the agent faces the value of not working:

    \begin{align*}
        N(a, \phi,\eta_1, t,e_{-1},p)  = max_{c, a^{\prime} \geq 0} \bigg\{u(c,0) +\Big (1- S(t)\Big)\mu(a^{\prime}) + \beta S(t) \mathbb{E} \Big[V^{Y}(a^{\prime},0, 0, \phi,\eta_1, t+1, e, p^{\prime})\Big]\bigg\}
    \end{align*}
   \par
   subject to:
   \begin{equation*}
      c+ a^{\prime} = a + Tr + B(t) + \mathcal{Y} \Big( ra, 0,\mathcal{I}_{p = 1} ssb(e)\Big)
    \end{equation*}

    Lastly, after age 80, agents no longer work and rely solely on interest from assets and pension benefits, which continue to depend on their past earnings.

     \begin{align*}
        V^{O}(a ,t,e_{-1}, p^{\prime})  = max_{c, a^{\prime} \geq 0} \bigg\{u(c,0) + \Big(1- S(t)\Big)\mu(a^{\prime})
        + \beta  S(t) \mathbb{E}\Big[V_{O}(a^{\prime}, t+1, e, p^{\prime}) \Big]\bigg\}
    \end{align*}

   \par
   subject to:
   \begin{equation*}
       c+ a^{\prime} = a + Tr + \mathcal{Y} \Big( ra, 0,\mathcal{I}_{p = 1} ssb(e)\Big)
   \end{equation*}

    \subsection{Representative Firm}
    I elaborate on the settings of the production sector. The representative firm demands capital, nonlinear labor, and linear labor.

    Labor inputs for nonlinear and linear occupations are denoted by $L_1$ and $L_2$, respectively, and are assumed to be complementary. The firm's capital demand is represented by $K$.

      \begin{equation*}
         max_{L_{1}, L_{2}, K} Y - w_{1}L_{1} - w_{2}L_{2} - (r + \delta ) K
     \end{equation*}

    ,where
     \begin{eqnarray*}
          Y &=& AK^{\alpha}(L)^{1 - \alpha} \\
          L &=& \Big[\nu L_{1}^{\frac{\psi - 1}{\psi}} + (1 - \nu )L_{2}^{\frac{\psi - 1}{\psi}}
          \Big]^{\frac{\psi}{\psi - 1}} \\
     \end{eqnarray*}

     The first-order conditions are as following:
     \begin{align*}
         r &= \alpha  \bigg(\frac{L}{K}\bigg)^{1-\alpha} - \delta \\
         w_1 &= (1-\alpha) \bigg(\frac{K}{L}\bigg)^{\alpha} \nu \bigg(\frac{L}{L_{1}}\bigg)^{\frac{1}{\psi}} \\
         w_2 &= (1-\alpha) \bigg(\frac{K}{L}\bigg)^{\alpha} (1-\nu) \bigg(\frac{L}{L_{2}}\bigg)^{\frac{1}{\psi}}
     \end{align*}

    \subsection{Stationary Competitive Equilibrium}
    The model follows a standard life-cycle framework.
    The initial asset $a_0$ is assumed to be zero, and individual states, $s_Y$ and $s_O$, are defined as $s_Y \equiv (a,x_{-1},j_{-1},\phi,\eta_1, t,e_{-1},p) \in \textbf{S}^{Y}$ and  $s_O \equiv (a, t,e_{-1}, p) \in \textbf{S}^{O}$, respectively, where the state spaces, $\textbf{S}^Y$ and $\textbf{S}^O$, are $\textbf{S}^{Y} \equiv \textbf{A} \times \textbf{X} \times \textbf{J} \times \Phi \times \textbf{$\eta$} \times \textbf{T} \times \textbf{E} \times \textbf{P}$ and $\textbf{S}^{O} \equiv \textbf{A}  \times \textbf{T} \times \textbf{E}\times \textbf{P}$.
     \par
     The equilibrium follows a standard definition, where both the capital and two labor markets are clear. In detail, the stationary competitive equilibrium consists of factor prices $(r,w_{1},w_{2})$, allocations to agents,$\{c_Y(s_Y),a_Y^{\prime}(s_Y),h_Y(s_Y)\}_{s_Y \in \textbf{S}^Y}$ and $\{c_O(s_O),a_O^{\prime}(s_O)\}_{s_O \in \textbf{S}^O}$, working decision rules of the agents $\{n(s_Y),j(s_Y)\}_{s_Y \in \textbf{S}^Y}$, value functions, $\{V^Y(s_Y),N(s_Y),W(s_Y)\}_{s_Y \in \textbf{S}^Y}$ and  $\{V^O(s_O)\}_{s_O \in \textbf{S}^O}$
     allocations to firms $(K^D,L_1^D, L_2^D)$, and probability measures , $F_Y(\cdot)$, on the Borel set $\mathrm{B}(\textbf{S}^Y)$ such that $F_Y(\cdot):  \mathcal{B}(\textbf{S}^Y) \to [0,1]$ and , $F_O(\cdot)$, on the Borel set $\mathcal{B}(\textbf{S}^O)$ such that $F_O(\cdot):  \mathcal{B}(\textbf{S}^O) \to [0,1]$.

    \begin{enumerate}
        \item Given ($r, w_{1}, w_{2}$), policy functions $\{c_Y(s_Y),a_Y^{\prime}(s_Y),h_Y(s_Y)\}_{s_Y \in \textbf{S}^Y}$, $\{c_O(s_O),a_O^{\prime}(s_O)\}_{s_O \in \textbf{S}^O}$ and $\{n(s_Y),j(s_Y)\}_{s_Y \in \textbf{S}^Y}$ and value functions $\{V^Y(s_Y),N(s_Y),W(s_Y)\}_{s_Y \in \textbf{S}^Y}$ and  $\{V^O(s_O)\}_{s_O \in \textbf{S}^O}$ solve the household problem.
        \par
        The working decision rules are determined by
        \begin{eqnarray*}
            \text{Extensive Margin}: n(s_Y) &=& argmax\{N(s_Y), W(s_Y)\} \\
            \text{Occupational Choice}: j(s_Y) &=& argmax \{J_{NL}(s_Y), J_{L}(s_Y)\}
        \end{eqnarray*}
        where $n(s_Y) \in \{NW, W\}$ and $j(s_Y) \in \{NL,L\}$.

        \item Given ($r, w_{1}, w_{2}$), $K^D$, $L_1^D$ and $L_2^D$ solve the firm's profit maximization problem as defined above.

        \item The government satisfies the balanced budget constraint by collecting income taxes, inheritance taxes, and social security payments, and distributing lump-sum transfers and pension benefits to individuals.

            \begin{align*}
             &\underbrace{Tax^{\text{Labor}}}_{\text{Labor Income Tax}} + \underbrace{Tax^{\text{Asset}}}_{\text{Financial Income Tax}} + \underbrace{Tax^{\text{Bequest}}}_{\text{Inheritance Tax}} + \underbrace{SSC}_{\substack{\text{Social Security}\\ \text{Contributions}}} \\
             &= \underbrace{\int  \Big(\underbrace{Tr}_{\substack{\text{lump-sum} \\ \text{transfer}}} +   \underbrace{ssb(e) \mathcal{I}_{p = 1}}_{\text{Pension benefit}}\Big) F_{Y}(ds_{Y})}_{\text{between 25 and 79 years old}}
             +  \underbrace{\int  \Big(Tr +   ssb(e) \mathcal{I}_{p = 1}\Big) F_{O}(ds_{O})}_{\text{between 80 and 104 years old}}
        \end{align*}

        \item All the bequest is allocated equally to all the individuals, and this equation holds.
        \begin{align*}
            \int \Big(1 - S(t)\Big) b(a')\, F(ds_Y)
            + \int \Big(1 - S(t)\Big) b(a')\, F(ds_O)
            = B \bigg(
                \int \Big( S(t) \Big)\, F_Y(ds_Y)
                + \int \Big( S(t) \Big)\, F_O(ds_O)
            \bigg)
        \end{align*}

        \item Both asset and labor markets are cleared.
        \begin{itemize}
            \item Asset Market Clearing Condition:
            \par
            \begin{equation*}
                K^{D} = \int a^{\prime}_Y(s_Y)F_Y(ds_Y) + \int a^{\prime}_O(s_O)F_O(ds_O)
            \end{equation*}
            \par
          \item Labor Market Clearing Condition:
          for each $j \in \{NL,L\}$,
            \begin{equation*}
                L^D_{j} = \int   \mathbbm{1}\Big\{j(s_Y)= j\Big\} m_j(\eta_1) g_{j}(h) z_{j}(x,j_{-1},t)f_{j}(t) h(s_Y) dF_Y(s_Y)
            \end{equation*}
        \end{itemize}

          \item The probability measures are consistent with the agent's optimal choices, and therefore, these equations hold.

          \begin{align*}
          \forall B_Y \in \mathcal{B}(\textbf{S}_Y), F_Y(B_Y) &= \frac{
          \int_{(a',\min \{x_{-1} + 1,n_j\},j,\phi,\eta_1,t+1,e) \in B_Y}
          \mathbbm{1}\Big\{n(s_Y) = W\Big\}  \mathbbm{1}\Big\{j(s_Y) = j_{-1}\Big\}S(t)F_Y(ds_Y)
          }
          {\int_{s_Y \in \textbf{S}_Y}S(t)F(ds_Y)}\\
          &+  \frac{\int_{(a',1,j,\phi,\eta_1,t+1,e) \in B_Y}
          \mathbbm{1}\Big\{n(s_Y) = W\Big\}  \mathbbm{1}\Big\{j(s_Y) \neq j_{-1}\Big\}S(t)F_Y(ds_Y)}
           {\int_{s_Y \in \textbf{S}_Y}S(t)F(ds_Y)}\\
           &+  \frac{\int_{(a',0,NW,\phi,\eta_1,t+1,e) \in B_Y}
          \mathbbm{1}\Big\{n(s_Y) = NW\Big\}  S(t)F_Y(ds_Y)}
           {\int_{s_Y \in \textbf{S}_Y}S(t)F(ds_Y)}\\
           \forall B_O \in \mathcal{B}(\textbf{S}_O), F_O(B_O) &= \frac{
          \int_{(a',t+1,e) \in B_O} S(t)F_O(ds_O)
          }
          {\int_{s_O \in \textbf{S}_O}S(t)F_O(ds_O)}
          \end{align*}
      \end{enumerate}

      \section{Calibration}
      \subsection{Externally Set Parameters}

      Calibration is performed to align the model with observed data from Japan and the corresponding calibrated moments. Certain parameters are externally calibrated to capture agents' retirement decisions, drawing on Japan’s economic institutions and relevant literature.

      The capital depreciation rate is set at 8.8\%. To determine the threshold between full-time and part-time work, $\mathcal{F}$, I analyze the distribution of hourly wages across working hours, noting sharp wage increase in wages beyond 35 hours per week. An upper bound on working hours, $\bar{h}$, of 105 hours per week.

      The tax and social security system reflects Japan's economic institutions from 2015 to 2019. National pension benefits and tax rates are based on regulations: the base pension benefit, $\underline{b}$,  is set at 65,008 JPY per month, and the financial tax rate, $\tau_r$, is 20.315\%.  I also incorporate Japan's progressive labor income tax and social insurance systems, including pensions and health insurance, despite their complexity. Agents begin receiving pensions at age sixty with a probability of 0.1372, while others start at age sixty-five. Pension eligibility is determined at age sixty, and the probabilities are independent of all other factors.

      For other parameters, I adopt values from existing literature. The capital share of income, $\alpha$, and the elasticity of substitution between nonlinear and linear occupations, $\psi$, are set to 0.36 and 0.67, respectively, following \cite{jang2022nonlinear}. For the Frisch elasticity, $\gamma$, I use the estimate of 1.50 from \cite{keane2022recent} after reviewing several studies on the parameter.

      I standardize prices using the mean hourly wage and adjust the total factor productivity of the representative firm so that the model's average hourly wage equals 1.

      \subsection{Calibration Result}

      \autoref{tab:cal_summary} summarizes the internally calibrated parameters, computed to match the target moments in the data.

\begin{figure}[H]
	    \centering
	    \includegraphics[width=1.10\linewidth]{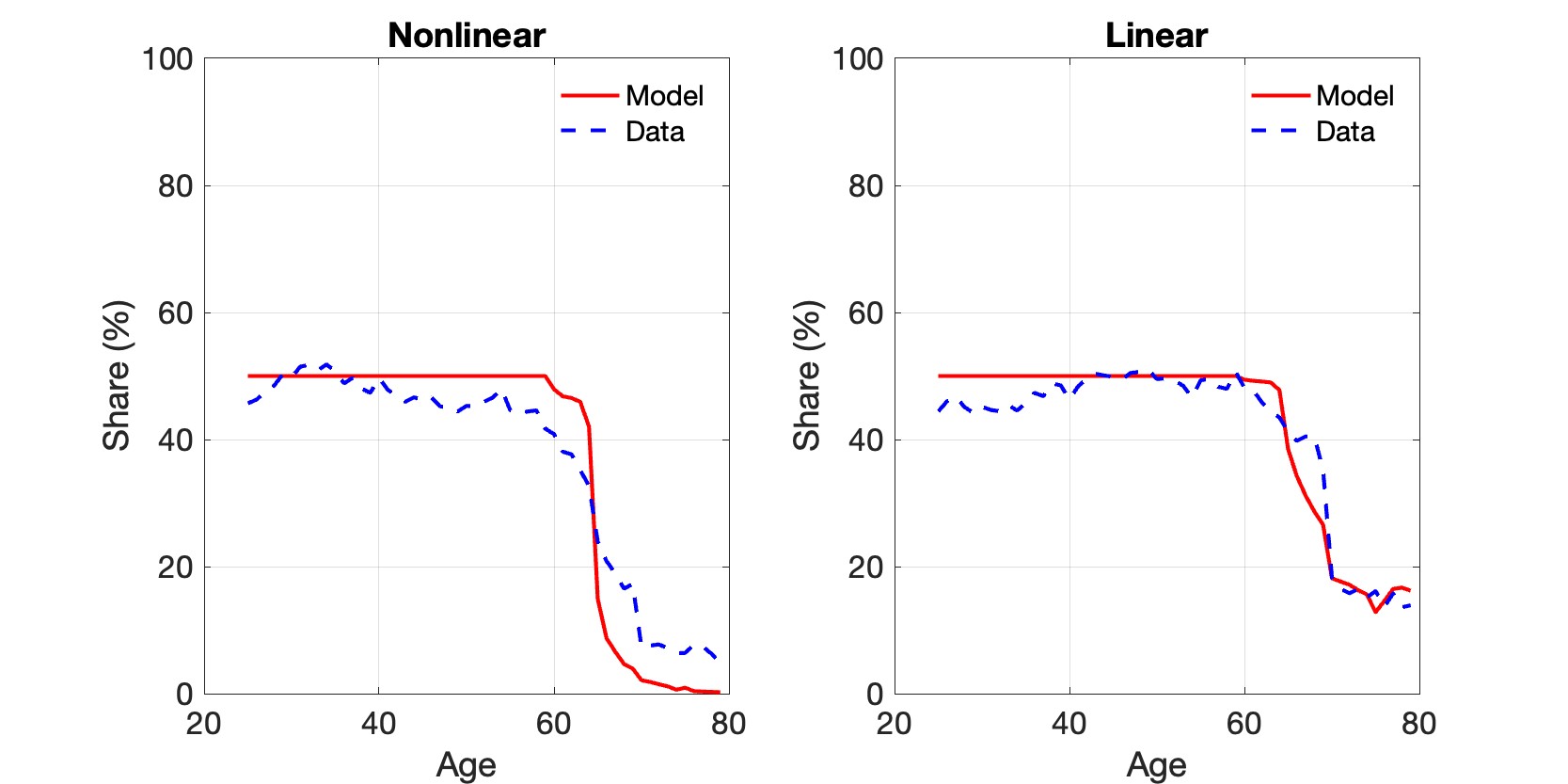}
        \caption{Unconditional Occupation Share in Japan (Model vs Data)}
	    \label{fig:Occ_sharw_undond_M_D}
    \end{figure}

\begin{figure}[H]
	    \centering
	    \includegraphics[width=1.10\linewidth]{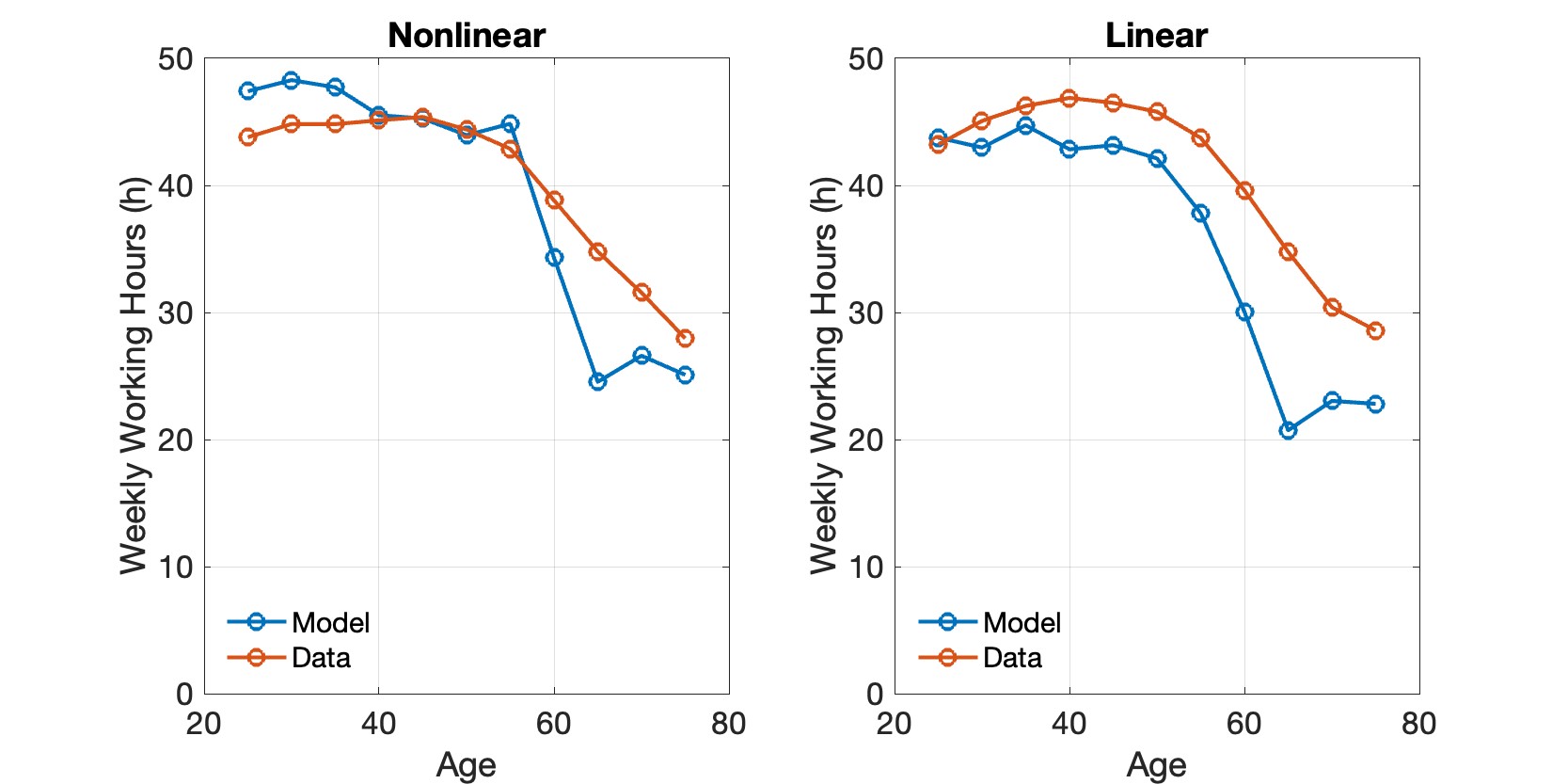}
        \caption{Working Hours per week in Japan (Model vs Data)}
	    \label{fig:wwh_M_D}
    \end{figure}

      \autoref{fig:Occ_sharw_undond_M_D} shows the share of workers in each occupation by age. As labor disutility increases independently of occupation, the model predicts a slight decline in the share of nonlinear occupations, while successfully replicating changes in the share of linear occupations. \autoref{fig:wwh_M_D} also reflects a similar trend between the data and the model. The presence of $\kappa$ motivates workers to reduce working hours after age 60, with the decline plateauing around age 70. At this point, workers are motivated to work  more than 20 hours per week to maintain or increase pension benefits. Working fewer hours results in slightly lower pension benefits, while full-time work subjects them to the earnings test,explaining the preference for part-time work among older individuals. In Appendix, I compare graphs of wage, labor force participation, and asset holdings from the baseline model with the data.

      \FloatBarrier
       \begin{table}[H]
            \centering
            \caption{Internally Calibrated Parameters}
            \label{tab:cal_summary}
            \scalebox{0.65}{
                \begin{threeparttable}
                    {
\def\sym#1{\ifmmode^{#1}\else\(^{#1}\)\fi}
            \begin{tabular}{lclccll}
                \hline\hline
               \multicolumn{3}{l}{\textbf{Parameter}} & \multicolumn{4}{l}{\textbf{Target Statistics}} \\ 
               & Value & Description & Data & Model & Description & Data Source \\ \hline  
            $\beta$            &  1.0094          & Discount Factor            & 0.0107  & 0.01082   & Real interest rate  & \cite{imf2017gdp}\\ 
            $\mu_{\phi}$       & 10.40        &  Mean of working hours     & 0.4226  & 0.4223   & $\mathbb{E}[h| 25 \leq t \leq 59]$ & JPSED\\ 
            $\delta_{\phi}$    & 1.900         &         S.D. of labor disutility   & 0.2665  & 0.2273  & sd(log(h)) (25-59 years old)  & JPSED\\ 
            $\kappa$           & 0.04850        &  Coefficient in $\Phi$ & 44.31 & 44.9187 & LFP rate between 60 and 79 & JPSED
            \\ 
            $\xi$           & 0.3900          &   Labor Force participation cost    & 5.6034  & 5.6090   & Proportion of workers with $h <20$ per week (\%) & JPSED\\ 
            $\nu$              & 0.5570         & Weight of NL laborforce  & 1.1605   & 1.1655   & $\mathbb{E}[wmh^{\theta}gzf|j=NL]/\mathbb{E}[wmh^{\theta}gzf|j=L]$ & JPSED\\ 
            $\sigma_{\eta_1}$  & 0.1600         &  Variance of $\eta_1$    & 0.8884  & 0.8950  & Share of all workers in NL   & JPSED\\ 
            $\theta_{1}$       & 0.4088         &  Curvature of $g_{NL}(\cdot)$    & 0.1408  &  0.1514  &  Part-time penalty (NL)  & JPSED \\ 
            $\theta_{2}$       & 0.2480         &    Curvature of $g_L(\cdot)$        & 0.07766  & 0.07740   & Part-time penalty (L) & JPSED\\ 
            $\Omega_{1}$       & 0.02130         &    Coefficient in $z_NL(\cdot) $     & 0.3868  & 0.3931   & Experience Premium in NL  & JPSED \\
            $\Omega_{2}$       & 0.01900         &    Coefficient in $z_L(\cdot) $      & .3018  & 0.3026   & Experience Premium in L  & JPSED\\ 
            $\pi_{1}$          & -0.1130        & Coefficient in $\log f_NL(\cdot) $      & 0.7564  & 0.7763    & Wage reduction after sixty in NL & JPSED\\ 
            $\pi_{2}$          & -0.06200        & Coefficient in $\log f_L(\cdot) $        & .7922  & 0.8082    &  Wage reduction after sixty in L & JPSED\\ 
            $\mu_{1}$          & -43.00         &     Concern about leaving bequests          & 0.01283  & 0.020216 & Inheritance rate   & JPSED\\ 
            $\mu_{2}$          & 1.700      &    Bequests as luxury goods  & 1.2753   & 1.1254   & 30th pct of Assets (80-105 years old)  & JHPS/KHPS\\
            $\rho$             & 0.3310         &  Coefficient in ssb(M)    & 0.1520  & 0.1519  & $\mathbb{E}[ssb(M)]$ & Ministry of Health, Labour and Welfare \\ \hline\hline
        \end{tabular}
    }

                    \begin{tablenotes}
                       \footnotesize
                        \raggedright
                        \item[1] The efficiency wage of each worker is $\frac{w_j m_j(\eta_1)g_j(h)z_j(x,t)f_j(t)}{h} = w_j m_j(\eta_1)h^{\theta_j}z_j(x,t)f_j(t)$.
\item[2] Part-time penalty in occupation j is defined by $\mathbb{E}[wmh^{\theta}zf|j,  h \geq \mathcal{F}] -\mathbb{E}[wmh^{\theta}zf|j=NL,  h <  \mathcal{F}]$. The difference is used as a measure of part-time penalty instead of ratios because it is a better way to gauge the curvature of the function, $g_j(\cdot)$.
\item[3] Experience Premium in occupation j is also defined by $\mathbb{E}[wmh^{\theta}zf|j, 50 \leq t \leq 59] -\mathbb{E}[wmh^{\theta}zf|j, 25 \leq t \leq 34]$. For the same reason as part-time penalty, I use the difference, not the ratio.
\item[4] Wage reduction after 60 years old in occupation j denotes $\mathbb{E}[wmh^{\theta}zf|j=NL, 60 \leq t \leq 69]/\mathbb{E}[wmh^{\theta}zf|j=NL, 50 \leq t \leq 59]$, which is the ratio of efficiency wage of workers between 60 and 69 years old to that of those between 50 and 59 years old. 
\item[5] LFP rate is an acronym for "Labor force participation rate".
                    \end{tablenotes}
                \end{threeparttable}
            }
        \end{table}
     \vspace{-1em}  
    \FloatBarrier

    \subsection{Source of nonlinearity}

    The parameters driving nonlinearity in the model are analyzed, focusing on the part-time penalty, experience premium, and wage reduction after retirement. Nonlinearity is characterized by three indicators: the wage gap between full-time and part-time workers, the wage gap between workers in their fifties and sixties, and the wage ratio between workers in their seventies and sixties.

    The analysis reduces the values of $\theta$, $\Omega$, and $\pi$ by half and examines the outcomes in partial equilibrium (\autoref{tab:source_NL}). A lower $\theta$ encourages part-time work, increasing both the extensive and intensive labor margins for older workers, particularly in nonlinear occupations, and boosting capital supply. In contrast, $\Omega$ has minimal impact on nonlinearity, while $\pi$ partially affects the part-time penalty.

    \FloatBarrier
    \begin{table}[H]
        \centering
        \caption{Source of nonlinearity (Partial equilibrium)}
        \label{tab:source_NL}
        \scalebox{0.70}{
            \begin{threeparttable}
                {
\def\sym#1{\ifmmode^{#1}\else\(^{#1}\)\fi}
            \begin{tabular}{lclccl}
                \hline\hline
                \textbf{} & \textbf{Baseline} & \textbf{$\theta \downarrow 50 \%$} & \textbf{$\Omega \downarrow 50 \%$} & \textbf{$\pi \downarrow 50 \%$} \\ \hline 
                \textbf{Supply side} &   &  &   &     \\
                 $\Delta$ Labor Supply(NL)(\%) & & +8.649  & -14.61 & +1.082\\
                 $\Delta$ Labor Supply(L)(\%) &   & +7.101 & -15.94 & +1.604\\
                 $\Delta$ Labor Supply(Age:60-79)(\%) &   & +61.55 & -30.80 &  +25.13\\
                 $\Delta$ Capital Supply(\%) & & -7.034 & -10.13 & -2.766 \\
                 $\Delta$ Tax Revenue(\%) & & +5.019 & -15.28 & 0.5190 \\ \hline
                  \textbf{Labor Market Indicators} &   &  &     & \\
                 LFP Rate(Age:60-79)(\%)  & 44.9187  &  77.04  &  37.09  &  53.46 \\
                  Part-time Rate (NL)(\%) & 10.9677  & 48.17 & 13.36 & 12.65\\
                 Part-time Rate (L)(\%) & 31.6372  & 57.18 & 39.01 & 34.56\\
                 Average Working Hours(Age:25-79) &  0.4223 &  0.3806 &  0.4192  &  0.4177 \\ 
                 NL/L: Population ratio   & 0.8950 &  0.9681  & 0.8901  & 0.9035  \\
                 NL/L: Wage ratio   & 1.1655 &  1.184 &  1.1605  &  1.1591 \\
                 Part-time penalty (NL)  & 0.1514 & 0.03233  & 0.1314  &  0.09217 \\
                 Part-time penalty (L)  & 0.07740 &  0.01602 & 0.05307  &  0.05151 \\
                 Experience Premium (NL) & 0.3931 & 0.4790  & 0.09224  & 0.3822  \\
                 Experience Premium (L) & 0.3026 & 0.3525  & 0.1028  & 0.3036 \\
                 Wage reduction after 60(NL) & 0.7763 & 0.7812  & 0.7946  &  0.8527 \\
                 Wage reduction after 60(L) & 0.8082 & 0.8480  &  0.8168  & 0.8561\\
                 \hline\hline
        \end{tabular}
    }

            \end{threeparttable}
        }
    \end{table}
   \footnotetext{NL/L:Population ratio is the ratio of the number of Nonlinear workers to that of Linear workers. Also, NL/L:Wage ratio means the ratio of the average efficiency wage of Nonlinear workers to that of Linear workers.}

\section{Counterfactual Experiment}
Several counterfactual experiments are conducted to evaluate both conventional and unconventional policy reforms. Although conventional policies are likely to reduce welfare, unconventional policies, including income tax reforms, increase welfare, while increasing output.

Note that in all the experiments, the real interest rate is about 1\% , while there is no population growth and no technological growth in the stationary equilibria: the equilibria are dynamically efficient.

\begin{figure}[H]
    	    \centering
    	    \includegraphics[width=0.70\linewidth]{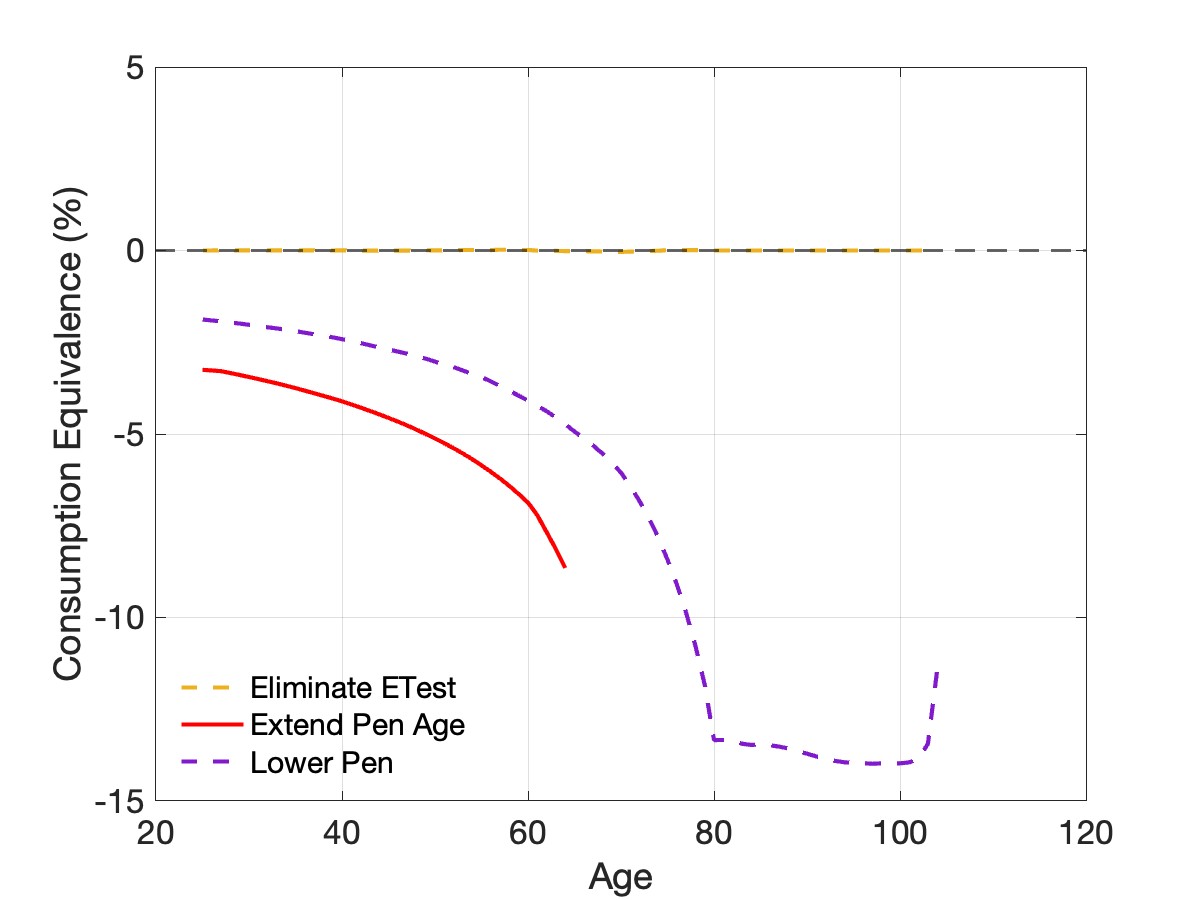}
            \caption{Short-term CEV by Age(Conventional Policies)}
    	    \label{fig:ConsEqv_conv}
        \end{figure}

First, as shown in \autoref{tab:Cexp_conv}, eliminating the earnings test positively impacts consumption equivalence: the short-term consumption equivalence (CEV)\footnote{
 The short-term CEV refers to the consumption equivalent variation for individuals aged 25 immediately following the policy reform, based on the distribution of agents in the baseline model. In contrast, the long-term CEV represents the consumption equivalent variation for individuals of the same age between the two stationary equilibria
 \par
 For example, CEV at age t, denoted as $CEV_t$, is defined as follows: Let $\{c_s^0,h_s^0,a_{s+1}^0\}_{s=t}^{T}$ and $\{c_s^1,h_s^1,a_{s+1}^1\}_{s=t}^{T}$ represent the household's allocation in an equilibrium of the baseline model and the compared allocation, respectively.
 \begin{align*}
     u\Big((1 + CEV_t)c_t^{0},h_t^0\Big) + \Big(1 - S(t)\Big)\mu (a_{t+1}^0) + \sum_{s=t+1}^{T} \beta^{s-t} S(s-1)\bigg\{u\Big((1 + CEV_t)c_s^0,h_s^0\Big) + \Big(1 - S(s)\Big)\mu (a_{s+1}^0)\bigg\} \\
     = u\Big(c_t^{1},h_t^1\Big) + \Big(1 - S(t)\Big)\mu (a_{t+1}^1) + \sum_{s=t+1}^{T} \beta^{s-t} S(s-1)\bigg\{u\Big(c_s^1,h_s^1\Big) + \Big(1 - S(s)\Big)\mu (a_{s+1}^1)\bigg\}
 \end{align*}
}
and long-term CEV at 25 years old are $0.001061\%$ and $0.01940\%$, respectively. Across all generations, the short-term CEV remains positive but small, as shown in \autoref{fig:ConsEqv_conv}. This reform does not affect the extensive margin of older individuals but increases the intensive margin by $2.539\%$. Additionally, it boosts aggregate output, labor supply, capital supply, and tax revenue, though these increases remain under $1\%$.

Second, extending the pension eligibility age by five years delays benefits from age 60 to 65 and from 65 to 70. This reform reduces welfare, particularly for those approaching retirement. While it increases labor supply among older workers in both nonlinear and linear occupations, it reduces capital supply and output. The expanded labor supply also causes younger and middle-aged individuals to work less, leading to wage declines that lower earnings and discourage savings in anticipation of extended working years, as shown in Appendix.
This result arises from capital adjustments driven by price changes and the intertemporal substitution of labor, as workers anticipate longer working lives. When comparing factors, these forces significantly contribute to the observed changes, whereas competition between younger and older individuals for labor supply has only a slight effect. Although \cite{imrohorouglu2012social} demonstrates that extending the normal retirement age increases capital as individuals save to smooth consumption during periods without pension eligibility, this model allows workers to adjust their extensive margin to maintain consumption levels, which consequently reduces the capital supply.

Third, reducing employees' pension benefits by half, resulting in a $28.66\%$ average decrease in pension benefits, lowers welfare in both the short and long term. Output declines by $0.0703\%$, and the increased labor supply from older individuals suppresses wages, discouraging labor supply and earnings for the working-age population. Workers in linear occupations, earning lower wages and holding fewer assets, experience a stronger income effect, increasing their labor supply in linear occupations more than in nonlinear ones.

Next, I examine the effects of unconventional policy reforms on welfare, output, and labor. Reducing $\theta_{NL}$ to the same level as $\theta_{L}$ boosts older individuals' labor supply by $19.47\%$ and raises the LFP rate for those in their 60s and 70s by $10.64\%$. This increase in nonlinear labor supply also enhances capital supply and output, yielding a long-term CEV of $0.8773\%$.

Moreover, I cut $\pi_1$ and $\pi_2$ by half, corresponding to the age-penalty in nonlinear and linear occupations, respectively. This experiment increases labor supply, capital supply, and output but results in a small negative CEV due to the extended working years required.

 Furthermore, enhancing tax credits by 1.5 times, including deductions like the basic and dependents' deductions, lowers the marginal tax rate for workers. This reform boosts older workers' labor participation by $5.35\%$ and increases hourly wages, encouraging continued employment. While tax revenue declines by $2.701\%$, the labor force expands across occupations, driving savings and increasing output by $2.207\%$.

 Lastly, exempting pensions from income tax increases disposable income, boosting labor supply by $8.958\%$ and raising the labor force participation rate by $6.92\%$. This reform enhances welfare and increases output by $0.7708\%$.

These experiments show that conventional policies like extending pension eligibility and cutting benefits reduce welfare (CEV) and face resistance from middle-aged and older individuals, delaying implementation. In contrast, unconventional policies boost welfare and encourage increased labor supply, including greater participation in nonlinear occupations.

\begin{table}[H]
    \centering
    \caption{Policy Experiments}
    \label{tab:Cexp_conv}
    \begin{threeparttable}
        \scalebox{0.7}{
            {
\def\sym#1{\ifmmode^{#1}\else\(^{#1}\)\fi}
\begin{tabular}{lcccccccc}
                \hline\hline
                \textbf{} & Baseline & \makecell{Eliminate \\ ETest} & \makecell{Extend Pen Age \\ by five years}  & \makecell{Lower \\ Pension}   & $\theta_{NL} \downarrow$ until $\theta_L$  & \makecell{$\pi \downarrow$ until 0.5 $\pi$}  & \makecell{Increase \\ Tax Credit} & \makecell{No Tax \\ on Pension} \\ \hline
                 \textbf{Consumption Equivalence} &   &  &   &  & & & &  \\
                  Short term CEV(Age:25)(\%)  &   & +0.001061  &  -4.152  &  -1.880 & -0.008137   &   -0.004552 &   +2.062 & +0.2610\\
                 Long term CEV(Age:25)(\%)  &   &  +0.01940  &  -0.05688  &  -4.441 & +0.8773     & -0.09088 &   + 0.896 &  + 0.9529\\ \hline
                 \textbf{Aggregate Change} &   &  &   &  &   \\
                 $\Delta$Output(\%) &   &  +0.3556 & -0.0811   &   -0.0703 &  +2.247    & +0.5620 & +2.207 &  +0.7708\\
                 $\Delta$ Labor Supply(NL)(\%) &   &  +0.04820 & +1.400 &  -1.1090 &  +5.504   & +0.1957  & +1.763 &  +0.4959\\
                 $\Delta$ Labor Supply(L)(\%) &   &  +0.03589 & +1.023 &   + 1.069&  -0.4785  & +0.9715 & +1.785  &  +0.9071\\
                 $\Delta$ Labor Supply(Age:60-79)(\%) &   &  +2.539 & +68.96  &  +48.92  & +19.47    & +9.324   & +6.010 &  +8.958\\ 
                  $\Delta$ Capital Supply(\%) & & +0.6505 & -2.366 & +0.02090 & +1.442 & +0.5843  & +2.982 & +0.9258 \\ 
                  $\Delta$ Tax Revenue(\%) & & +0.1932 &  -1.256 &  -4.241& +1.574 & +1.574   & -2.701 & -0.1220 \\ \hline
                  \textbf{Labor Market Indicators} &   &  &   &  &   \\
                 Real interest rate (\%)& 1.082  &  1.082 & 1.201 &  1.091 & 1.185 &  1.091 & 1.020 &  1.081 \\
                 LFP Rate(Age:60-79)(\%)  & 44.92  &  44.78 & 62.55  &  79.63 & 55.56  & 48.33 & 49.27 &  51.84\\
                 Part-time Rate (NL)(\%) & 10.97  & 9.800  &  24.24 &  31.17 & 41.69    & 10.40  & 11.98 &  11.89\\
                 Part-time Rate (L)(\%) & 31.64  & 31.25  &  39.10  &  50.23  & 32.11     & 33.56  & 34.69 &  35.93\\
                 AVG Working Hours(Age:25-79) &  0.4223  & 0.4225  &   0.4009  &  0.4029  & 0.4039  & 0.4213 & 0.4258 & 0.4218\\
                 NL/L: Population ratio   & 0.8950  & 0.8929  & 0.9314   &   0.8992 & 1.018    & 0.8924 & 0.8718 & 0.8811\\
                 NL/L: Wage ratio   & 1.1655  &  1.166  & 1.152   &  1.154  & 1.198     & 1.164  & 1.167 & 1.165 \\ 
                 Part-time penalty (NL)  & 0.1514  & 0.1303  & 0.02576   & 0.1457  &  0.03690  & 0.1231 & 0.1531 &  0.1501\\
                 Part-time penalty (L)  & 0.07740  & 0.07689  &  0.04881  &  0.06888 &  0.07138  & 0.06779   & 0.08553 & 0.08731 \\
                 Experience Premium (NL) & 0.3931  &  0.3785 & 0.3552   & 0.3746  & 0.4262   & 0.3734 & 0.4025 & 0.3791\\
                 Experience Premium (L) & 0.3026  & 0.3050  & 0.2878   & 0.2942 & 0.3043   & 0.3007 &  0.3128 & 0.3010\\
                 Wage reduction after 60(NL) & 0.7673  &  0.7960 & 0.9192   &  0.7212  & 0.7787  & .8121  &  0.7645 & 0.7668\\
                 Wage reduction after 60(L) & 0.8082  &  0.8095  &  0.9094  &  0.8057 & 0.8173   &0.8258  & 0.7988 &  0.8003\\
                 \hline\hline
        \end{tabular}
    } 
        }
        \begin{tablenotes}
            \footnotesize
            \raggedright
            \item[1] Apart from the short-term CEV, all indicators reflect either the changes in the stationary equilibrium relative to the baseline model \\
or the levels within the stationary equilibrium achieved in the experiment.
\item[2] "Eliminate ETest" denotes eliminating earnings test.
\item[3] "Extend Pend Age by five years" means extending pension eligibility age by five years.

        \end{tablenotes}
    \end{threeparttable}
\end{table}
\vspace{-1em}
          \FloatBarrier
       \footnotetext{
The CEV for the 'Extend Pension Age' policy is plotted only up to age 64 because individuals aged 65 to 69 lose pension eligibility under this reform, which is embedded within the value functions.}
     \FloatBarrier
    \begin{figure}[H]
    	    \centering
    	    \includegraphics[width=0.70\linewidth]{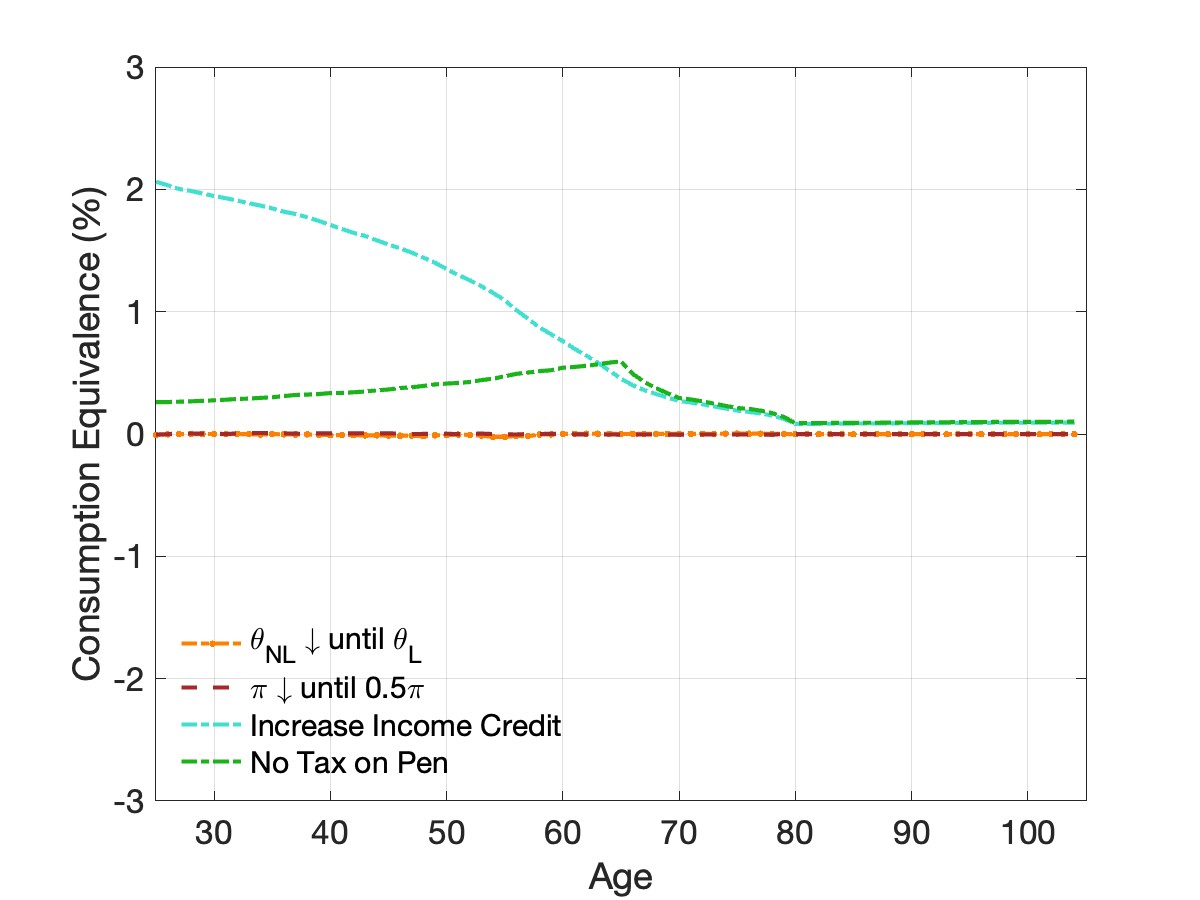}
            \caption{Short-term CEV by Age(Unconventional Policies)}
    	    \label{fig:ConsEqv_org}
        \end{figure}

        \section{Conclusion}

        This study reveals that workers in nonlinear occupations tend to retire earlier than those in linear occupations, based on analyses using JPSED and JHPS/KHPS data. This disparity arises from the high part-time penalties in nonlinear occupations, further magnified by the presence of pensions and assets, which elevate the reservation wage for workers in these occupations.

        This study also demonstrates that while reducing pension benefits decreases labor supply in nonlinear occupations, tax-based reforms—such as increasing tax credits and exempting pension benefits from income taxation—effectively enhance both the intensive and extensive labor supply margins among older workers. These heterogeneous effects stem from the interplay between income and substitution effects across different occupational types. Therefore, policy design should carefully account for such occupational heterogeneity in behavioral responses.

        For future research, it is important to examine additional factors that influence age-friendliness across occupations, such as working conditions and health status. It is also necessary to justify the use of linear or nonlinear occupational structures on the production side. Conceptually, workers in nonlinear occupations tend to earn a premium for working full-time because their tasks—such as responding to client calls at any time or attending meetings—are difficult for firms to reallocate or substitute. A more comprehensive assessment of these dimensions could inform the design of more effective and targeted policy interventions.

        Moreover, examining the labor supply of older female workers has become increasingly important. In 2023, while 45.56\% of women aged 25–59 were employed part-time, their labor force participation rate stood at approximately 81.91\%. As women continue to comprise a growing share of the labor market and approach retirement age, understanding their retirement behavior alongside that of men is essential for designing inclusive and comprehensive labor market policies.

\bibliographystyle{abbrvnat}
\bibliography{../ref}
The data for this secondary analysis, "Japanese Panel Study of Employment Dynamics(2016-2023)\footnote{This questionnaire is conducted every January and it asks people about the last year. Thus, I regard the data of a certain year as the information of the previous year.}, Recruit Works Institute," was provided by the Social Science Japan Data Archive, Center for Social Research and Data Archives, Institute of Social Science, The University of Tokyo.\\
The data for this analysis, Japan Household Panel Survey (JHPS/KHPS), was provided by the Panel Data Research Center, Institute for Economic Studies, Keio University.

\section*{Appendix}
  \subsection*{A.1. Data Description in Japan}
    \begin{figure}[H]
        \centering
        \includegraphics[width=0.70\linewidth]{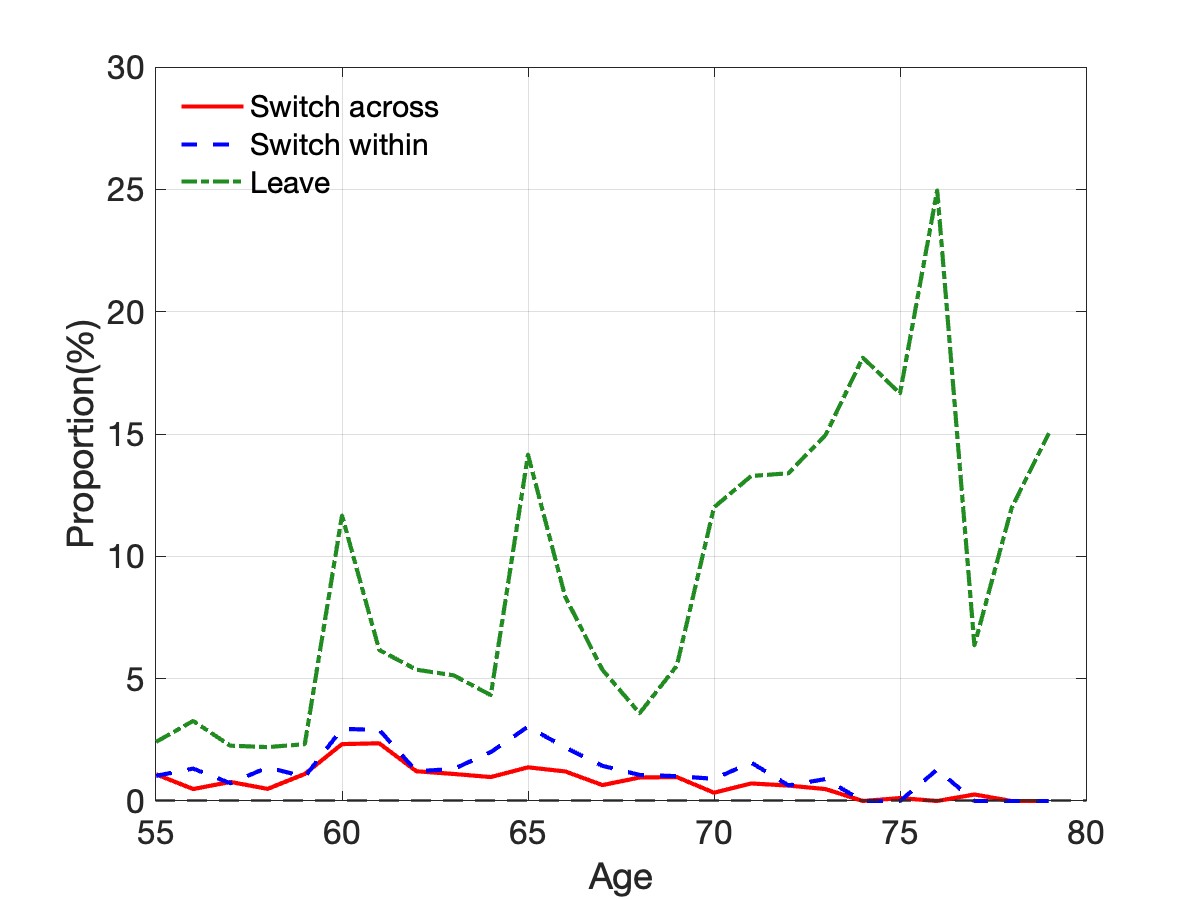}
        \caption{Unconditional Proportion of Workers' Choices After Quitting Jobs in Japan: Male, 2015-2019}
        \label{fig:job_switch}
    \end{figure}
   \footnotetext{"Leave" represents the proportion of workers who exit the labor force after leaving their occupation. "Switch across" refers to transitions between different occupation types; for example, a nonlinear worker moving to a linear occupation. "Switch within" indicates job changes within the same occupation category; for instance, a nonlinear worker moving to another nonlinear job.}
   \vspace{-1em}
   \FloatBarrier
   \begin{figure}[H]
    \centering
    \includegraphics[width=0.70\linewidth]{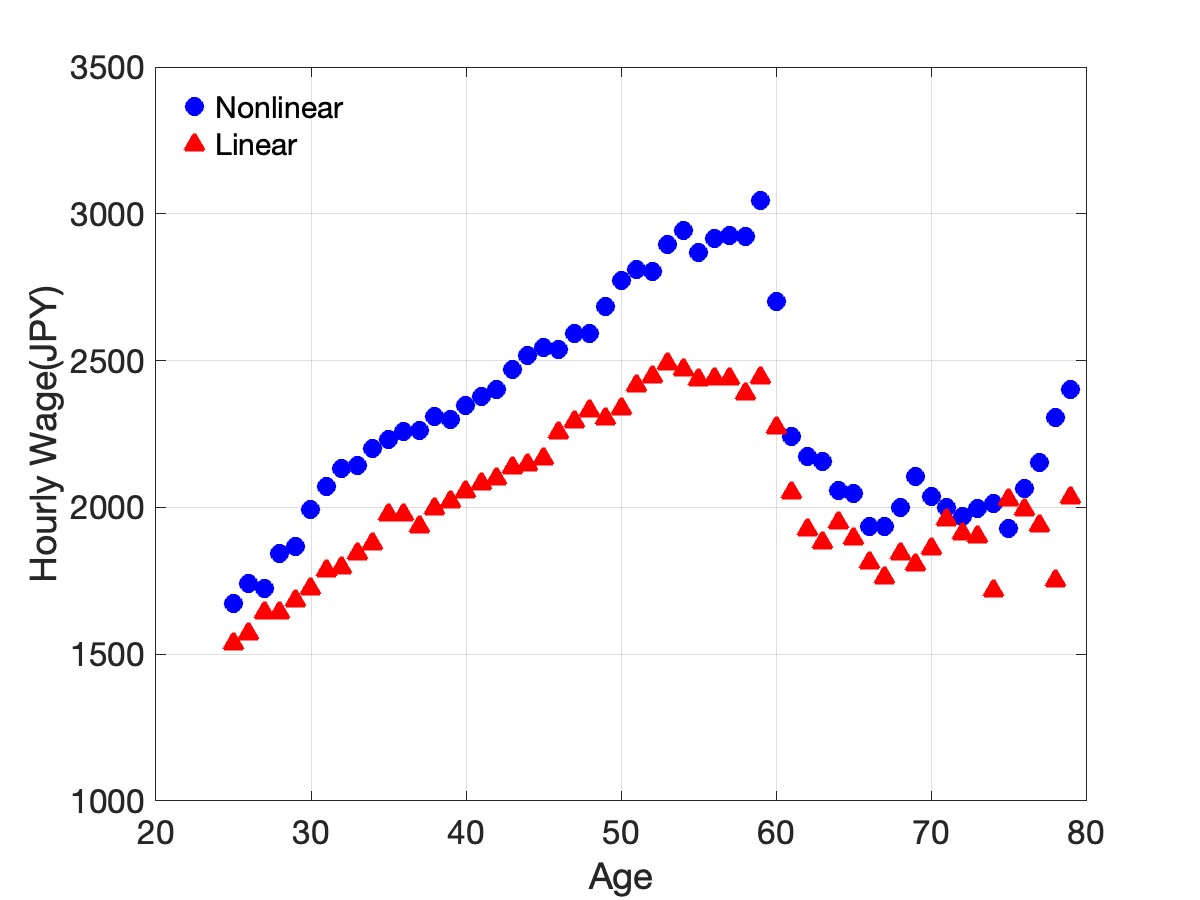}
    \caption{Lifetime hourly wage (nonlinear vs. linear) in Japan: Male, 2015-2019}
    \label{fig:Occ_wage_JP}
\end{figure}
 \vspace{-1em}
   \FloatBarrier
   \begin{figure}[H]
    \centering
    \includegraphics[width=0.70\linewidth]{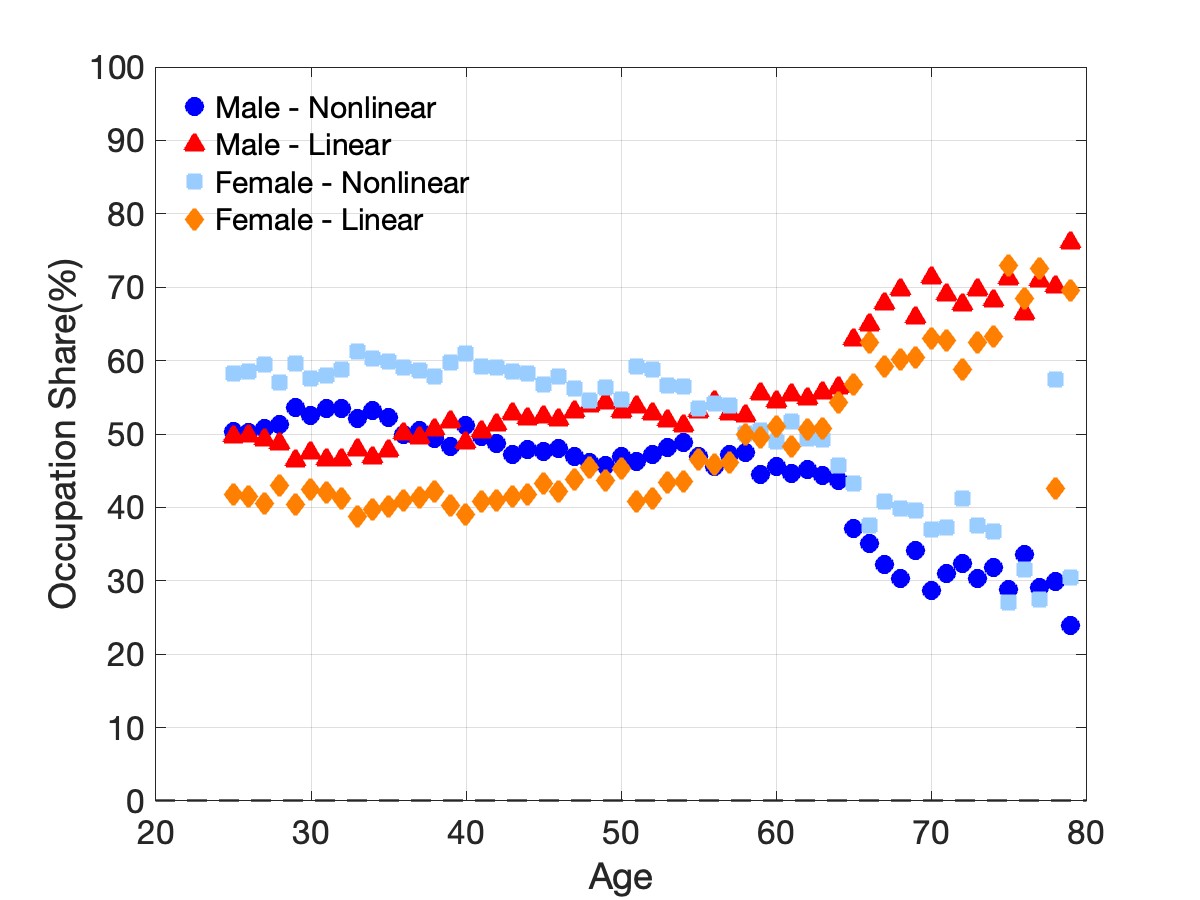}
    \caption{Change of Conditional Occupation Share over Age(nonlinear vs. linear) in Japan: Male and Female, 2009-2019}
    \label{fig:Occ_share_sex_JP}
\end{figure}
\subsection*{A.2. Data Description in US}
  \begin{figure}[H]
    \centering
    \includegraphics[width=0.70\linewidth]{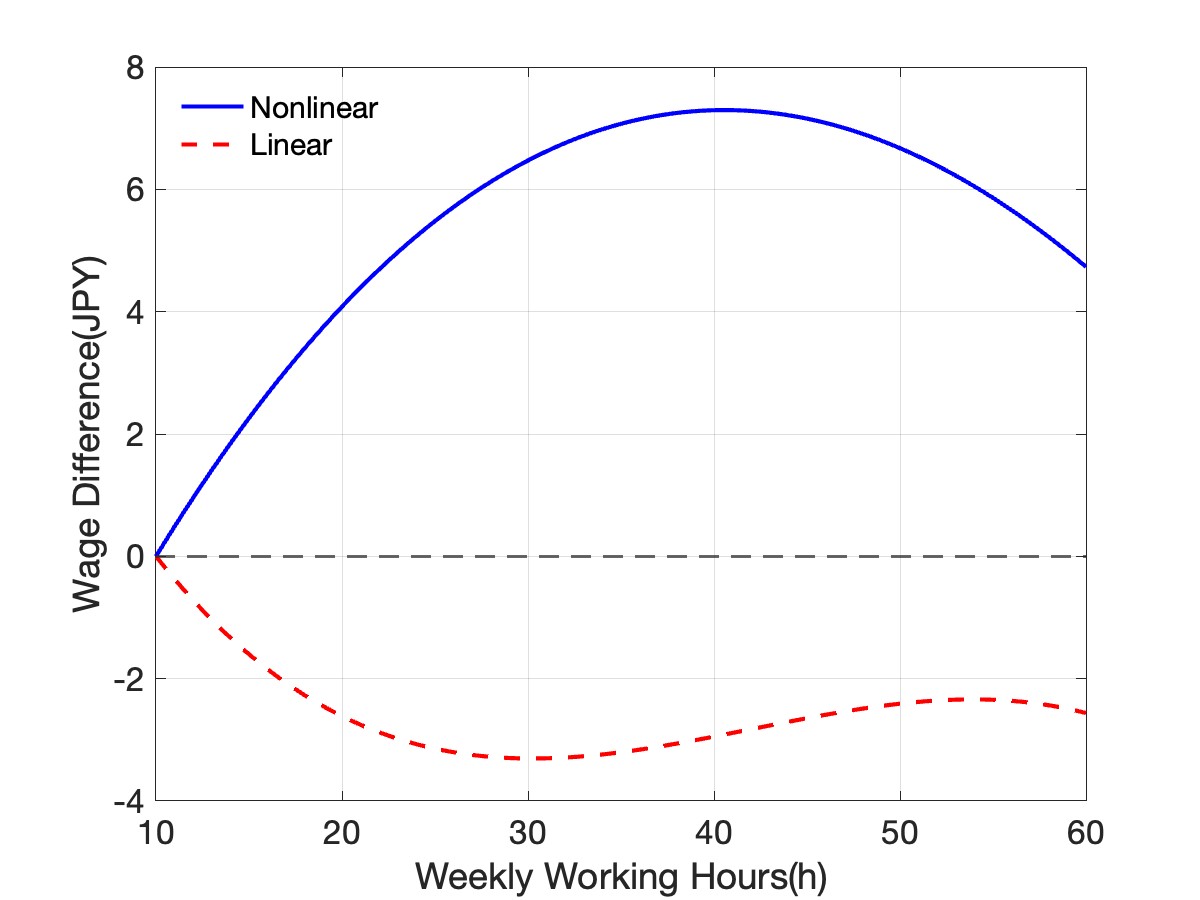}
    \caption{Hourly Wage Change over Working Hours(nonlinear vs. linear) in US: Male, 2009-2019}
    \label{fig:Occ_wwh_US}
\end{figure}
\vspace{-1em}  
\FloatBarrier
   \begin{figure}[H]
    \centering
    \includegraphics[width=0.70\linewidth]{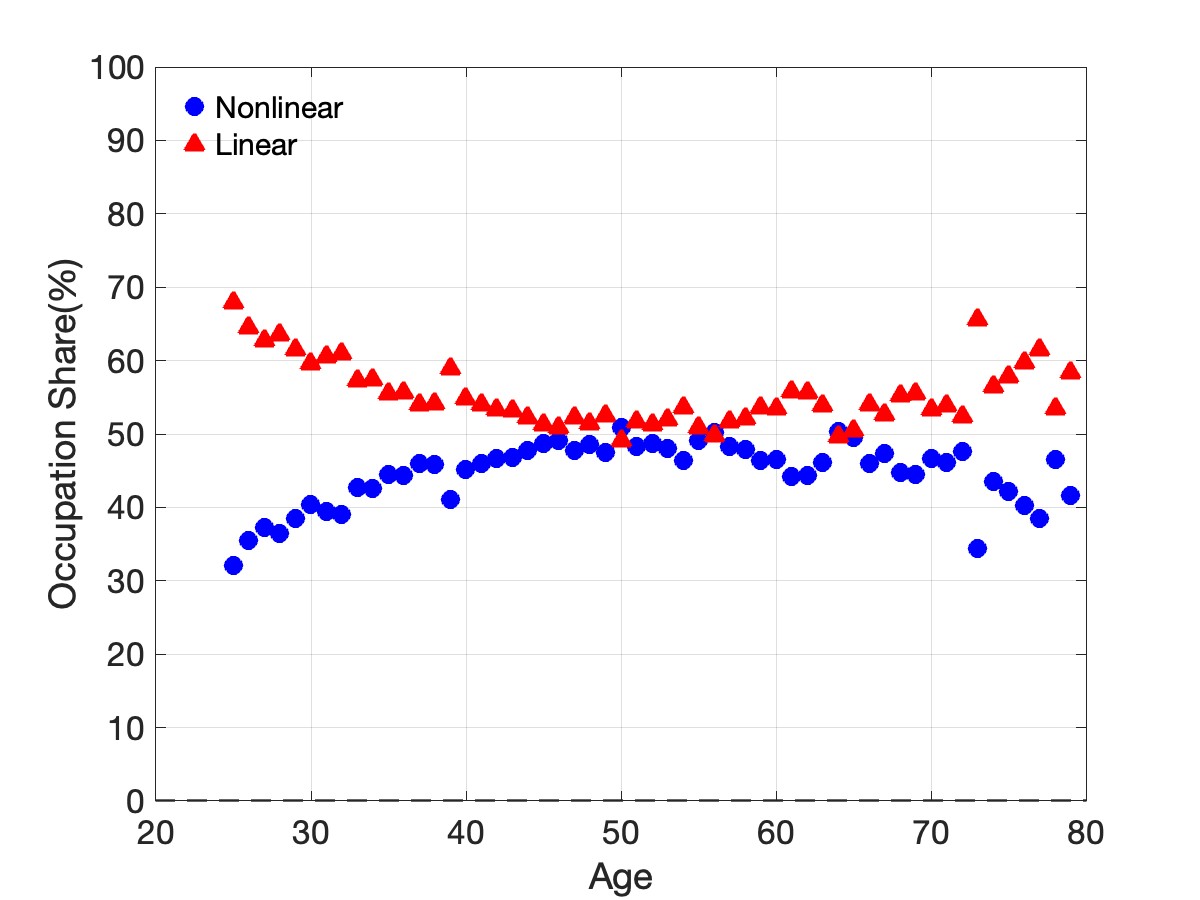}
    \caption{Change of Conditional Occupation Share over Age(nonlinear vs. linear) in US: Male, 2009-2019}
    \label{fig:Occ_share_US}
\end{figure}
\vspace{-1em}  
\FloatBarrier
   \begin{figure}[H]
    \centering
    \includegraphics[width=0.70\linewidth]{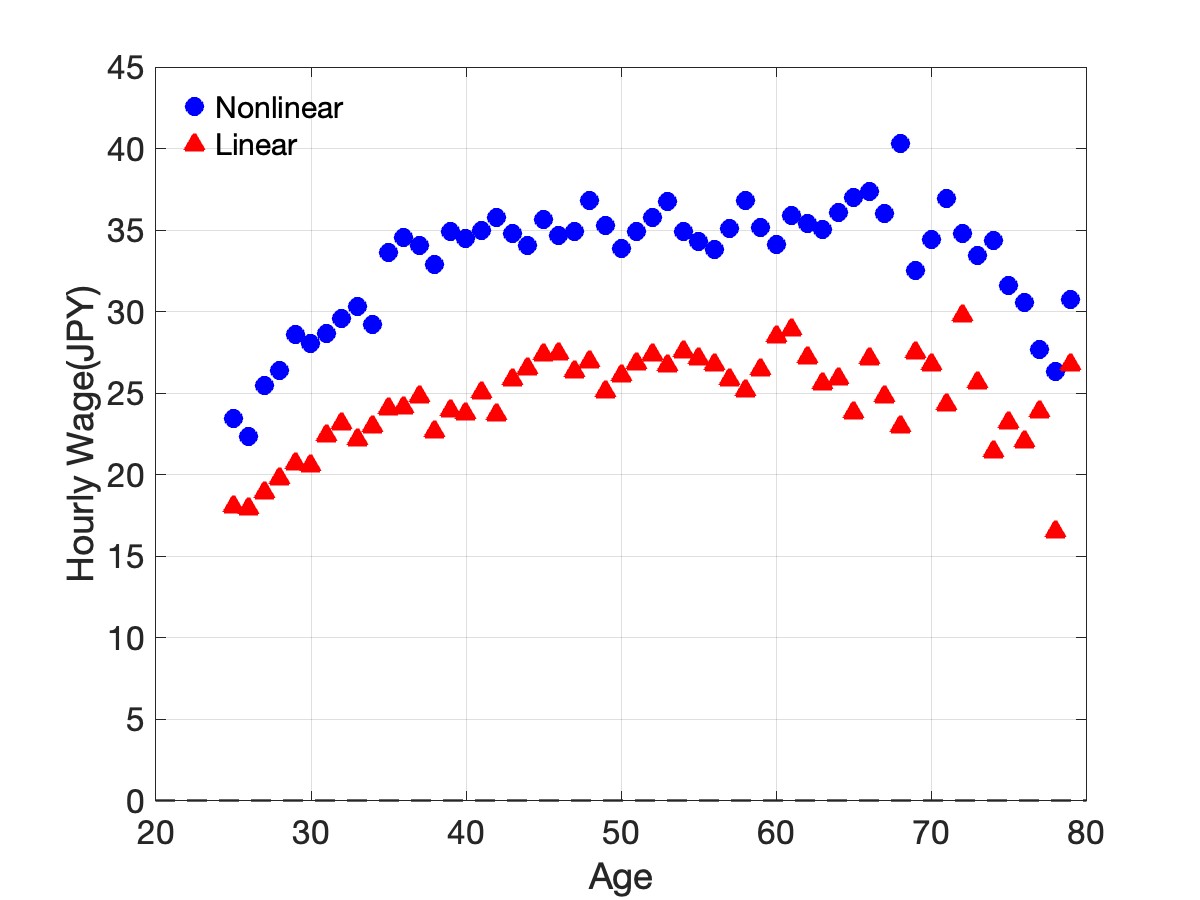}
    \caption{Lifetime hourly wage (nonlinear vs. linear) in US: Male, 2009-2019}
    \label{fig:Occ_wage_US}
\end{figure}
\subsection*{A.3. Cross-country Data Description}

   \subsection*{A.4. Calibration result}
   \FloatBarrier
\begin{figure}[H]
    \centering
    \includegraphics[width=1.10\linewidth]{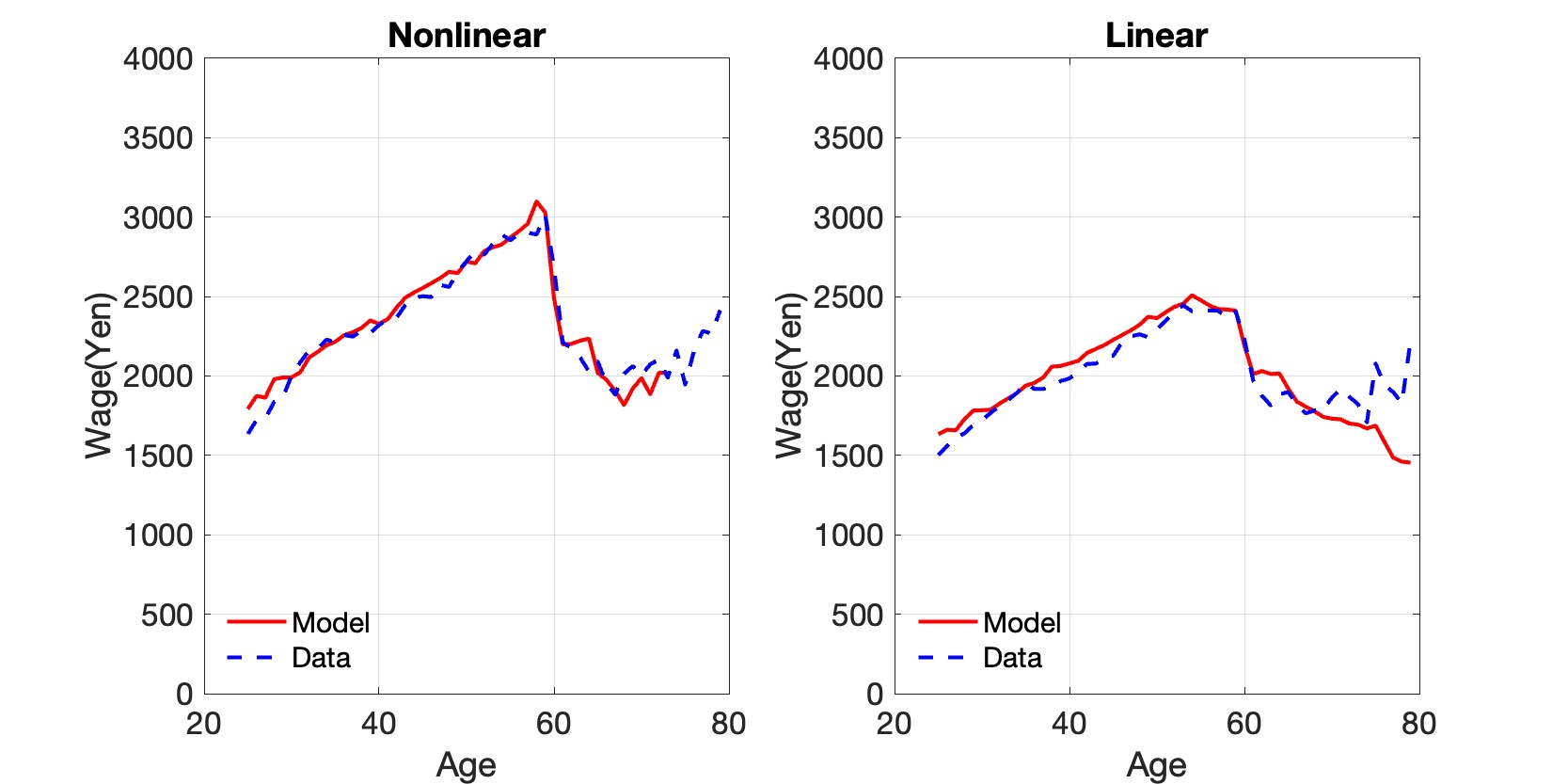}
    \caption{Wage difference in Japan (Model vs Data)}
    \label{fig:Occ_wage_M_D}
\end{figure}
\vspace{-1em}  
\FloatBarrier
\begin{figure}[H]
    \centering
    \includegraphics[width=0.70\linewidth]{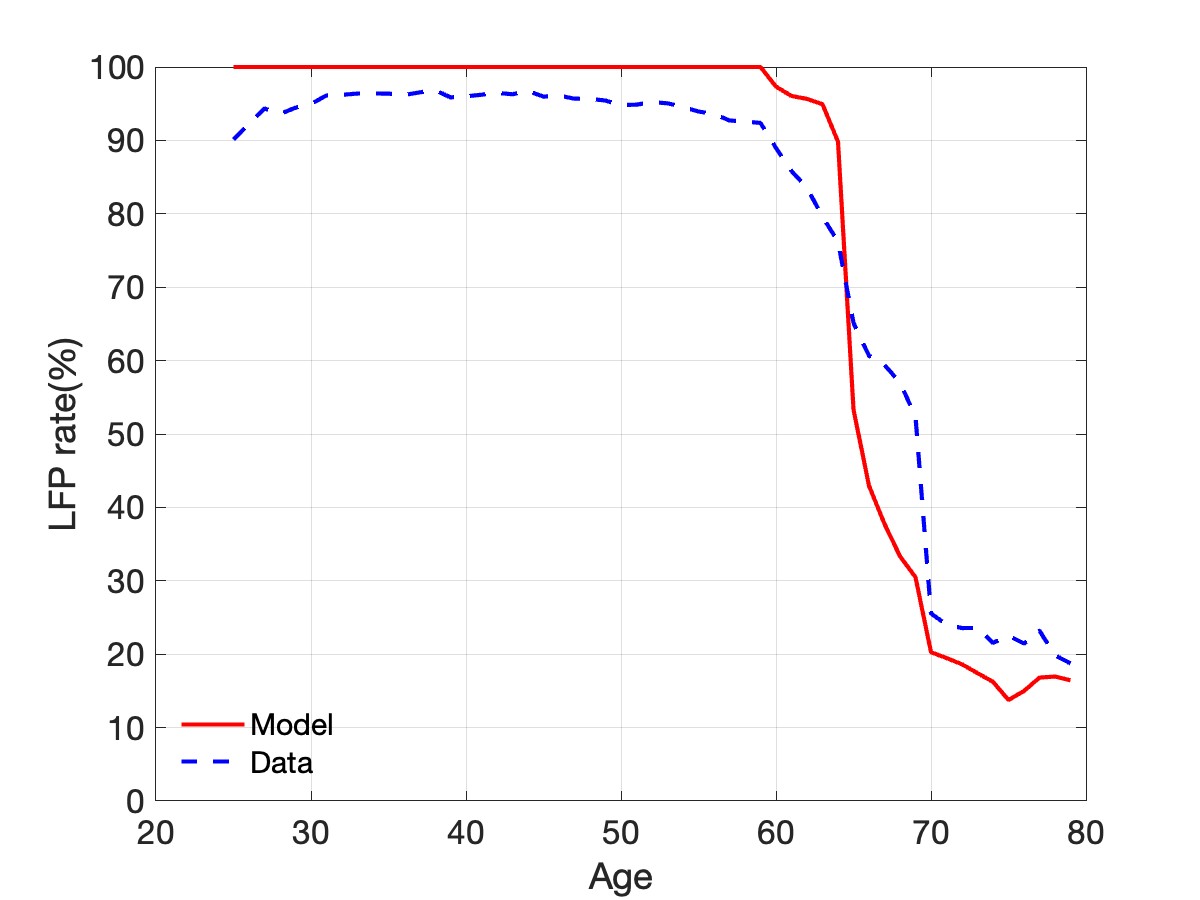}
    \caption{LFP rate in Japan (Model vs Data)}
    \label{fig:LFP_M_D}
\end{figure}
\vspace{-1em}  
\FloatBarrier
\begin{figure}[H]
    \centering
    \includegraphics[width=0.70\linewidth]{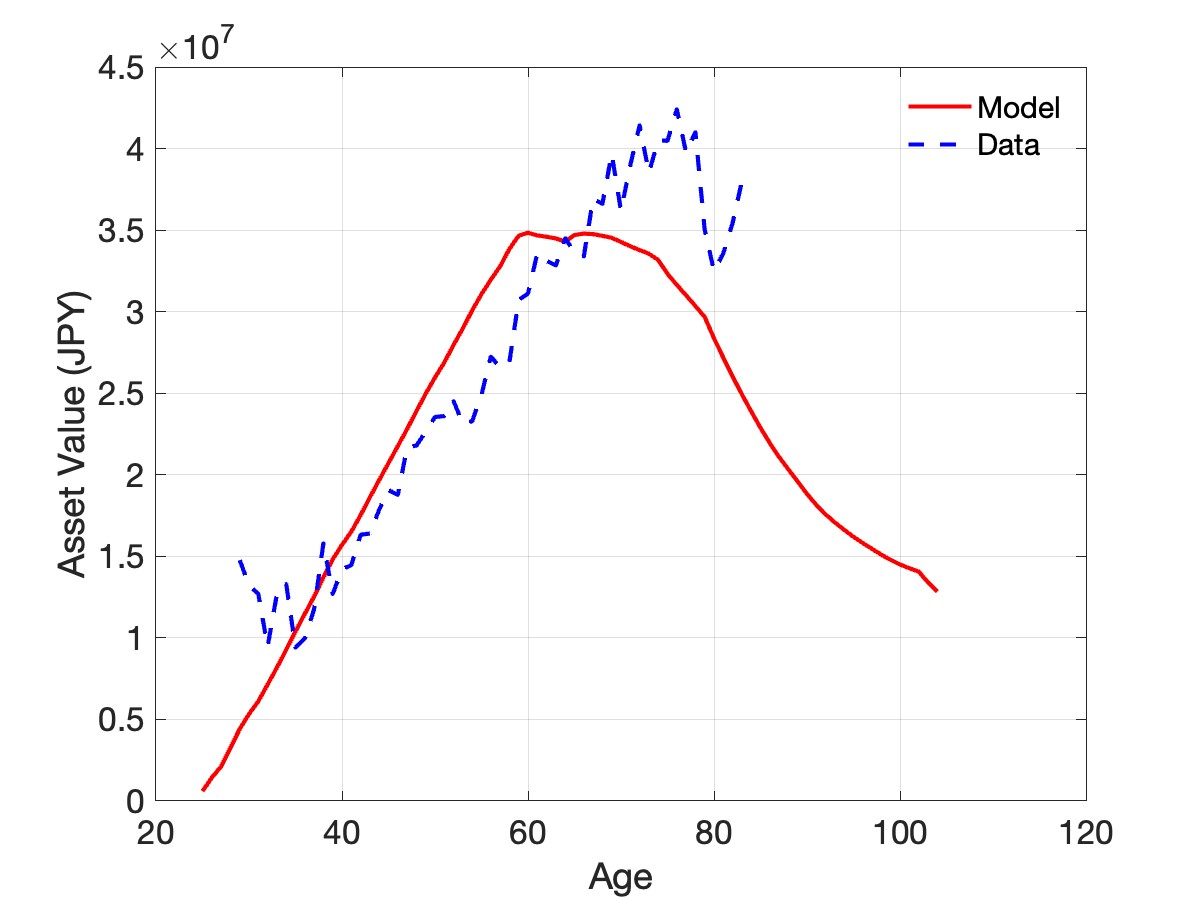}
    \caption{Asset in Japan (Model vs Data)}
    \label{fig:Asset_M_D}
\end{figure}

\subsection*{A.5. Source of nonlinearity}
\vspace{-1em}  
         \FloatBarrier
        \begin{figure}[H]
    	    \centering
    	    \includegraphics[width=1.10\linewidth]{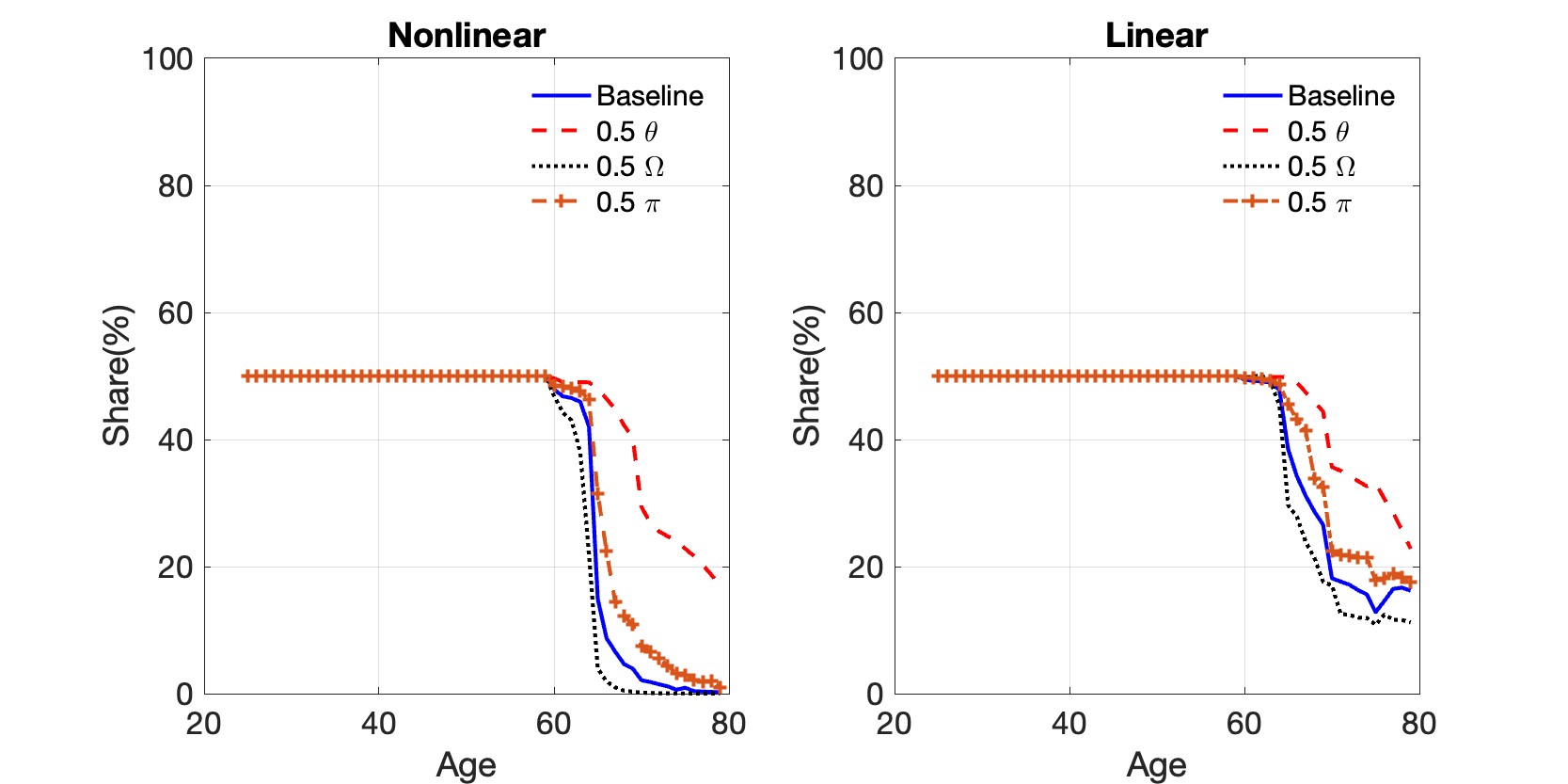}
            \caption{Unconditional Occupation Share (Source of nonlinearity)}
    	    \label{fig:Uncond_occ_share_scheck}
        \end{figure}

         \vspace{-1em}  
\begin{figure}[H]
    \centering
    \includegraphics[width=1.10\linewidth]{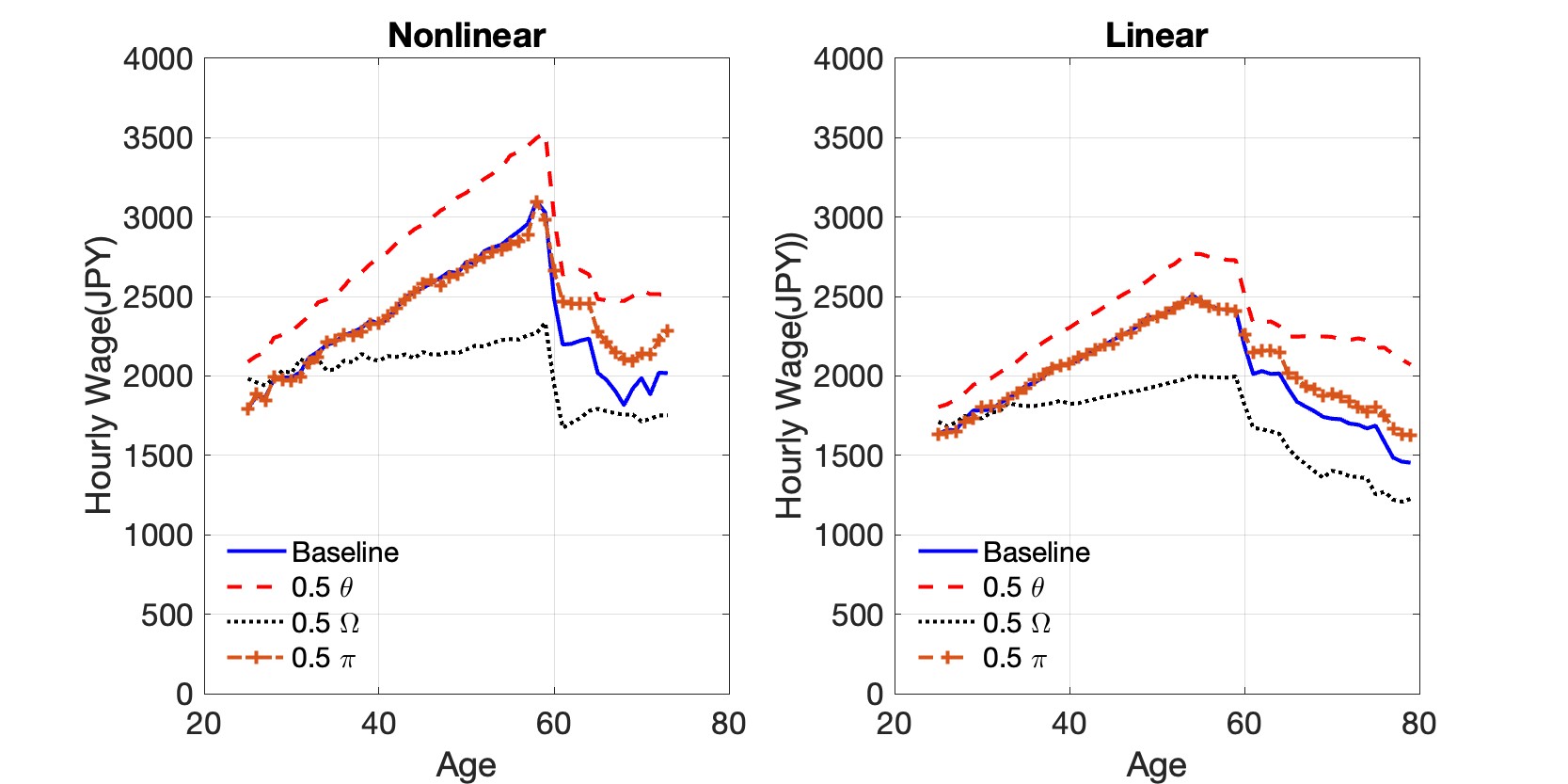}
    \caption{Wage difference (Source of nonlinearity)}
    \label{fig:Scheckck_wage}
\end{figure}
\vspace{-1em}  
         \FloatBarrier
        \begin{figure}[H]
    	    \centering
    	    \includegraphics[width=1.10\linewidth]{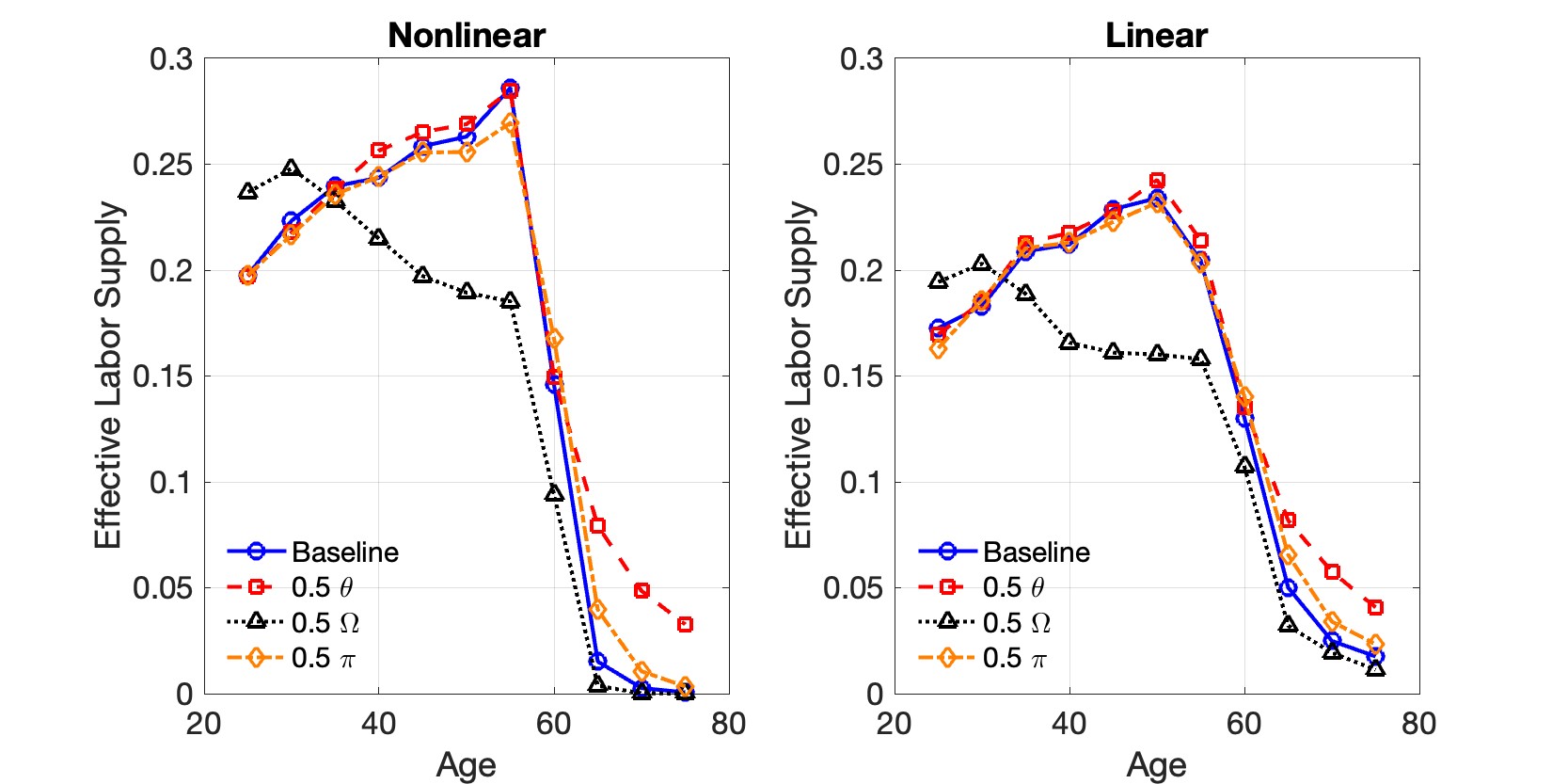}
            \caption{Effective Labor Supply (Source of nonlinearity)}
    	    \label{fig:EffLs_scheck}
        \end{figure}
\vspace{-1em}
\begin{figure}[H]
    \centering
    \includegraphics[width=1.10\linewidth]{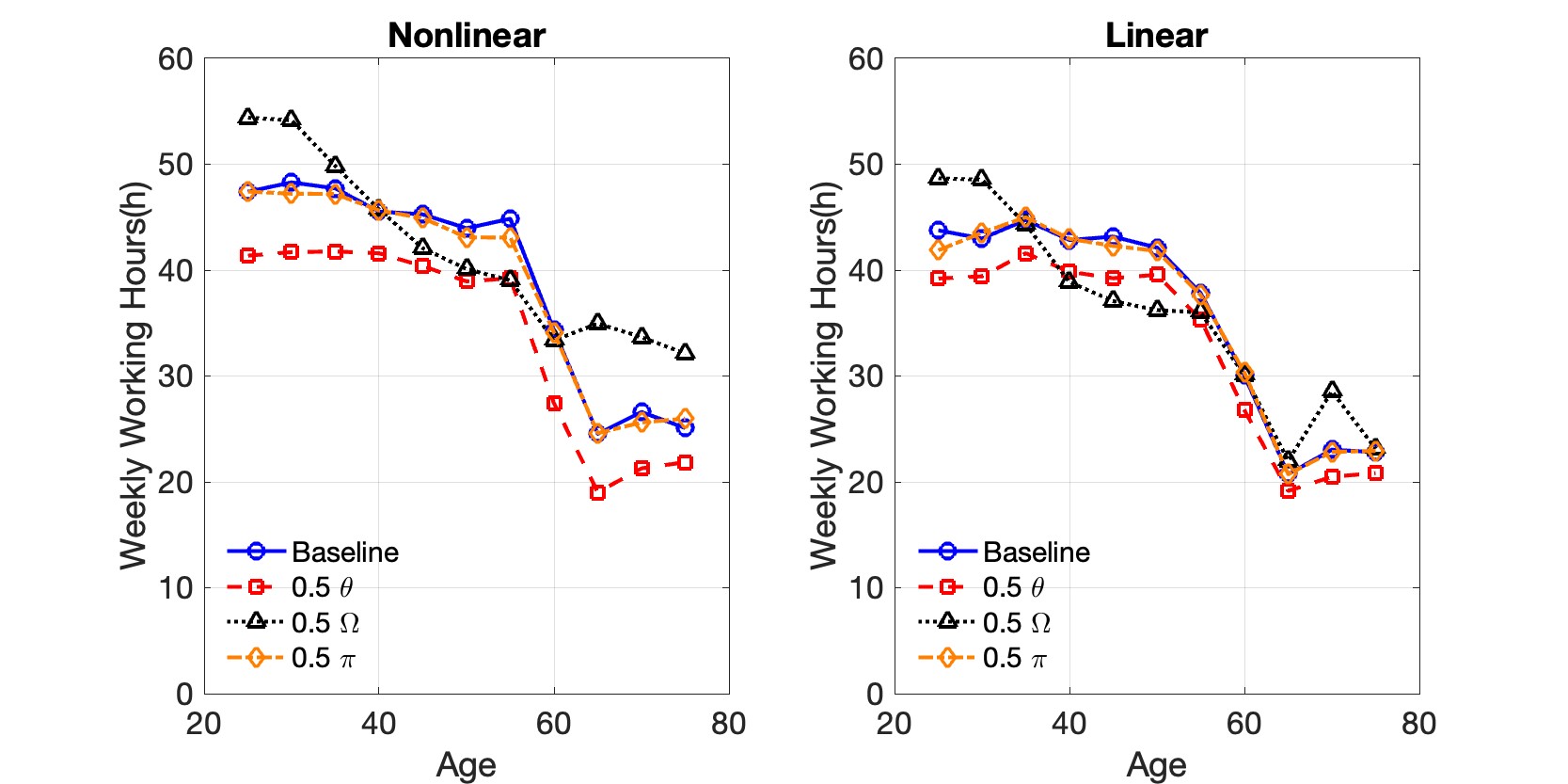}
    \caption{Working Hours per week (Source of nonlinearity)}
    \label{fig:Scheckck_wwh}
\end{figure}
\vspace{-1em}
\begin{figure}[H]
    \centering
    \includegraphics[width=0.70\linewidth]{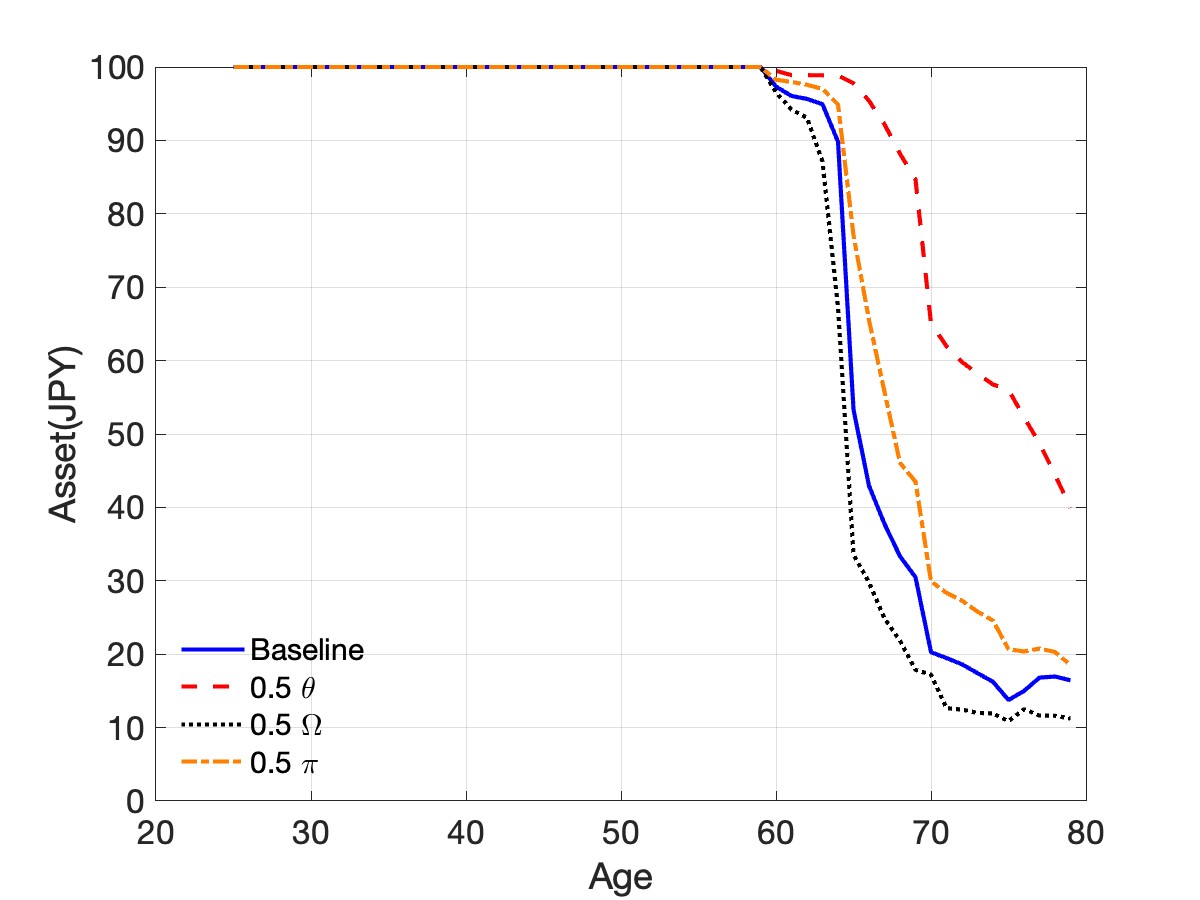}
    \caption{LFP (Source of nonlinearity)}
    \label{fig:Scheckck_LFP}
\end{figure}
\vspace{-1em}
\begin{figure}[H]
    \centering
    \includegraphics[width=1.10\linewidth]{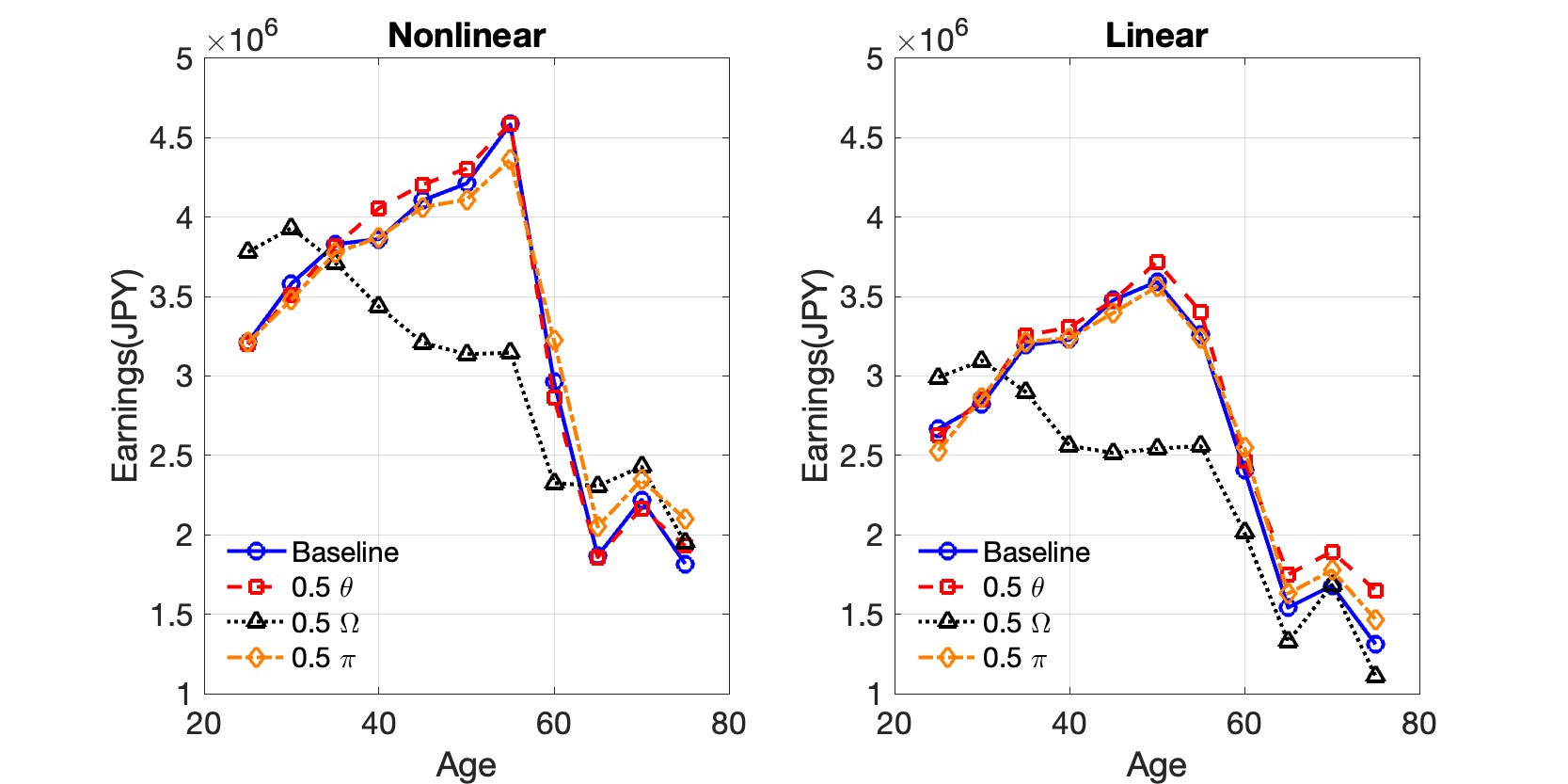}
    \caption{Earnings (Source of nonlinearity)}
    \label{fig:Scheckck_earn}
\end{figure}
\vspace{-1em}
\begin{figure}[H]
    \centering
    \includegraphics[width=0.70\linewidth]{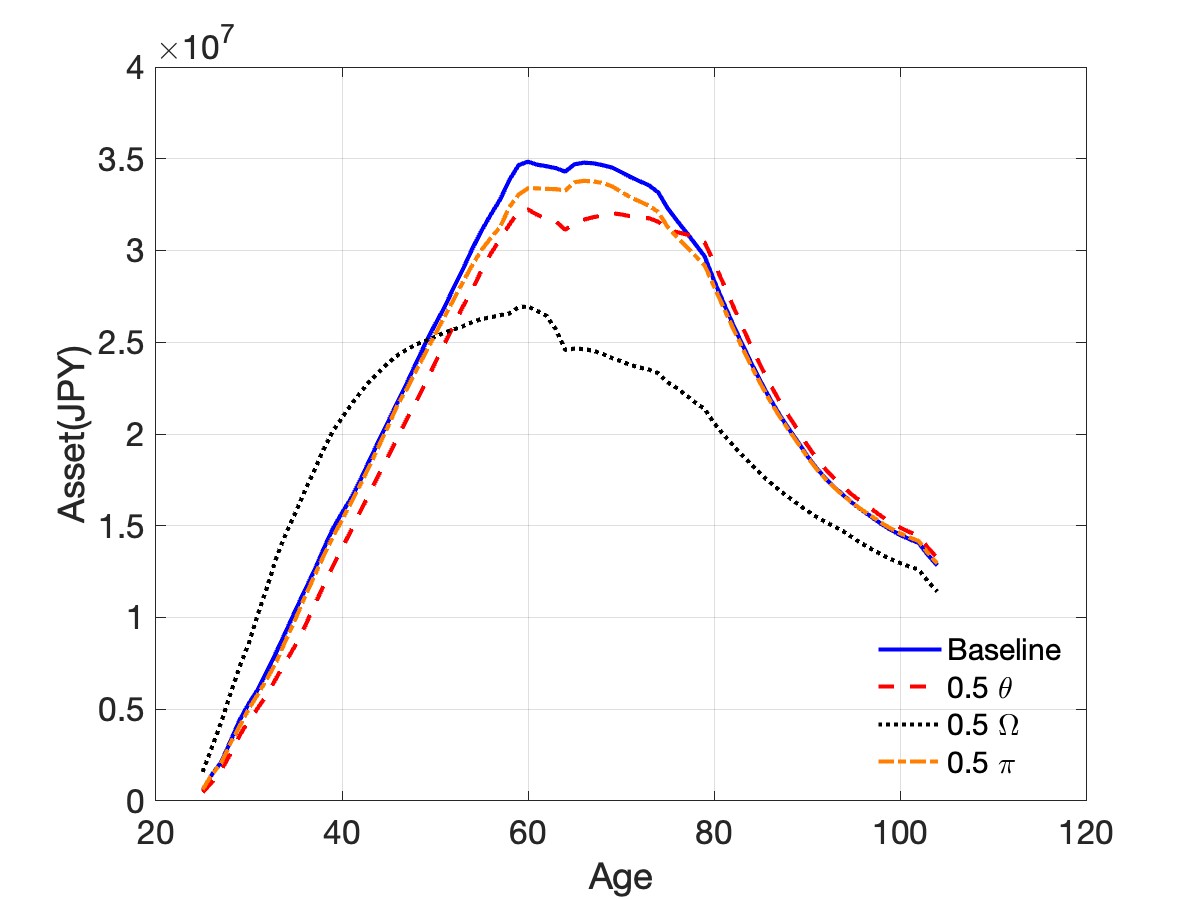}
    \caption{Asset (Source of nonlinearity)}
    \label{fig:Scheckck_asset}
\end{figure}
\subsection*{A.6. Counterfactual Experiment(Conventional Policy)}
\begin{figure}[H]
    \centering
    \includegraphics[width=1.10\linewidth]{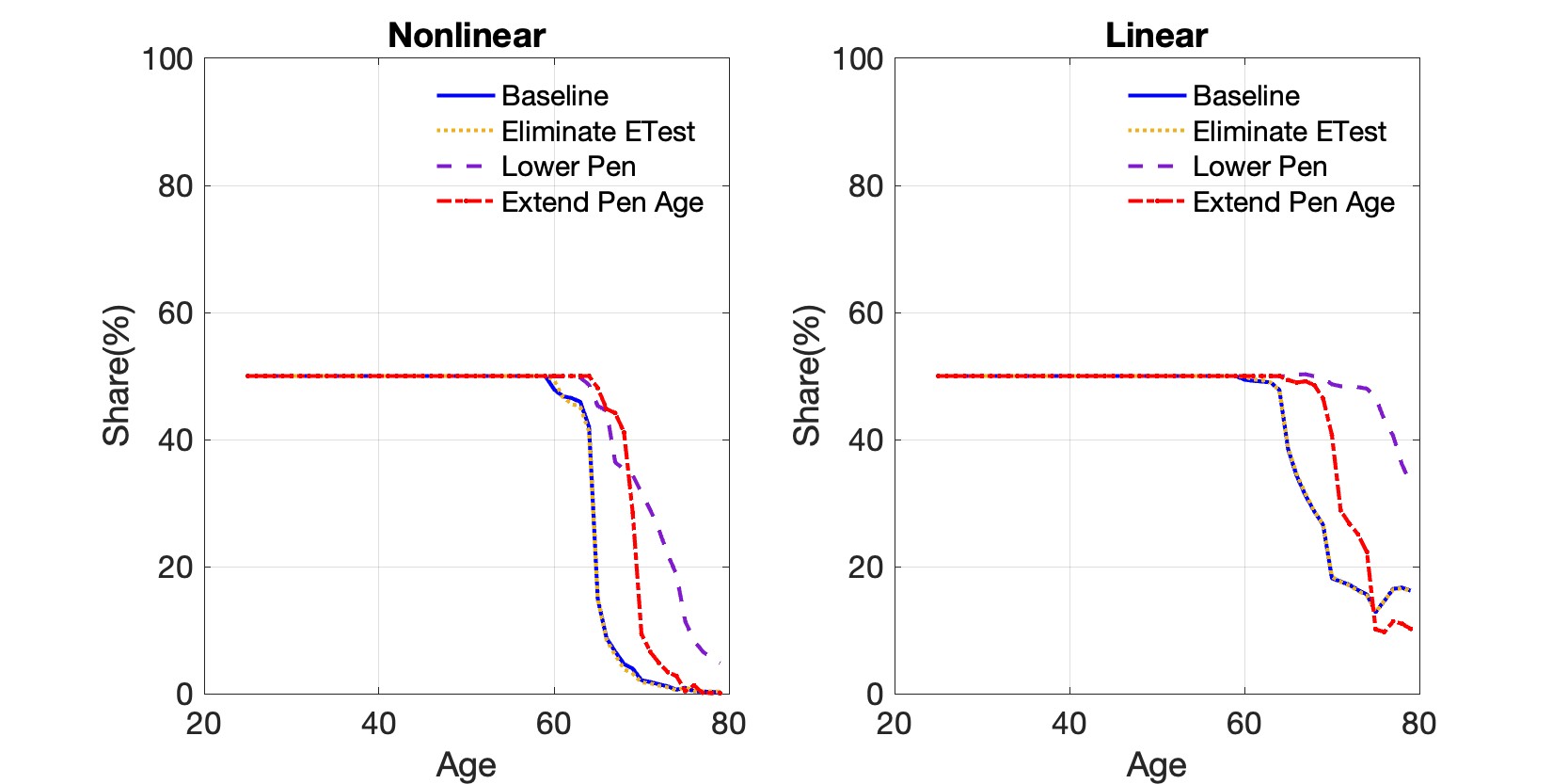}
    \caption{Unconditional Occupation Share(Conventional Policy)}
    \label{fig:Exp_occ_conv_1}
\end{figure}
\vspace{-1em}
\begin{figure}[H]
    \centering
    \includegraphics[width=1.10\linewidth]{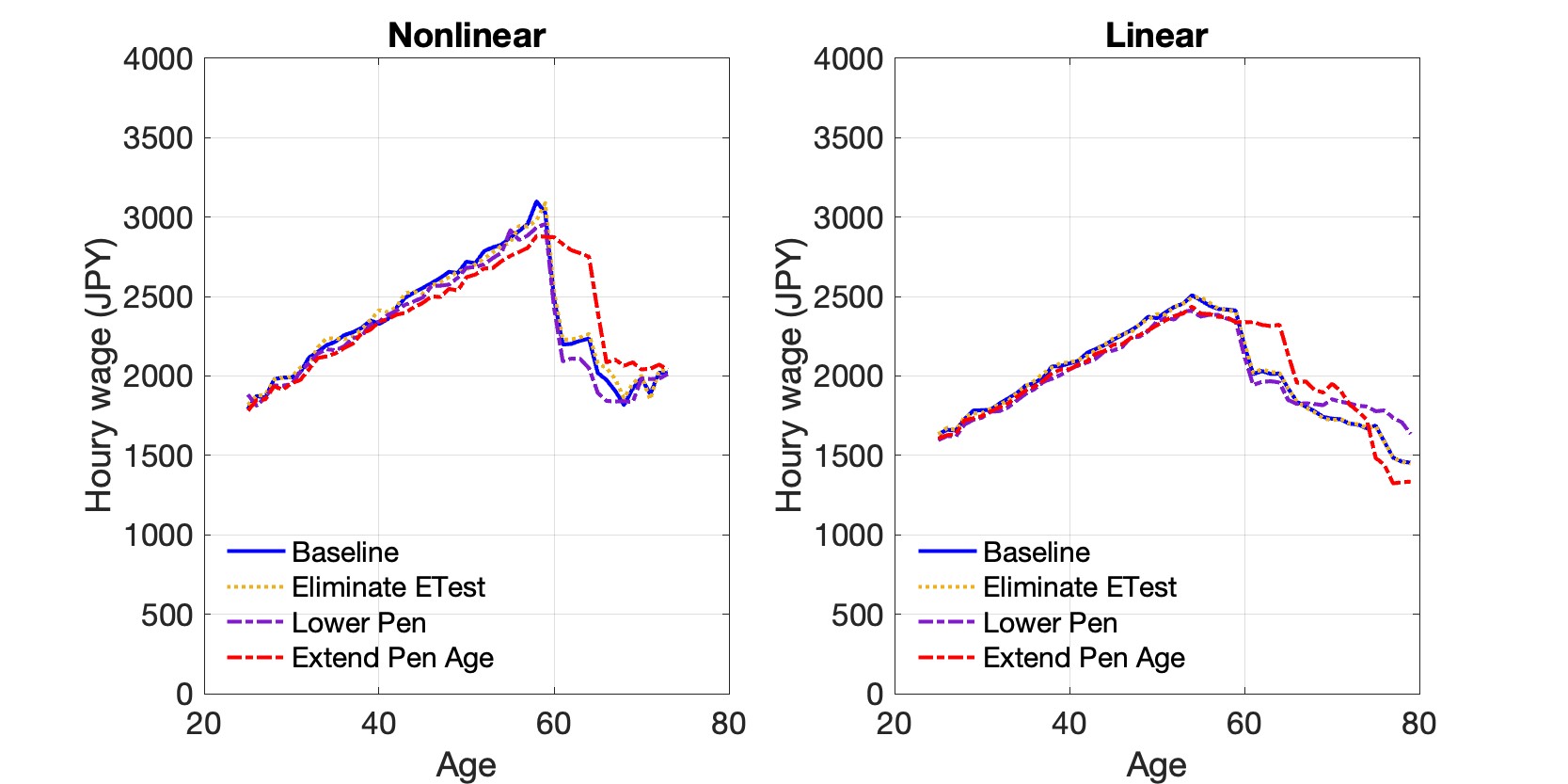}
     \caption{Wage difference (Conventional Policy)}
     \label{fig:Exp_wage_conv_1}
\end{figure}
\vspace{-1em}
        \FloatBarrier
    \begin{figure}[H]
    	    \centering
    	    \includegraphics[width=1.10\linewidth]{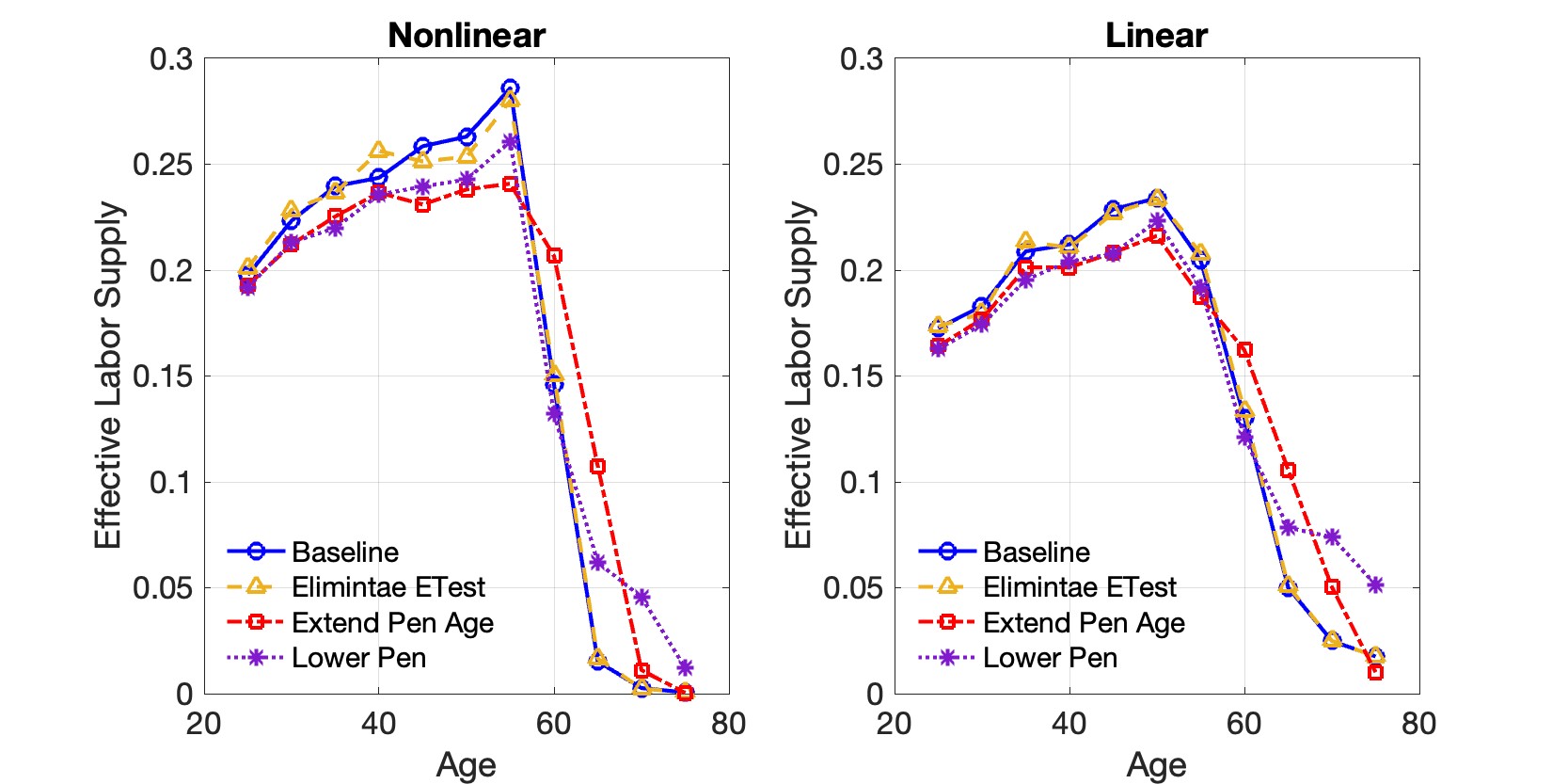}
            \caption{Effective Labor Supply(Conventional Policies)}
    	    \label{fig:EffLs_conv}
        \end{figure}
\vspace{-1em}
\begin{figure}[H]
    \centering
    \includegraphics[width=1.10\linewidth]{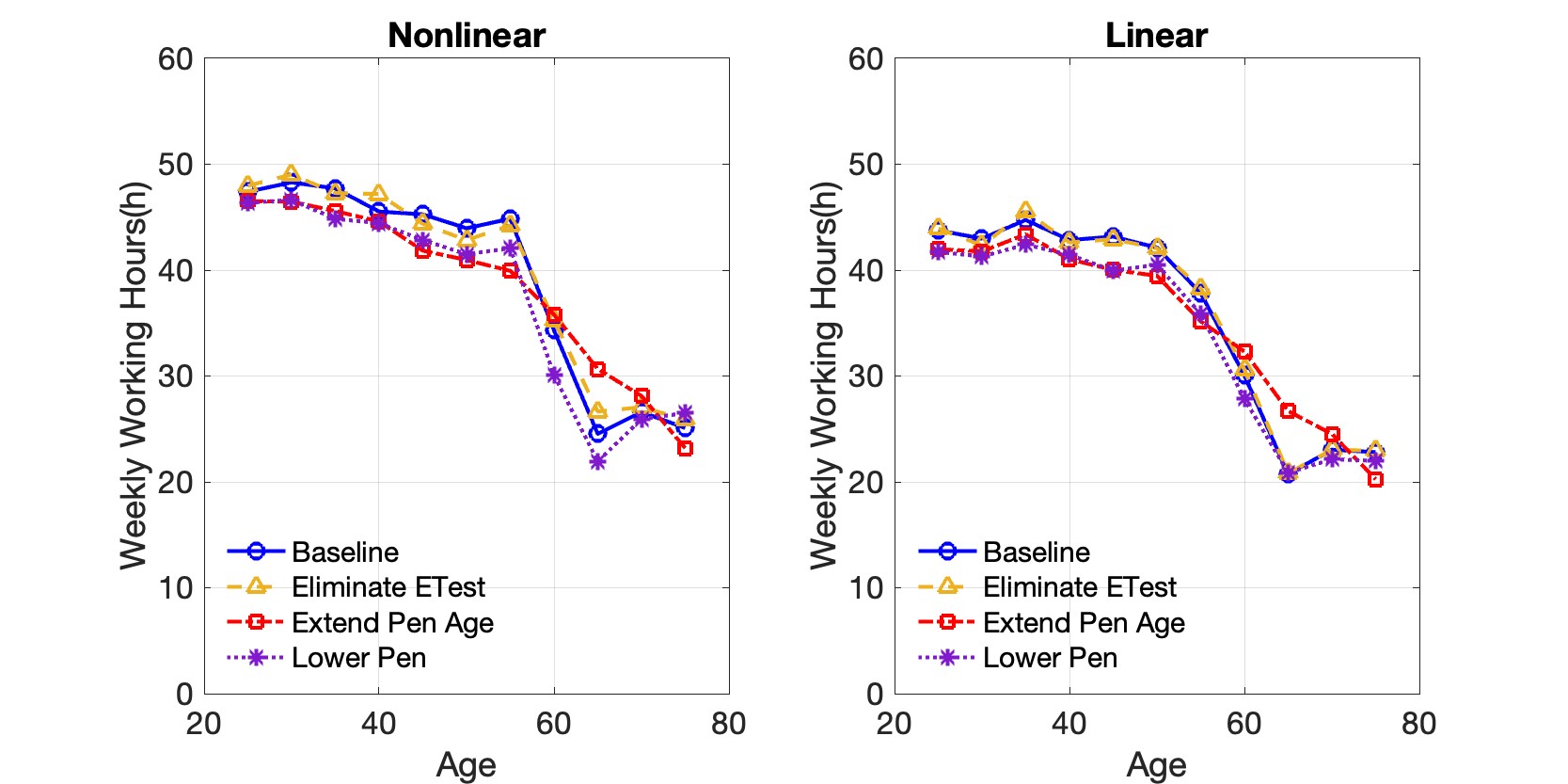}
     \caption{Weekly Working Hours (Conventional Policy)}
     \label{fig:wwh_conv}
\end{figure}
\vspace{-1em}
\begin{figure}[H]
    \centering
    \includegraphics[width=0.70\linewidth]{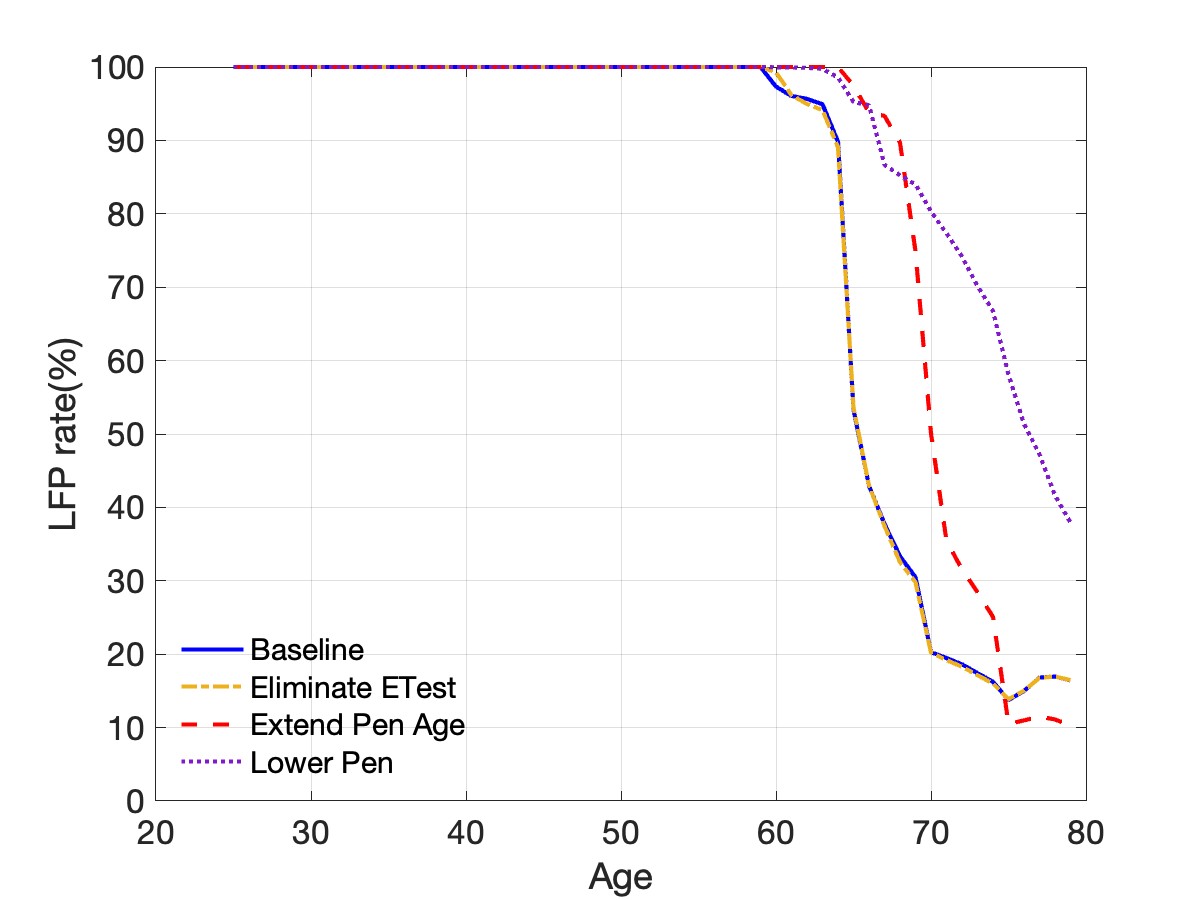}
     \caption{LFP rate (Conventional Policy)}
     \label{fig:LFP_conv}
\end{figure}
\vspace{-1em}
\begin{figure}[H]
    \centering
    \includegraphics[width=1.10\linewidth]{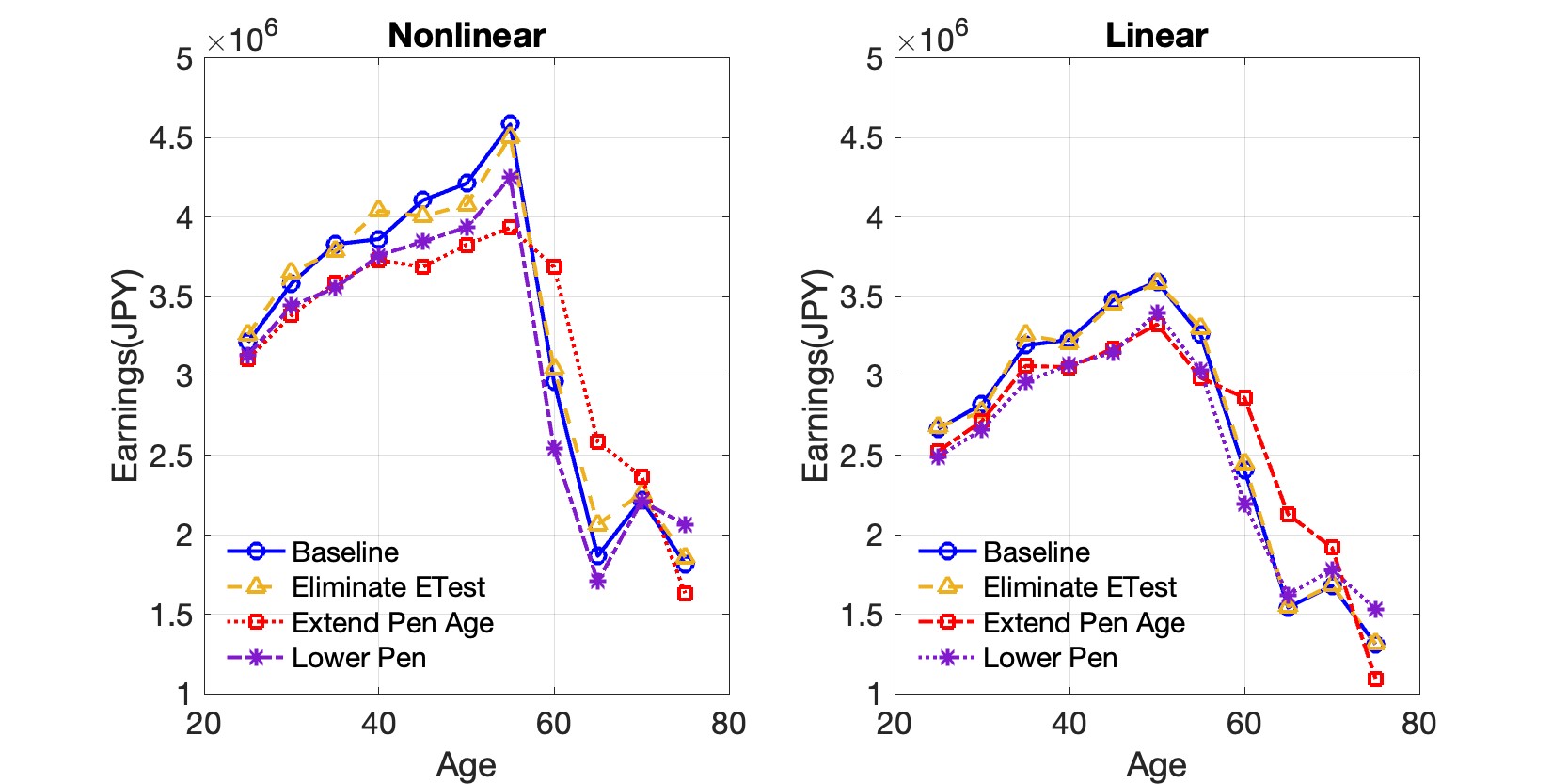}
     \caption{Earnings (Conventional Policy)}
     \label{fig:earn_conv}
\end{figure}
\vspace{-1em}
\begin{figure}[H]
    \centering
    \includegraphics[width=0.70\linewidth]{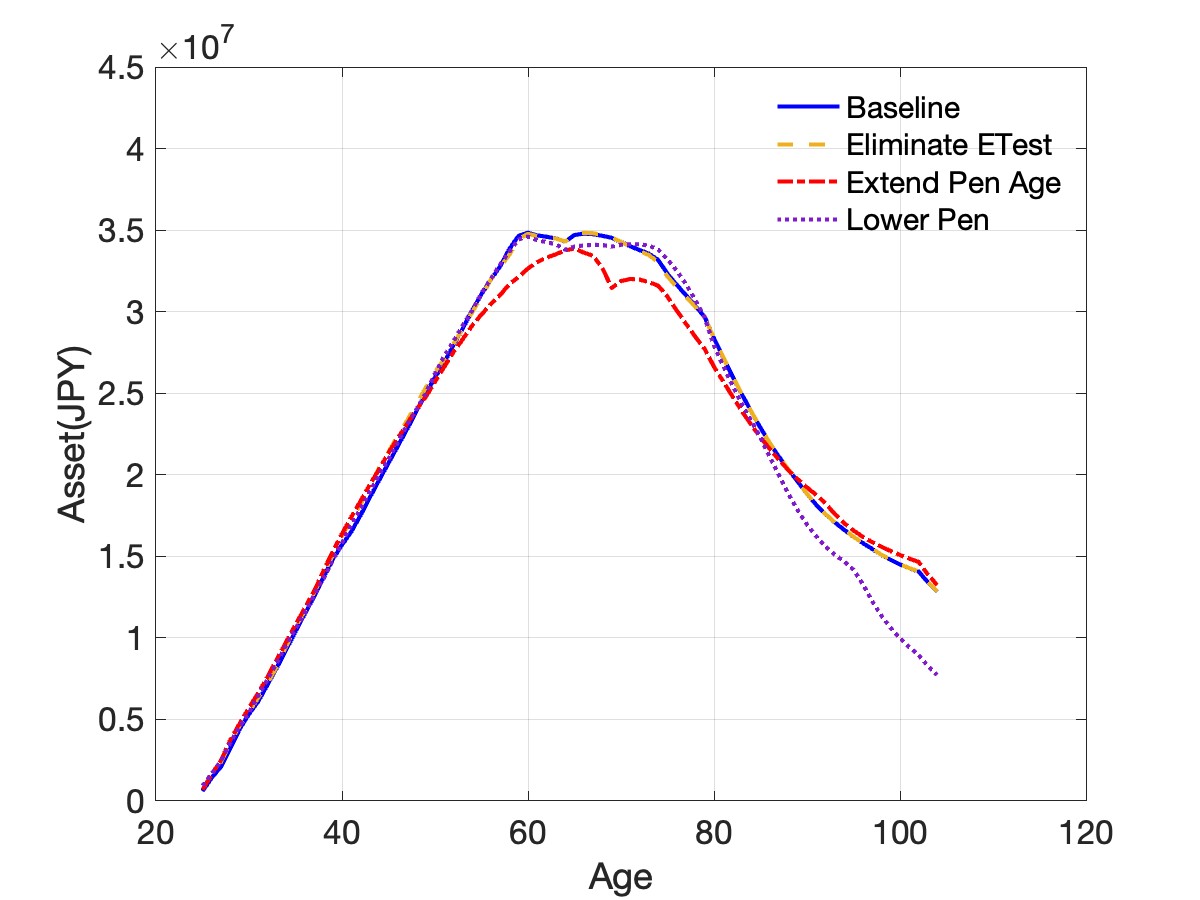}
     \caption{Asset (Conventional Policy)}
     \label{fig:asset_conv}
\end{figure}
\subsection*{A.7. Counterfactual Experiment(Unconventional Policy)}
\begin{figure}[H]
    \centering
    \includegraphics[width=1.10\linewidth]{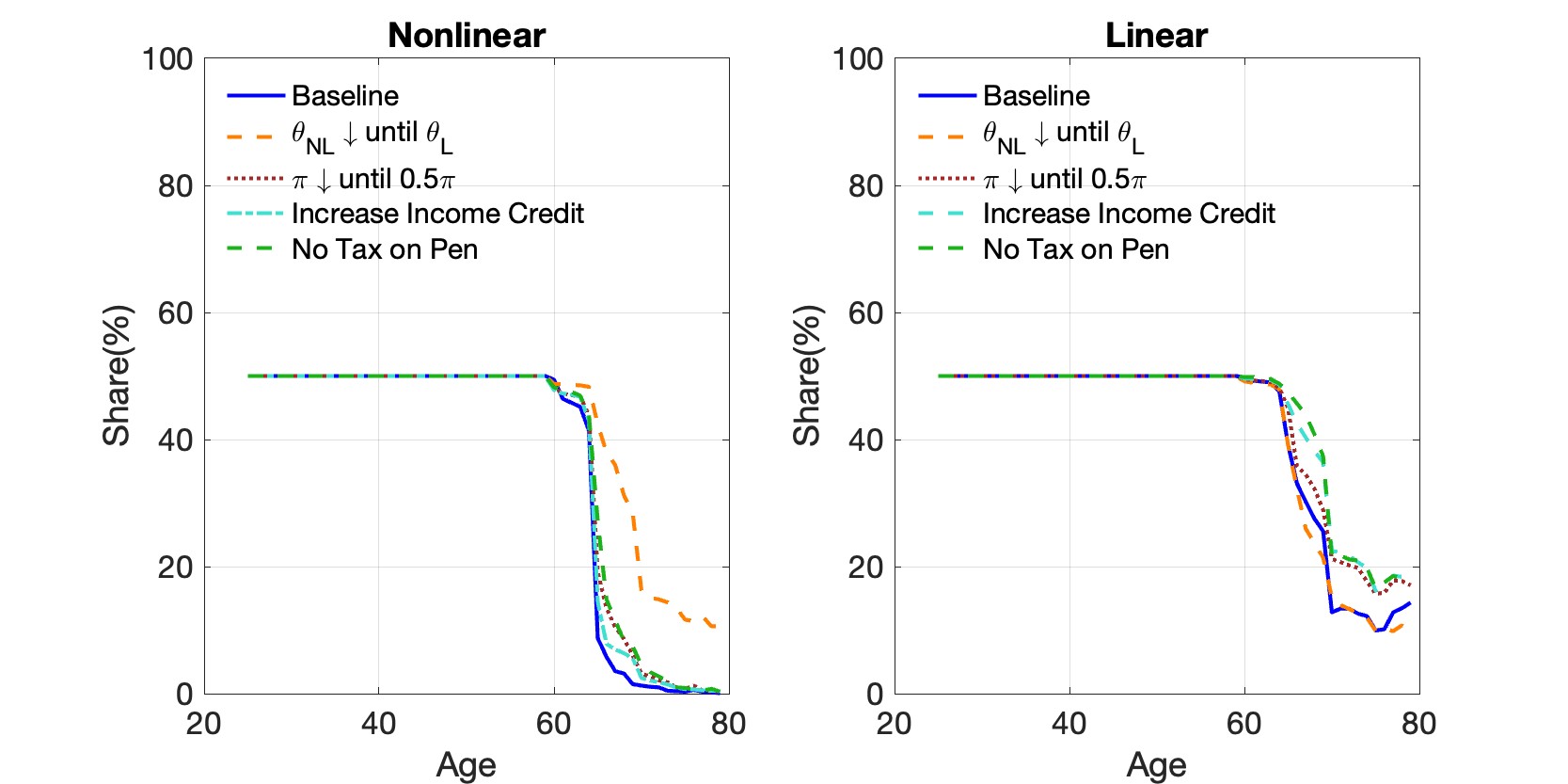}
    \caption{Unconditional Occupation Share (Unconventional Policy)}
     \label{fig:Occ_share_org_ALL}
\end{figure}
\vspace{-1em}
\begin{figure}[H]
    \centering
    \includegraphics[width=1.10\linewidth]{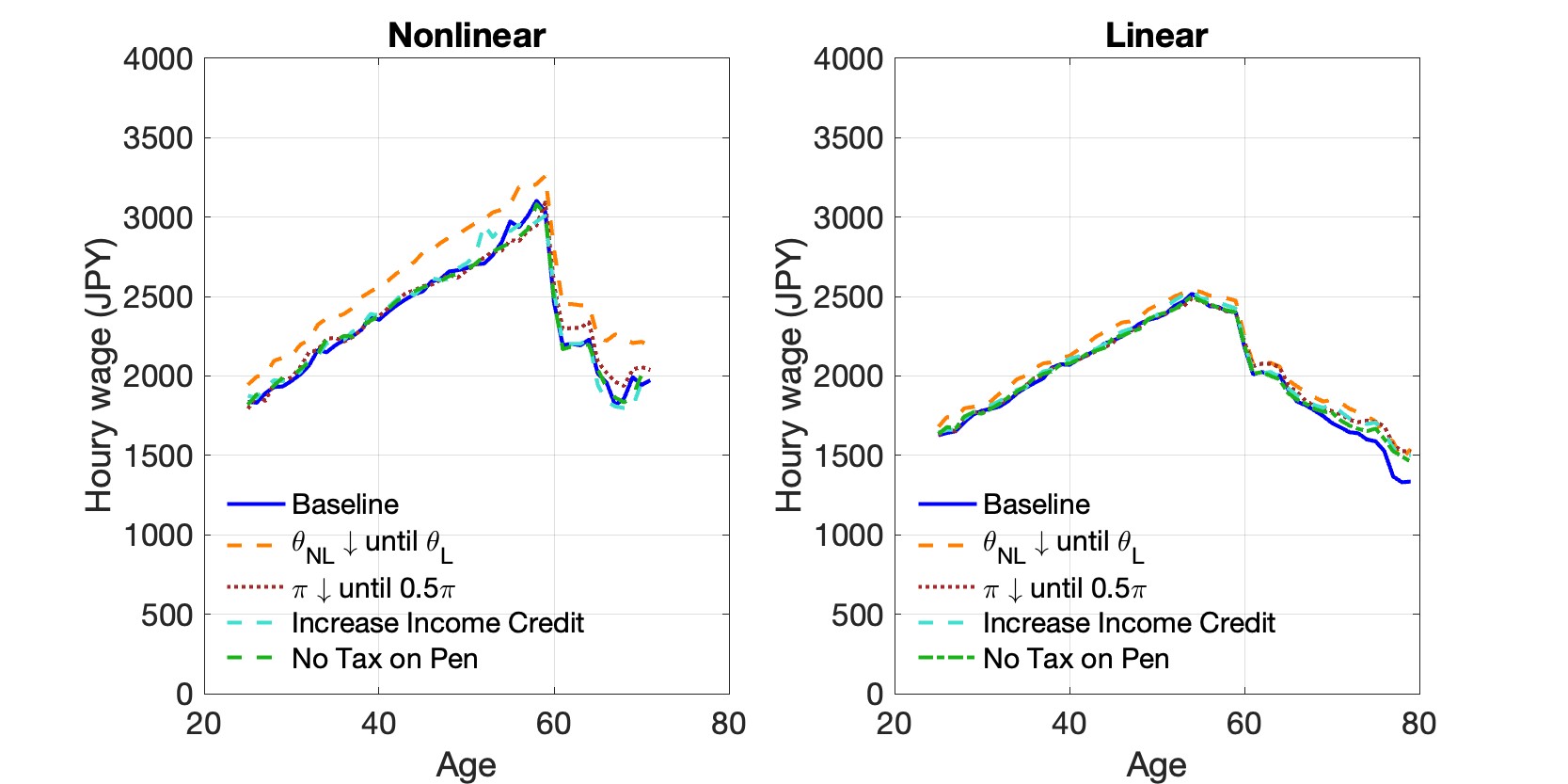}
    \caption{Wage difference (Unconventional Policy)}
     \label{fig:Exp_wage_org_ALL}
\end{figure}
\footnotetext{When I reduce $\theta_{NL}$ to $\theta_{L}$, since both of $\frac{K}{L}$ and $\frac{EL}{E_1}{L_1}$ decrease, leading to a reduction in $w_1$, which represents the payment per unit of efficiency labor in nonlinear occupations. However, nonlinear workers with high labor disutility increase their productivity and receive a higher hourly wage, while those with lower labor disutility reduce their working hours due to the weakened nonlinearity. As a result, the overall wage level for a nonlinear worker increases.}
\vspace{-1em}
     \FloatBarrier
    \begin{figure}[H]
    	    \centering
    	    \includegraphics[width=1.10\linewidth]{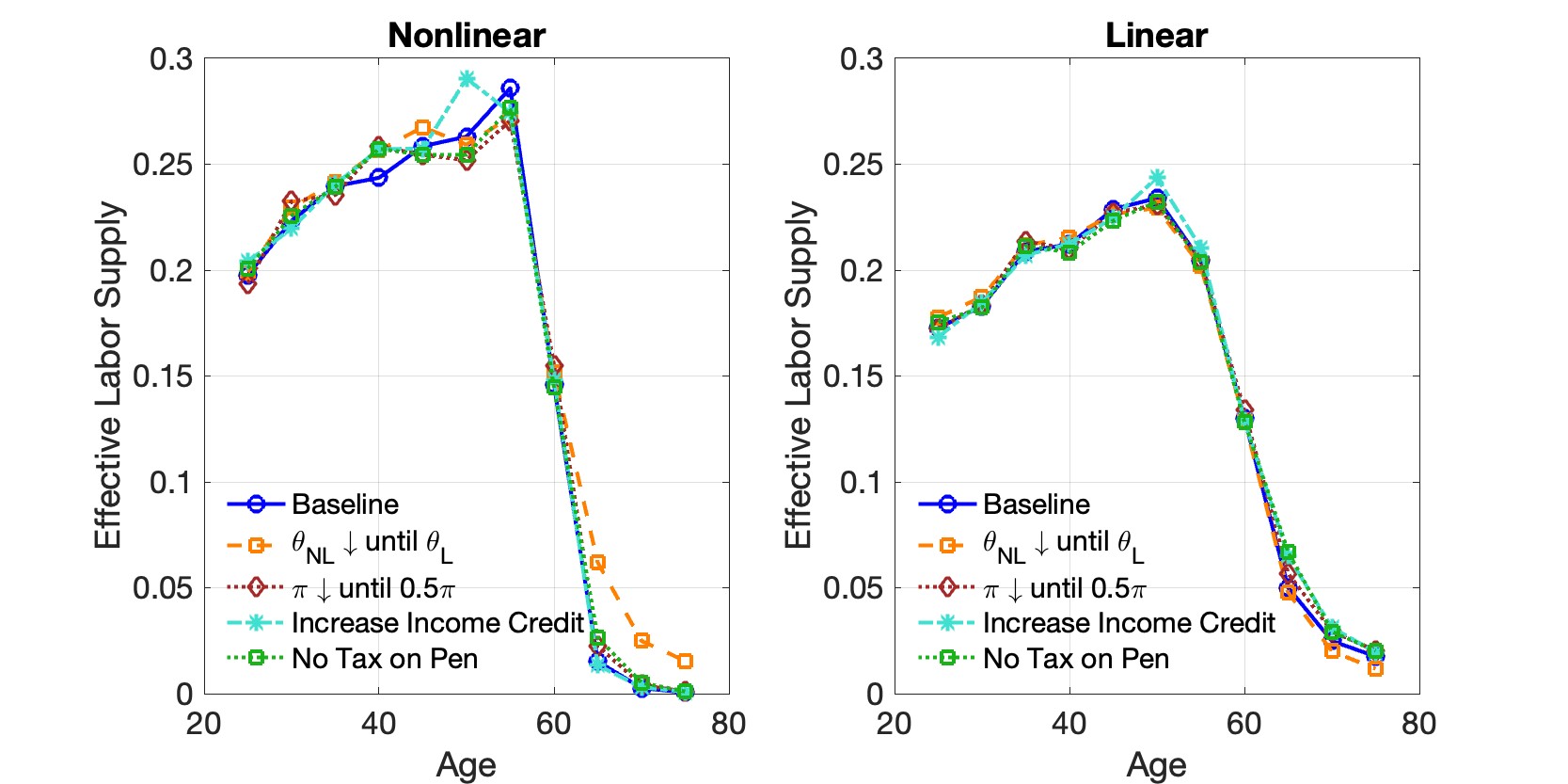}
            \caption{Effective Labor Supply(Unconventional Policy)}
    	    \label{fig:EffLs_NL_org_ALL}
        \end{figure}
        \vspace{-1em}
\begin{figure}[H]
    \centering
    \includegraphics[width=1.10\linewidth]{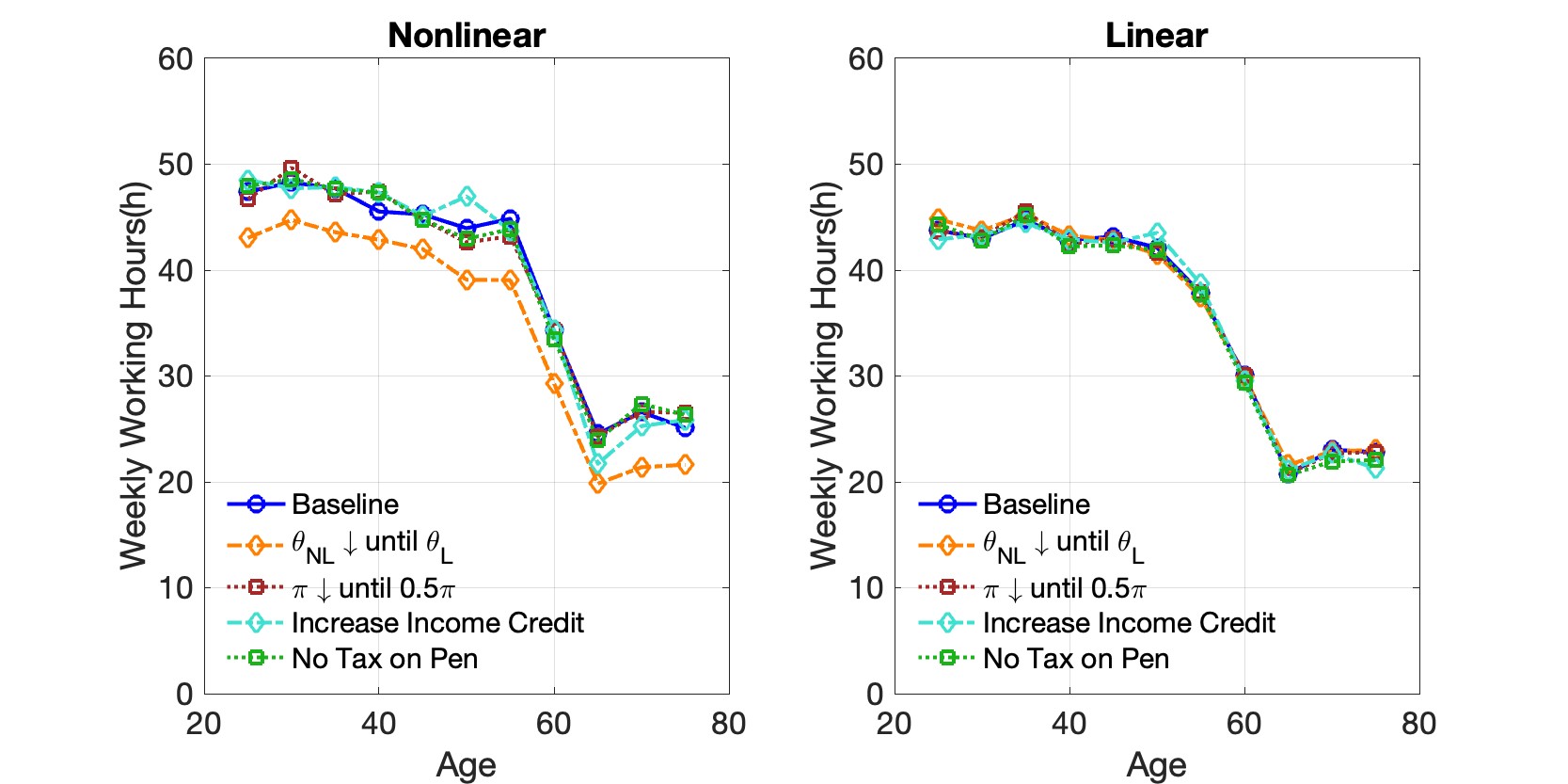}
    \caption{Working Hours per week (Unconventional Policy)}
     \label{fig:wwh_org_ALL}
\end{figure}
\vspace{-1em}
\begin{figure}[H]
    \centering
    \includegraphics[width=0.70\linewidth]{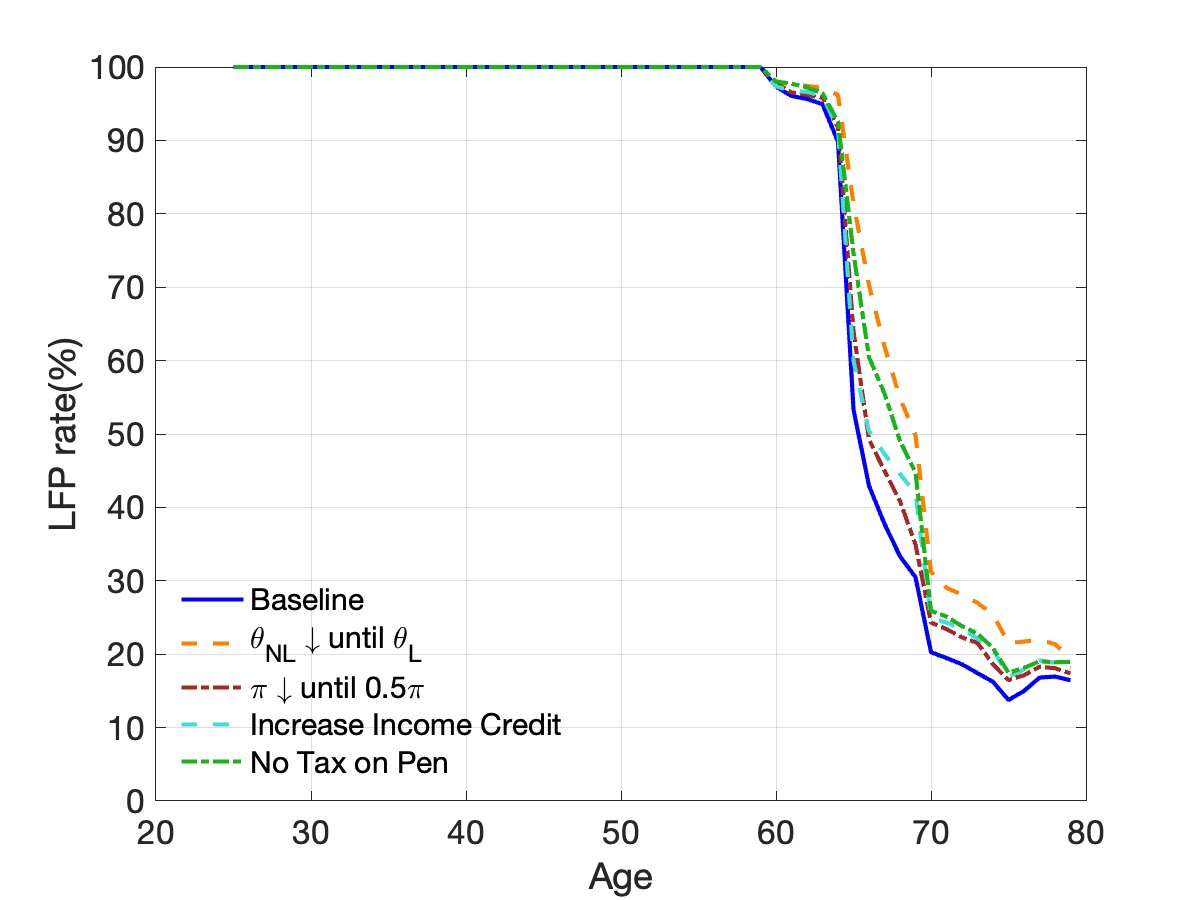}
    \caption{LFP (Unconventional Policy)}
     \label{fig:LFP_org_ALL}
\end{figure}
\vspace{-1em}
\begin{figure}[H]
    \centering
    \includegraphics[width=1.10\linewidth]{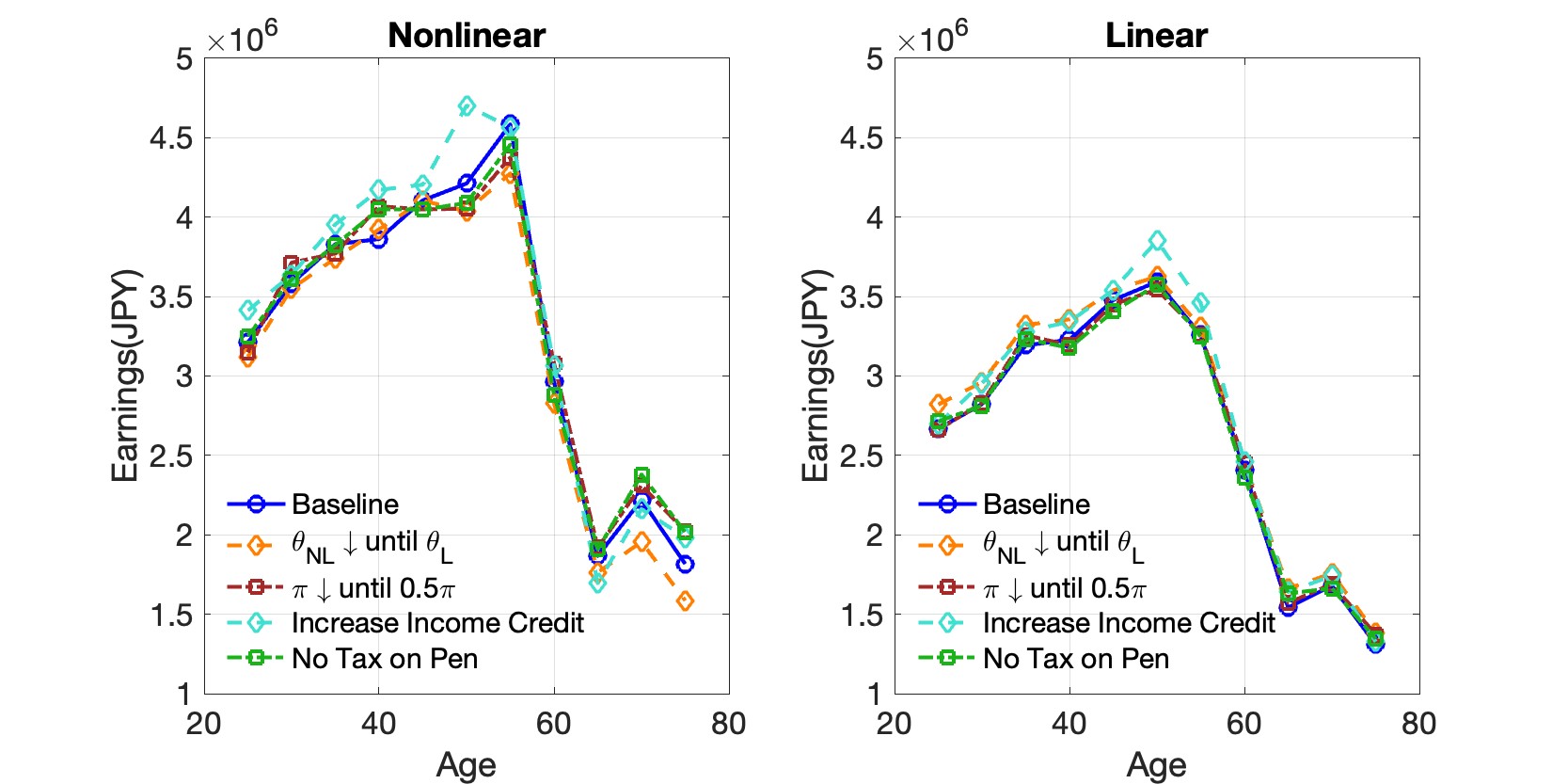}
    \caption{Earnings (Unconventional Policy)}
     \label{fig:earn_org_ALL}
\end{figure}
\vspace{-1em}
\begin{figure}[H]
    \centering
    \includegraphics[width=0.70\linewidth]{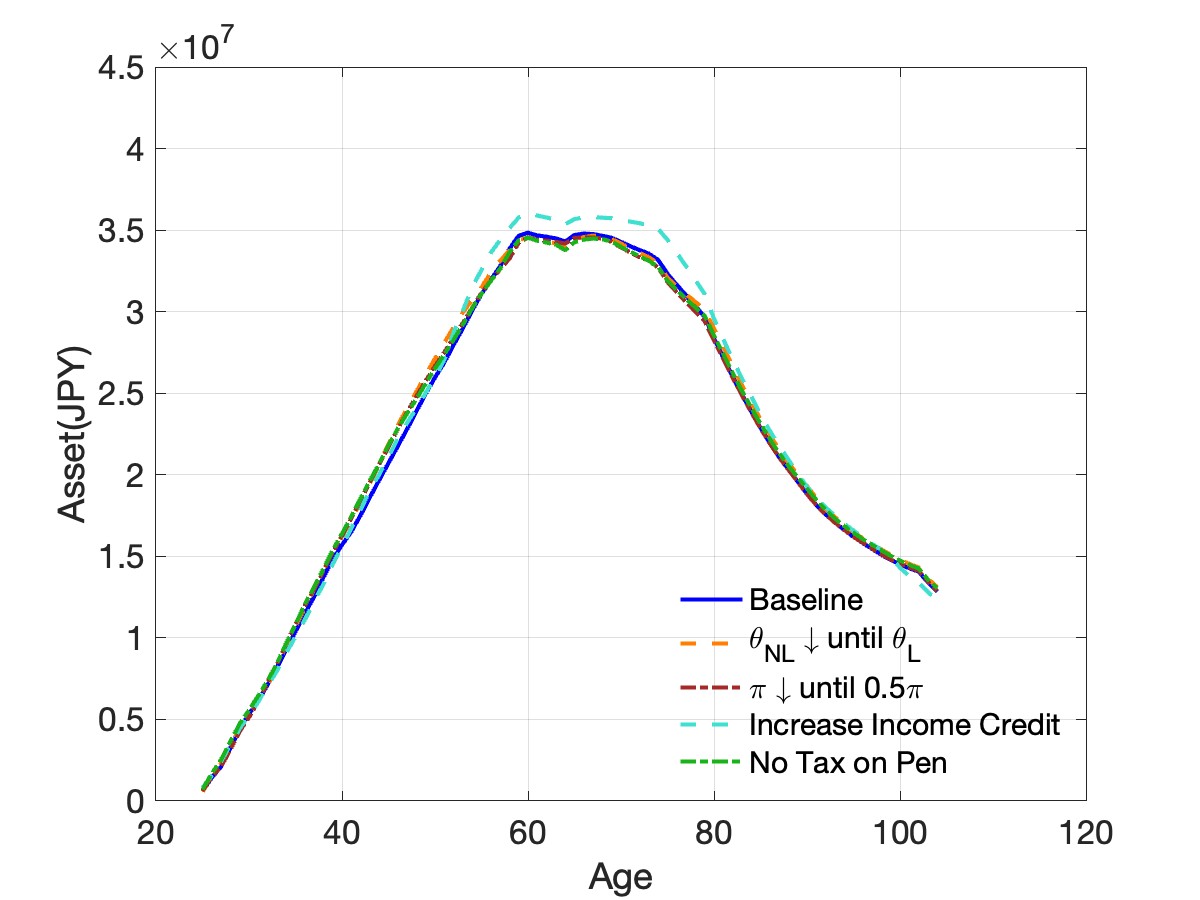}
    \caption{Asset (Unconventional Policy)}
     \label{fig:asset_org_ALL}
\end{figure}

\clearpage  

\begingroup
\refstepcounter{table}
\label{tab:Occ_list_JP}
\begingroup
\renewcommand{\arraystretch}{0.9}
\begin{longtable}{r p{9cm} p{4cm}}
\caption{Nonlinear/Linear Occupation List in Japan\label{tab:Occ_list_JP}} \\
\toprule
No. & Occupation (Small Category) & Nonlinear/Linear \\
\midrule
\endfirsthead
\toprule
No. & Occupation (Small Category) & Nonlinear/Linear \\
\midrule
\endhead
\bottomrule
\endfoot
\bottomrule
\endlastfoot
1 & Database SE & Nonlinear \\
2 & Pharmaceutical Sales & Nonlinear \\
3 & Banking Sales & Nonlinear \\
4 & Research and Development(Chemistry) & Nonlinear \\
5 & Product Development, Merchandiser & Nonlinear \\
6 & Infrastructure Engineer & Nonlinear \\
7 & Customer Engineer & Nonlinear \\
8 & Machinery Sales & Nonlinear \\
9 & System, IT Consultant & Nonlinear \\
10 & Public Health Nurse, Midwife & Nonlinear \\
11 & Human Resources & Nonlinear \\
12 & Web Designer & Nonlinear \\
13 & Research and Development(Biotechnology) & Nonlinear \\
14 & Technical Development(Construction, Civil Engineering, Plant, Equipment) & Nonlinear \\
15 & Doctor, Dentist, Veterinarian & Nonlinear \\
16 & Marketing & Nonlinear \\
17 & Business Planning & Nonlinear \\
18 & Store Development, Other Planning, Promotion Office Professional & Nonlinear \\
19 & Pharmacist & Nonlinear \\
20 & Purchasing, Materials & Nonlinear \\
21 & Communication, Network Engineer & Nonlinear \\
22 & Railway Operator, Telephone Operator, Mail Delivery, etc. & Nonlinear \\
23 & Programmer & Nonlinear \\
24 & Civil Engineering Design & Nonlinear \\
25 & Manager(Professional) & Nonlinear \\
26 & Other Financial Specialist Occupation & Nonlinear \\
27 & Manager(Service) & Nonlinear \\
28 & Planning, Sales Promotion & Nonlinear \\
29 & Support Engineer(Software) & Nonlinear \\
30 & Legal Affairs & Nonlinear \\
\pagebreak \\[0pt]
31 & Other Internet-related Technical Occupation & Nonlinear \\
32 & Other SE & Nonlinear \\
33 & Teacher, Lecturer, Instructor, Interpreter, etc. & Nonlinear \\
34 & Manager(Administrative) & Nonlinear \\
35 & Nurse(including Assistant Nurse) & Nonlinear \\
36 & Radiologic Technologist, Clinical Laboratory Technician, Dental Technician, Physical Therapist, etc. & Nonlinear \\
37 & Other Office Worker & Nonlinear \\
38 & Web Application Development & Nonlinear \\
39 & General Affairs & Nonlinear \\
40 & Food Sales & Nonlinear \\
41 & Author, Journalist, Editor, Proofreader, etc. & Nonlinear \\
42 & Public Relation & Nonlinear \\
43 & Business & Nonlinear \\
44 & Inventory Management & Nonlinear \\
45 & Electrical Circuit Design & Nonlinear \\
46 & Research and Development(Food) & Nonlinear \\
47 & Product Management & Nonlinear \\
48 & Other Research and Development & Nonlinear \\
49 & Welfare Counseling Specialist, Childcare Worker, Caregiver, etc. & Nonlinear \\
50 & Other Advertising, Publishing, Media-Related Professional Position & Nonlinear \\
51 & Receptionist & Nonlinear \\
52 & Product Planning & Nonlinear \\
53 & Architectural Design & Nonlinear \\
54 & Other Technical Occupation & Nonlinear \\
55 & Research and Development(Machinery) & Nonlinear \\
56 & Administrative Management & Nonlinear \\
57 & Metal, Machinery, Electrical, Automobile Manufacturing, Production, Repair Worker & Nonlinear \\
58 & Real Estate Sales & Nonlinear \\
59 & Medical Administration & Nonlinear \\
60 & Other Building, Civil Engineering, Surveying Technicians & Nonlinear \\
\pagebreak \\[0pt]
61 & Self-Defense Officer, Police Officer, Security Guard, etc. & Nonlinear \\
62 & Web Producer, Director, Planner & Nonlinear \\
63 & Telecommunications Technician & Nonlinear \\
64 & Sales Administration & Nonlinear \\
65 & Other General Office Professional & Nonlinear \\
66 & Development(Software-related Occupation) & Linear \\
67 & Unclassified Occupation & Linear \\
68 & Certified Public Accountant, Tax Accountant, etc. & Linear \\
69 & Keypuncher, Computer Operator, etc. & Linear \\
70 & Other Unclassified Service Professional & Linear \\
71 & Other Manufacturing Worker & Linear \\
72 & Cleaning, Delivery, Warehouse Work, etc. & Linear \\
73 & Finance, Accounting & Linear \\
74 & Manager(Business) & Linear \\
75 & Manager(Sales) & Linear \\
76 & Service Staff(Gas Station) & Linear \\
77 & Labor Affairs & Linear \\
78 & Other Electrical, Electronic, Mechanical Design-Related Professions & Linear \\
79 & Building, Parking Lot, Condominium, Boiler Management & Linear \\
80 & Nutritionist, Masseur, Counselor, etc. & Linear \\
81 & Other Customer Service and Serving Occupation & Linear \\
82 & Lawyer, Patent Agent, Judicial Scrivener, etc. & Linear \\
83 & Secretary & Linear \\
84 & Other Engineer & Linear \\
85 & Housekeeper, Home Helper, etc. & Linear \\
86 & Telephone Operator & Linear \\
87 & Telecommunications Sales & Linear \\
88 & DTP Operator & Linear \\
89 & Other Sales & Linear \\
90 & Store Clerk, Fashion Advisor, Cashier & Linear \\
\pagebreak \\[0pt]
91 & Food, Daily Necessities Manufacturing, Production Worker & Linear \\
92 & Driver(Van, Wagon) & Linear \\
93 & Other Cooking Professionals, Bartender & Linear \\
94 & Other Printing-related Specialist Occupation & Linear \\
95 & Store Manager & Linear \\
96 & Other Life and Hygiene Service Professional & Linear \\
97 & Other Construction, Civil Engineering, Mining Worker & Linear \\
98 & Manager(Technical) & Linear \\
99 & Management and Accounting Consultant, etc. & Linear \\
100 & Facility Construction Site Management, Site Supervisor, Construction Management & Linear \\
101 & Systems Sales & Linear \\
102 & Arrangement Operation & Linear \\
103 & Procurement & Linear \\
104 & Fashion-related Occupation & Linear \\
105 & Driver(Bus) & Linear \\
106 & Manager(Other) & Linear \\
107 & Mechanical Design & Linear \\
108 & Driver(Truck) & Linear \\
109 & Research and Development(Electrical, Electronic) & Linear \\
110 & Accommodation Services & Linear \\
111 & Machinery Maintenance & Linear \\
112 & Waiter, Waitress & Linear \\
113 & Farming, Landscaping, Livestock Worker, Forestry, Fisheries Worker & Linear \\
114 & Insurance Sales & Linear \\
115 & Construction Worker(Construction Worker) & Linear \\
116 & Real Estate Mediator, Salesperson, Insurance Agent, etc. & Linear \\
117 & Photographer & Linear \\
118 & Barber, Beautician & Linear \\
119 & Esthetician & Linear \\
120 & Driver(Taxi, Limousine) & Linear \\
\pagebreak \\[0pt]
121 & Western Cuisine Chef & Linear \\
122 & Construction Worker(Facility Construction Worker) & Linear \\
123 & Civil Engineering Site Management, Site Supervisor, Construction Supervision & Linear \\
124 & Other Design & Linear \\
125 & Electrical, Electronic Equipment Sales & Linear \\
126 & Automobile, Motorcycle Mechanic & Linear \\
127 & Hall Staff(Pachinko, Amusement Arcade) & Linear \\
128 & Trade Administration & Linear \\
129 & Japanese Cuisine Chef, Sushi Chef & Linear \\
130 & Construction Site Management, Site Supervisor, Construction Supervision & Linear \\
131 & Supervisor & Linear \\
132 & Character, CG Designer & Linear \\
133 & Construction Worker(Civil Engineer) & Linear \\
134 & Control SE & Linear \\
135 & Control Design & Linear \\
\end{longtable}
\endgroup

\endgroup

\end{document}